\documentclass[twocolumn,aps,pra,showpacs,raggedbottom,nobalancelastpage,amssymb,superscriptaddress,longbibliography]{revtex4-1}

\usepackage{graphicx}
\usepackage{amsmath}
\usepackage{amssymb}
\usepackage{bm}
\usepackage[usenames]{color}
\usepackage{amsfonts}
\usepackage{appendix}

\usepackage{comment}

\usepackage[colorlinks=true , citecolor=blue,urlcolor=blue]{hyperref}

\newcommand{\ie}{{i.e.}}

\newcommand{\e}{\textrm{e}}

\newcommand{\qbf}{\mathbf{q}}
\newcommand{\qp}{\mathbf{q}_{\perp}}
\newcommand{\ah}{\hat{a}}
\newcommand{\ahd}{\hat{a}^{\dagger}}
\newcommand{\Eav}{E_{\mathrm{av}}}
\newcommand{\Jeff}{J_{\mathrm{eff}}}

\newcommand{\Jnature}{Nature (London)}
\newcommand{\Jnatphys}{Nature Phys.}

\newcommand{\Jscience}{Science}

\newcommand{\Jprl}{Phys. Rev. Lett.}

\newcommand{\Jpra}{Phys. Rev. A}
\newcommand{\Jprb}{Phys. Rev. B}

\newcommand{\Jprd}{Phys. Rev. D}

\newcommand{\Jprx}{Phys. Rev. X}
\newcommand{\Jrmp}{Rev. Mod. Phys.}

\newcommand{\Jepl}{Europhys. Lett.}
\newcommand{\Jnjp}{New J. Phys.}

\newcommand{\JRepProgPhys}{Rep. Prog. Phys.}

\newcommand{\Jadvphys}{Adv. Phys.}

\definecolor{Nathanpurple}{rgb}{0.5,0.,0.5}

\definecolor{Nathanblue}{rgb}{0.96,0.24,0.00}

\definecolor{Nathanred}{rgb}{0.06,0.24,0.90}

\def\be{\begin{equation}}
\def\ee{\end{equation}}

\def\bs#1{\boldsymbol{#1}}

\begin{document}
\title{Parametric Instability Rates \\ in Periodically-Driven Band Systems}
\author{S. Lellouch}
\email[]{samuel.lellouch@ulb.ac.be}
\affiliation{Center for Nonlinear Phenomena and Complex Systems, Universit\'e Libre de Bruxelles, CP 231, Campus Plaine, B-1050 Brussels, Belgium}
\author{M. Bukov}
\affiliation{Department of Physics, Boston University, 590 Commonwealth Ave., Boston, MA 02215, USA}
\author{E. Demler}
\affiliation{Department of Physics, Harvard University, Cambridge, MA 02138, USA}
\author{N. Goldman}
\email[]{ngoldman@ulb.ac.be}
\affiliation{Center for Nonlinear Phenomena and Complex Systems, Universit\'e Libre de Bruxelles, CP 231, Campus Plaine, B-1050 Brussels, Belgium}
\begin{abstract}
This work analyses the dynamical properties of periodically-driven band models. Focusing on the case of Bose-Einstein condensates, and using a meanfield approach to treat inter-particle collisions, we identify the origin of dynamical instabilities arising due to the interplay between the external drive and interactions. We present a widely-applicable generic numerical method to extract instability rates, and link parametric instabilities to uncontrolled energy absorption at short times. Based on the existence of parametric resonances, we then develop an analytical approach within Bogoliubov theory, which quantitatively captures the instability rates of the system, and provides an intuitive picture of the relevant physical processes, including an understanding of how transverse modes affect the formation of parametric instabilities. Importantly, our calculations demonstrate an agreement between the instability rates determined from numerical simulations, and those predicted by theory. To determine the validity regime of the meanfield analysis, we compare the latter to the weakly-coupled conserving approximation. The tools developed and the results obtained in this work are directly relevant to present-day ultracold-atom experiments based on shaken optical lattices, and are expected to provide an insightful guidance in the quest for Floquet engineering.
\end{abstract}

\date{\today}
\maketitle

\section{Introduction}

	Periodically-driven systems have been widely studied over the past decades, revealing rich and exotic phenomena in a large class of physical systems. Some of the original applications include examples from classical mechanics, e.g.~the periodically-kicked rotor~\cite{casati1979} and the Kapitza pendulum~\cite{kapitza1951}, which display fascinating effects, such as dynamical stabilization/destabilization~\cite{landau1969} or integrability to chaos transitions. More recently, studies of their quantum counterparts have led to remarkable applications in a wide range of physical platforms, such as ion traps \cite{Leibfried:2003}, photonic crystals~\cite{Lu:2014Review}, irradiated graphene~\cite{cayssol2013,oka_09,kitagawa_11} and ultracold gases in optical lattices~\cite{Goldman:2016Review,Eckardt:2016Review}.

	A major reason for the renewed interest in periodically-driven quantum systems is the fact that they constitute the essential ingredient for \textit{Floquet engineering}~\cite{cayssol2013,goldman2014b,bukov2014,Goldman:2016Review,Eckardt:2016Review}. To see this, recall that periodically-driven quantum systems obey Floquet's theorem~\cite{shirley_65}, which implies that the time-evolution operator $\hat U(t,0)$ of a system, whose dynamics is generated by the Hamiltonian $\hat H(t+T)\!=\! \hat H(t)$, can be written as~\cite{rahav_03_pra,goldman2014b}
		\begin{align}
			\label{eq:Floquet_thm}
			\hat U(t,0) &= \mathcal{T} \mathrm{exp}\left(-i\int^t_0\mathrm{d}t' \hat H(t')\right) \\
			&= \mathrm{e}^{-i \hat K_\mathrm{kick}(t)}\mathrm{e}^{-it \hat H_\mathrm{eff}}\mathrm{e}^{i \hat K_\mathrm{kick}(0)},\notag
		\end{align}  
		where $\mathcal{T}$ denotes time-ordering. Here we introduced the time-independent effective Hamiltonian $\hat H_\mathrm{eff}$, as well as the time-periodic ``kick" operator $\hat K_\mathrm{kick}(t+T)\!=\! \hat K_\mathrm{kick}(t)$, which has zero average over one driving period~\cite{goldman2014b}.
		It then follows that the stroboscopic time-evolution ($t\!=\!NT$, $N \in \mathbb{Z}$) of such systems is governed by
		\begin{align}
			&\hat U(NT,0) = \mathrm{e}^{-i \hat K_\mathrm{kick}(0)}\mathrm{e}^{-i N T \hat H_\mathrm{eff} }\mathrm{e}^{i \hat K_\mathrm{kick}(0)} =  \mathrm{e}^{-i N T \hat H_\mathrm{F} }, \notag \\
			& \hat H_\mathrm{F}=\mathrm{e}^{-i \hat K_\mathrm{kick}(0)} \hat H_\mathrm{eff}\mathrm{e}^{i \hat K_\mathrm{kick}(0)}.\notag
		\end{align}
		Hence, up to a change of basis~\footnote{This change of basis is important to keep in mind when preparing the system in a given initial state.}, the stroboscopic time-evolution is completely described by the effective Hamiltonian $\hat H_\mathrm{eff}$, which can be designed by suitably tailoring the driving protocol~\cite{goldman2014b,goldman2015a,bukov2014,eckardt2015,mikami_15}. Moreover, we note that the dynamics taking place within each period of the drive, the so-called \emph{micro-motion}, is entirely captured by the kick operator $\hat K_\mathrm{kick}(t)$. On the theoretical side, perturbative methods to compute both the effective Hamiltonian and the micromotion operators have been developed in the form of an inverse-frequency expansion~\cite{rahav_03_pra,rahav_03,goldman2014b,goldman2015a}, variants of which include the Floquet-Magnus~\cite{blanes_09,bukov2014}, the van Vleck~\cite{eckardt2015}, and the Brillouin-Wigner~\cite{mikami_15} expansions.

	On the experimental side, Floquet engineering has been used to explore a wide variety of physical phenomena, ranging from dynamical trapping of ions in Paul traps~\cite{Leibfried:2003} to one-way propagating states in photonic crystals~\cite{Lu:2014Review}. In the context of ultracold quantum gases, this concept soon resonated with the idea of quantum simulation~\cite{Georgescu:2014}, as it became clear that Floquet engineering could offer powerful schemes to reach novel models and properties, typically inaccessible in conventional condensed matter materials~\cite{Goldman:2016Review,Eckardt:2016Review}. For instance, shaken optical lattices have been used to explore dynamical localization~\cite{dunlap_86,dunlap_88,holthaus1992,lignier2007,eckardt2009}, photon-assisted tunneling~\cite{eckardt2007,sias2008} and driven-induced superfluid-to-Mott-insulator transitions~\cite{eckardt2005,ciampini_11,zenesini_09,zenesini_10}. In the field of disordered quantum systems, the periodically-kicked rotor has been exploited as a flexible and successful simulator for Anderson localization~\cite{fishman1982,casati1989,lopez2012}. More recently, a series of experiments implemented shaken or time-modulated optical-lattice potentials in view of designing artificial gauge fields for neutral atoms~\cite{aidelsburger2011,aidelsburger2013,miyake2013,aidelsburger2014,jotzu2014,kennedy2015,creffield2016}, some of which led to non-trivial topological band structures~\cite{Goldman:2016Review}, or frustrated quantum models~\cite{eckardt2010,struck2011}. Other recent applications of Floquet engineering in cold atoms include the experimental realization of state-dependent lattices~\cite{Jotzu:2015PRL} and dynamical topological phase transitions~\cite{flaschner2016}, as well as proposals to dynamically stimulate quantum coherence~\cite{robertson2011}, or to create sub-wavelength optical lattices~\cite{nascimbene2015}, synthetic dimensions~\cite{price2016} or pure Dirac-type dispersions~\cite{goldman2014b,budich2016}. In an even broader context, Floquet engineering is at the core of photo-induced superconductivity~\cite{goldstein2015, fausti2011, knap2015} and topological insulators~\cite{lindner2011}, and time crystals~\cite{sacha2015,else2016,zhang2016}.
	
	\begin{figure}[t!]
		\centering
		\includegraphics[width=0.5\textwidth]{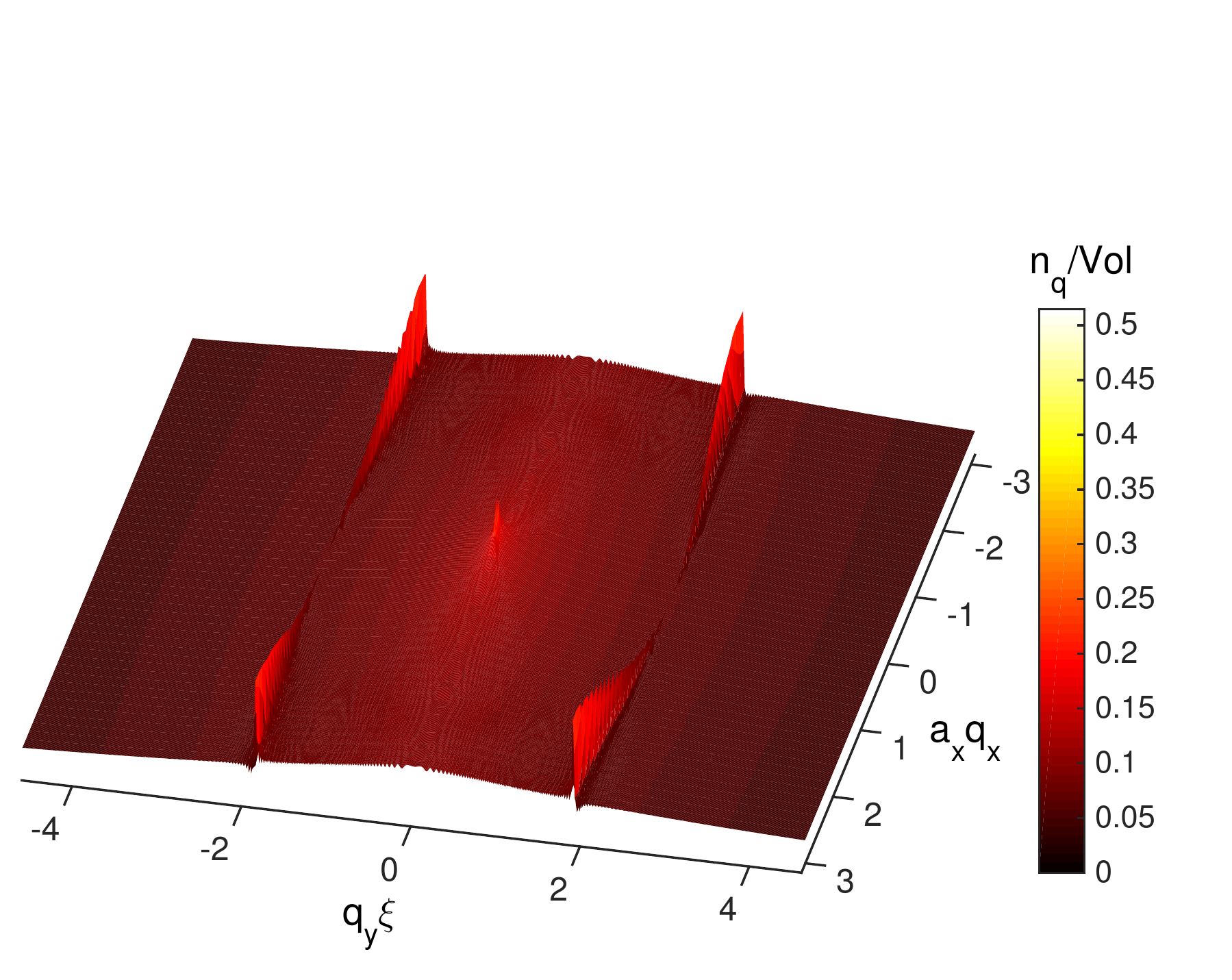}
		\caption{\label{fig:nk_intro} The quasiparticle momentum distribution $n_q$ of a 2D shaken BEC trapped in a one-dimensional lattice (along $x$), and allowed to move along a continuous (tube) transverse direction ($y$), develops unique resonance structures in the presence of a periodic drive. Dominating the dynamics in the early stage of evolution, they signal the onset of instabilities and provide a clear experimental signature of the parametric resonance phenomenon explained in this work. The momentum distribution is shown after $60$ driving periods ($t=60T$). Here, the distribution $n_q$ is divided by the system's volume Vol; $\xi$ denotes the BEC healing length and $a_x$ is the lattice spacing constant. The precise model and the corresponding parameters are the same as in Fig.~\ref{fig:nk_WCCA_vs_lin}.}  
	\end{figure}
	
	Today, the theory of periodically-driven quantum systems (including the resulting artificial gauge fields and topological band structures~\cite{Goldman:2016Review,Eckardt:2016Review}) is well-established for single particles, and there is a strong need for incorporating inter-particle interactions in the description of such systems. Indeed, inter-particle interactions constitute an essential ingredient in simulating condensed-matter systems~\cite{Georgescu:2014}, and the relevance of many theoretical proposals strongly relies on the extent to which the systems of interest remain stable away from the non-interacting limit.  Moreover, in their quest for realizing new topological states of matter, experiments based on driven quantum systems are more than ever eager to consider interacting systems~\cite{anisimovas2015}; indeed, building on the successful optical-lattice implementation of flat energy bands with non-trivial Chern numbers~\cite{aidelsburger2014}, mastering interactions in this context would allow for the engineering of intriguing strongly-correlated states, such as fractional Chern insulators~\cite{grushin_14}. In this framework, we point out that a useful tool to derive the effective low-energy physics of strongly-correlated systems, namely the Schrieffer-Wolff transformation, was generalized to periodically-driven systems~\cite{bukov_15_SW}.

	However, until now, the presence of inter-particle interactions has led to complications. Recent experiments on time-modulated (``shaken") optical lattices~\cite{aidelsburger2011,aidelsburger2013,miyake2013,aidelsburger2014,pweinberg_15} have reported severe heating, particle loss and dissipative processes, whose origin presumably stems from a rich interplay between the external drive and inter-particle collisions. In this sense, understanding the role of interactions in driven systems, and its relation to heating processes and instabilities, is both of fundamental and practical importance. On the theory side, several complementary approaches have been considered to tackle the problem. 
	On the one hand, a perturbative scattering theory, leading to a so-called ``Floquet Fermi Golden Rule"~\cite{bilitewski_14,bilitewski2015}, has been developed to estimate heating and loss rates, and showed good agreement with the band-population dynamics reported in Ref.~\cite{aidelsburger2014}. Extensions to Bose-Einstein condensates (BEC) have been proposed~\cite{choudhury2014,choudhury2015a}, however, to the best of our knowledge, the resulting loss rates have not yet been confirmed by experiments nor through numerical simulations. 
 On the other hand, various studies analyzed the dynamical instabilities that occur in BEC trapped in moving, shaken or time-modulated optical lattices~\cite{fallani2004,modugno2004,zheng2004,Kramer:2005,tozzo2005,desarlo2005,creffield2009}. For instance, Ref.~\cite{creffield2009} analyzed such dynamical instabilities through a numerical analysis of the Bogoliubov-de Gennes equations; however, no link was made with heating processes and the approach lacked a conclusive comparison with analytics. Importantly, several studies pointed out the \textit{parametric} nature of those instabilities~\cite{Kramer:2005,tozzo2005,bukov2015,citro2015}. Interestingly, we note that parametric instabilities, and their impact on energy absorption and thermalization processes, have also been studied theoretically in a wider context, ranging from Luttinger liquids~\cite{bukov_12} to cosmology~\cite{berges2014,berges2015,pineiro2015}. In the context of driven optical lattices, the parametric amplification of scattered atom pairs, through phase-matching conditions involving the band structure and the drive, has also been identified~\cite{gemelke2005,campbell2006,anderson2016}.
	Altogether, there is a strong need for a combined analytical and numerical study of driven optical-lattice models, which would provide analytical estimates of instability and heating rates supported by numerical simulations.

	\subsection*{Scope of the paper}
	
	The object of the present paper is to explore the physics of periodically-driven bosonic band models, where inter-particle interactions are treated within mean-field (Bogoliubov) theory. Indeed, as in Refs.~\cite{creffield2009,choudhury2014,choudhury2015a,citro2015}, we will assume that the driven atomic gas forms a Bose-Einstein condensate (BEC), and we will analyze the properties of the corresponding Bogoliubov excitations in response to the external drive. To do so, we develop a generic and widely-applicable method to investigate the short-time dynamics of periodically-driven bosonic systems and to extract the corresponding instability rates, both numerically and analytically.
	
	We will first focus on a shaken one-dimensional (1D) lattice model, and extensively explore the occurrence of instabilities and heating in this system, before considering more elaborate models. In a previous study on a similar model~\cite{creffield2009}, a dynamical instability of the condensate was found to occur above a critical interaction strength, which was numerically calculated as a function of the drive amplitude $K$ and frequency $\omega$. This allowed the author to numerically map out the stability boundaries between the stable and unstable phases. More recently, Ref.~\cite{bukov2015,citro2015} suggested that such boundaries could be understood in terms of a simple energy-conservation criterion, which could be formulated within a parametric-instability analysis.  In this work, we offer a significant advance in the description and understanding of the instabilities that affect bosonic periodically-driven systems, both qualitatively and quantitatively, as we summarize below:
	
	\begin{figure}[!t]
		\begin{center}
			\includegraphics[width=9cm]{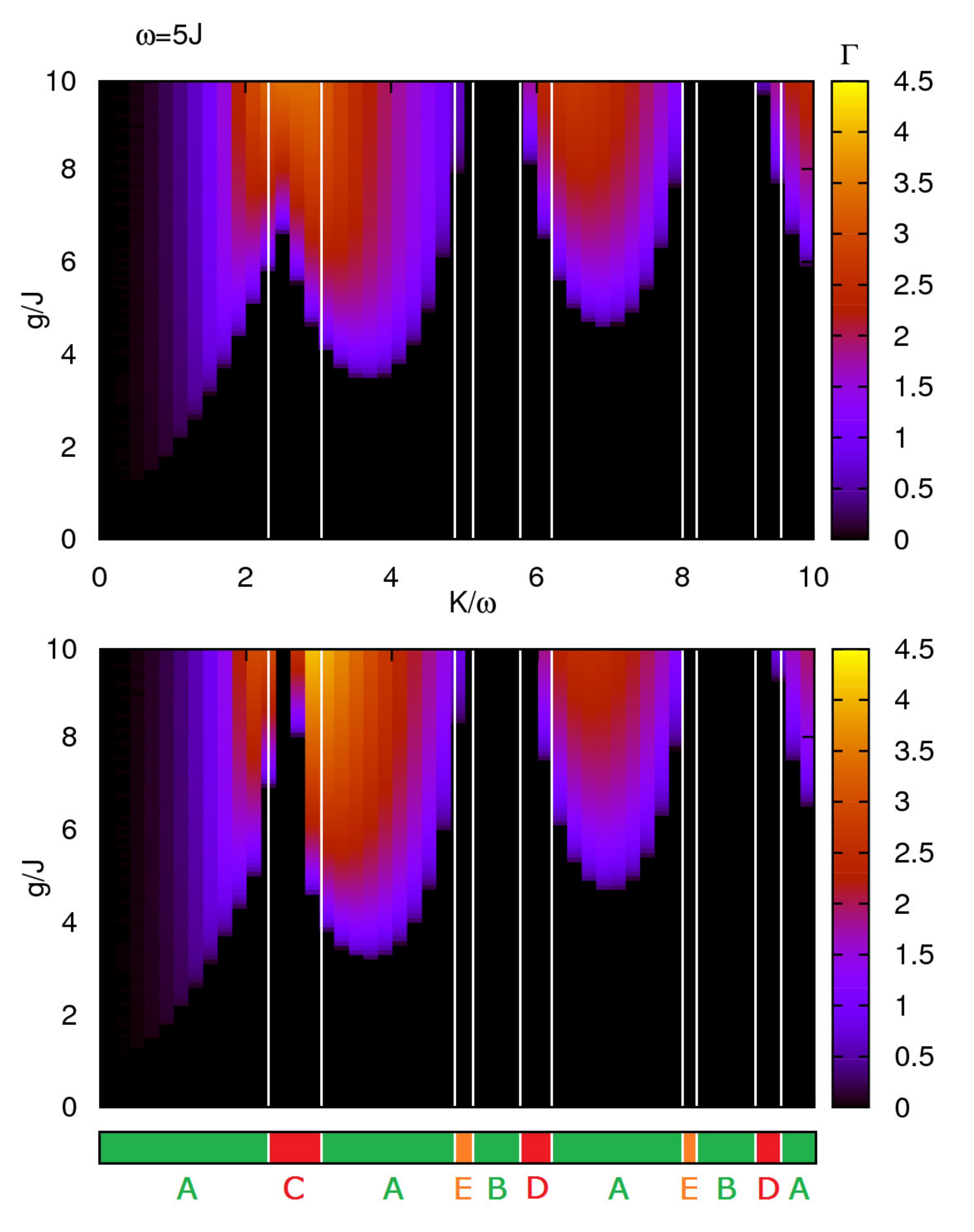}
		\end{center}
		\caption{Numerical (top) and analytical (bottom) stability diagram for the model described by Eq.~\eqref{eq:GPE}. The instability rate $\Gamma$ (expressed in units of the hopping amplitude $J$, see text) is plotted as a function of the modulation amplitude $K/\omega$ [defined in Eq.~\eqref{eq:GPE}] and interaction strength $g$, for a driving frequency $\omega\!=\!5J$.
			The vertical lines and letters summarize the agreement and disagreement regions, as discussed in Sec.~\ref{sec:AnaRes}:
			A: Away from the zeros of $\mathcal{J}_0(K/\omega)$ and $\mathcal{J}_2(K/\omega)$, the analytics successfully capture the instability rates extracted from numerics.
			B: Agreement zone near the (close) zeros of $\mathcal{J}_0(K/\omega)$ and $\mathcal{J}_2(K/\omega)$, where both analytics and numerics predict a stable behaviour.
			C/D: Close to a zero of $\mathcal{J}_0(K/\omega)$, the analytical perturbative approach breaks down.
			E: Potential quantitative disagreement when the contribution of the second harmonic has a weak amplitude:~the instability is then partly due to higher harmonics (see Sec.~\ref{sec:anaEff}).
			\label{fig:2Diags}
		}
	\end{figure}
		\begin{enumerate}
		\item We show how to obtain the instability rates as a function of the model parameters, both numerically and analytically, and we establish their relation to the growth of physical quantities in the system, such as the energy, thus providing a link between dynamical instability and heating. Moreover, we identify different instability regimes in the system, which are associated with different timescales and characterized by a different behaviour in the growth of physical quantities. We introduce several observables that reveal clear signatures of these instabilities, among which the non-condensed (depleted) fraction and the momentum distribution of quasiparticle modes [see Fig.~\ref{fig:nk_intro}]; we discuss how they could be observed in current ultracold-atom experiments.
		
		\item We present a full numerical solution of the mean-field problem, determining the stability diagram of the model from first principles [see Fig.~\ref{fig:2Diags}]. We stress that the quantitative character and versatility of our procedure make it general enough to be applicable to a wide class of periodically-driven systems, including those involving resonant modulations, higher dimensional lattices, various geometries (e.g.~square or honeycomb lattices, continuous space), and spin-dependent lattices. Moreover, it can also be enriched to incorporate a full experimental sequence, e.g.~including adiabatic state preparation~\cite{pweinberg_15}. Our results are expected to determine experimentally-favorable regimes by providing \textit{ab initio} numerical (and analytical) estimates for the instability rates in a variety of experimental configurations.

		\item We perform an analytical treatment of the problem based on the existence of parametric resonances in the Bogoliubov-De Gennes equations. In Ref.~\cite{bukov2015}, following a weak-coupling conserving approximation, it was argued that a driven-lattice model was stable against parametric resonance provided that the drive frequency satisfies $\omega\!>\!2W_{\mathrm{eff}}$, with $W_{\mathrm{eff}}$ the bandwidth associated with $\hat H_\mathrm{eff}$ in the Bogoliubov approximation. Intuitively, this can be understood by recalling that the elementary excitations of the system (i.e.~the Bogoliubov phonons) are always created in pairs, and thus, whenever the drive frequency exceeds twice the maximum possible excitation energy, stability is ensured by energy conservation. However, our rigorous analytical derivation reveals that this simple criterion needs to be revised:~As we demonstrate, an accurate qualitative explanation requires taking into account both higher-order photon-absorption processes, as well as the  detuning from  resonance within the parametric-resonance treatment~\cite{book_LL}. This analysis allows us, for the first time, to derive the functional dependence of the instability growth rate on the model parameters, and ultimately, to understand all features of the stability diagram, see~Fig.~\ref{fig:2Diags}.

		\item We extend the stability analysis to two-dimensional (2D) models and study the effects due to transverse directions, both considering  the case of a transverse lattice (i.e.~a discrete transverse degree of freedom), and that of transverse tubes (i.e.~a continuous transverse degree of freedom). In the case of continuous degrees of freedom, a theoretical simplification in the equations allows one to obtain very simple analytical formulas for instability rates, which are in perfect agreement with our full numerical simulations. Moreover, this study provides a clear physical picture of how instabilities are enhanced by the presence of transverse modes in the system, as already anticipated in Refs.~\cite{choudhury2015a,bilitewski2015}. More generally, we argue that our theory should provide guidance for experiments; it illustrates, for instance, the advantage to work at high frequency and the necessity of using a strong transverse confinement to reduce parametric instabilities. We also include a discussion of finite-size effects, making a link between the physics of double-wells and that of optical lattices.  
	\end{enumerate}

	\subsection*{Outline}
	
	This paper is organized as follows. In Sec.~\ref{sec:Imodel}, we derive the general mean-field equations that will constitute the basis of our analysis. Section \ref{sec:Num} is devoted to the numerical solution of those equations, including details on the general procedure used and a presentation of the results, among which the stability diagram of the model under consideration, the identification of several timescales and instability regimes in the problem and the dynamics of various physical observables. In Sec.~\ref{sec:ana}, we perform an analytical treatment of the problem:~mapping the Bogoliubov equations on a parametric oscillator (Sec.~\ref{sec:anaPO}), we build an effective model from which analytical instability rates can be inferred (Sec.~\ref{sec:anaEff}). The analytical results are presented in Sec.~\ref{sec:AnaRes}, including a discussion of the validity regimes of the approach. The case of finite-size systems is presented in Sec.~\ref{sec:FiniteSize}. We discuss in Sec.~\ref{sec:2D} the case where a transverse direction is present (a lattice or a continuous one); this includes simple analytical formulas for instability rates as well as a physical understanding of the enhancement of instabilities by transverse modes.
	Finally, we discuss in Sec.~\ref{sec:BH} the application of our results to the weakly-interacting Bose-Hubbard model in the meanfield regime; in order to understand the role of non-linear processes and study the regime of validity of our linearized analysis, we employ a weak-coupling conserving approximation to study the leading-order (in the interaction strength) features of particle-conserving dynamics; we identify clear signatures of the instabilities (among which the non-condensed fraction and the momentum distribution of quasiparticles) that could be directly probed by current ultracold-atom experiments in modulated optical lattices.

	\section{Mean-field Equations and Stability Analysis}
	\label{sec:Imodel}
	
	Consider a Bose gas in a shaken 1D lattice, with mean-field interactions, governed by the time-dependent Gross-Pitaevskii equation (GPE)~\cite{Eckardt:2016Review,creffield2009}
		\begin{equation}
			i \partial_t a_n=-J \left (a_{n+1}+a_{n-1} \right)+K\cos(\omega t)na_n+U|a_n|^2 a_n,
			\label{eq:GPE}
		\end{equation}
		where $n$ labels the lattice sites, $J>0$ denotes the tunneling amplitude for nearest-neighbour processes, $U>0$ is the on-site interactions strength, and where the time-periodic modulation has an amplitude $K$ and a frequency $\omega=2\pi/T$. Note that we set $\hbar\!=\!1$ throughout this work.
		
		Equation~\eqref{eq:GPE} describes a wide variety of physical systems. On the one hand, it is expected to provide a good description for the shaken 1D Bose-Hubbard model~\cite{Eckardt:2016Review}, when treated in the weakly-interacting regime:~in this case, the time-modulated system in Eq.~\eqref{eq:GPE} can be realized by mechanically modulating an optical lattice filled with weakly-interacting bosonic atoms~\cite{Eckardt:2016Review}; see also Sec.~\ref{sec:BH} for further discussion. On the other hand, some physical systems are ``\textit{exactly}" described by Eq.~\eqref{eq:GPE}:~for instance, non-linear optical systems~\cite{carusotto2013}, including helical photonic crystal~\cite{Lu:2014Review}. In all cases, Eq.~\eqref{eq:GPE} defines a close self-consistent problem, which constitutes the core of the present study. \\
	
	Similar to the analysis of Ref.~\cite{creffield2009}, we study the dynamical instabilities of a BEC described by Eq.~\eqref{eq:GPE}. To do so, we proceed along the following steps:\\
	\begin{itemize}
		\item[(i)] First, we determine the time-evolution of the condensate wavefunction $a_n^{(0)}(t)$, by solving the full time-dependent GPE and assuming that the initial state $a_n^{(0)}(t=0)$ forms a BEC. More precisely, our choice for the initial state corresponds to the solution of the static (effective) GPE:
			\begin{align}
				-\Jeff \left (a_{n+1}^{(0)}+a_{n-1}^{(0)} \right)&+U|a_n^{(0)}|^2 a_n^{(0)}=\mu a_n^{(0)}, \label{eq:GPEeff} \\
				 \Jeff\!&=\!J\mathcal{J}_0(K/\omega),\notag		
			\end{align}
			where we introduced  the effective tunneling amplitude $\Jeff$ renormalized by the drive~\cite{Eckardt:2016Review}, and where $\mathcal{J}_0$ denotes the zeroth-order Bessel function. For the model under consideration, we find that $a_n^{(0)}(t=0)$ is the Bloch state $e^{ip_0n}$ of momentum $p_0\!=\!0$ if $\mathcal{J}_0(K/\omega)\!>\!0$ (homogeneous condensate), and $p_0\!=\!\pi$ for $\mathcal{J}_0(K/\omega)\!<\!0$. 
			Note that such a choice takes into account the initial kick due to the launching of the modulation~\cite{goldman2014b}  \footnote{This is generically done by applying the kick operator~\cite{goldman2014b}  at time $t\!=\!0$. However, for this model, the action of the kick operator $K_\mathrm{kick}(0)$ reduces to a multiplication by a trivial phase when acting on a Bloch state, so that there is, in fact, no need to ``rotate" the initial state in this specific case. See also Sec.~\ref{sec:BH} for details about the effective Hamiltonian and the kick operator in the Bose-Hubbard model.}. Altogether, this choice for the initial state features both the effects of tunneling renormalization  and initial launching of the drive, which are both present in our model~\eqref{eq:GPE}. We emphasize that this prescription for the initial state is the only step of our calculations that relies on the existence of a well-defined high-frequency limit, as provided by the inverse-frequency expansion~\cite{rahav_03_pra,goldman2014b,goldman2015a}; indeed, all subsequent results are based on the full time-dependent equations. Note that it is not the purpose of this paper to discuss how to prepare the system in this initial state; one possibility is to use Floquet adiabatic perturbation theory, see Ref.~\cite{pweinberg_15,novicenko2016}, which is also the experimentally-preferred strategy.
			 
		\item[(ii)] Given the time-dependent solution for the condensate wave function $a_n^{(0)}(t)$, we analyse its stability by considering a small perturbation 
		\begin{equation}
		a_n(t)\!=\!a_n^{(0)}(t)[1\!+\!\delta a_n(t)],\label{para_text}
		\end{equation} 
		and linearizing the Gross-Piteavskii equation~\eqref{eq:GPE} in $\delta a_n$;  we refer to Appendix~\ref{sec:appBogoparam} for a discussion on the specific parametrization chosen in Eq.~\eqref{para_text}. This yields the time-dependent Bogoliubov-de Gennes equations, which take the general form
		\begin{equation}
			i \left( \begin{matrix} \dot{\delta a_n}  \\ \dot{\delta a}_n^{*} \end{matrix} \right) =\left( \begin{matrix} \mathcal{F}_n(t) & U |{a_n^{(0)}}|^{2} \\ -U |{a_n^{(0)}}|^{2} & -\mathcal{F}_n(t)^{*} \end{matrix} \right)\left( \begin{matrix} \delta a_n  \\ \delta a_n^{*} \end{matrix} \right),
			\label{eq:BdGE}
		\end{equation}
		where we introduced the operator $\mathcal{F}_n(t)$, whose action on $\delta a_n$ is defined by
		\begin{align}
			\mathcal{F}_n(t){\delta a_n}\equiv &-J\dfrac{\hat{L}(a^{(0)}{\delta a_n})_n}{a_n^{(0)}} + 2U|a_n^{(0)}|^2{\delta a_n} \notag \\
			&+K\cos(\omega t)n{\delta a_n}-i\dfrac{\dot{a}_n^{(0)}}{a_n^{(0)}}{\delta a_n},
			\label{eq:Fnt}
		\end{align}
		and where the discrete operator $\hat{L}$ is defined by $\hat{L}(\cdot)_n\equiv (\cdot) _{n+1}+(\cdot)_{n-1}$. 
		
	\end{itemize}
	
	At this stage, let us emphasize that the Bogoliubov equations~\eqref{eq:BdGE} contain the complete time dependence of the problem, which includes the effects related to the micro-motion (note that the BEC wavefunction $a_n^{(0)}(t)$ is computed exactly, not stroboscopically). Importantly, and as will become apparent below in Section~\ref{sec:ana}, it is the micro-motion (and not the time-averaged dynamics) that determines the stability of the system.
	
	The Bogoliubov equations of motion~\eqref{eq:BdGE} can be time-evolved over one driving period $T$, which allows one to determine the associated ``time-evolution" (propagator) matrix $\Phi(T)$. From this, we extract the ``Lyapunov" exponents $\epsilon_q$, which are related to the eigenvalues $\lambda_q$ of $\Phi(T)$ by $\lambda_q=\e^{-i\epsilon_q T}$. The appearance of Lyapunov exponents with positive imaginary parts thus indicates a dynamical instability~\cite{tozzo2005,creffield2009}, \ie~an exponential growth of the associated modes at the rate $s_q=\mathrm{Im}\; \epsilon_q$, where $s_q$ denotes the growth rate of the momentum mode $q$.
	As we shall see later, a quantitative indicator of the instability is the maximum growth rate of the spectrum, 
	\begin{equation}
		\Gamma(J,U,K,\omega)\equiv \max_{q} s_q,
		\label{eq:GammaDef}
	\end{equation}
	which, in the following, will be referred to as the \textit{instability rate} $\Gamma$. This choice will be justified in Sec.~\ref{sec:NumEn}.\\
	
	For the model under investigation, one can considerably simplify the problem by working in a rotating frame, in which translational invariance is manifest. This is achieved by applying the following gauge transformation~\footnote{This transformation is formally reminiscent of a standard gauge transformation in electromagnetism, i.e.~when going from a gauge where the electric field is written as $\boldsymbol{E}\!=\!-\boldsymbol{\nabla}V$ to a gauge where $\boldsymbol{E}\!=\!\text{d}\boldsymbol{A}/\text{d}t$, where $V$ and $\boldsymbol{A}$ denote the scalar and vector potentials.}
	\begin{equation}
		a_n(t)\longrightarrow \mathrm e^{-in(K/\omega)\mathrm{sin}(\omega t)}a_n(t).
		\label{eq:Frame}
	\end{equation}
	In this frame, and going to momentum space, the system of coupled Bogoliubov-de Gennes equations reduces to a $2\times2$ matrix~\cite{creffield2009}, and the dynamics of the mode with momentum $q\in\mathrm{BZ}$ (first Brillouin zone) is described by~\footnote{The quantity $q$ represents the momentum of the corresponding Bogoliubov quasi-particle \textit{with respect to the ground-state} (as becomes clear from the equation $a_n(t)=a_n^{(0)}(t)[1+\delta a_n(t)]$). This will  always be tacitly implied in the following,  when referring to a \textit{mode q}.} 
	\begin{equation}
		i \partial_t \left( \begin{matrix} u_q  \\ v_q \end{matrix} \right)=\left( \begin{matrix}  \varepsilon(q,t)+g & g        \\ -g & -\varepsilon(-q,t)-g \end{matrix} \right)\left( \begin{matrix} u_q  \\ v_q \end{matrix} \right),
		\label{eq:BdGEk_pre}
	\end{equation}
	where 
	\begin{equation}
	\varepsilon(q,t)=4J\sin\left(\dfrac{q}{2}\right)\sin\left(\dfrac{q}{2}+p_0-\dfrac{K}{\omega}\sin(\omega t)\right).\label{epsilon_av}
	\end{equation}
	with $p_0$ denoting the momentum of the initial BEC wavefunction.  Throughout this paper, we set the lattice constant to unity ($a_x\!=\!1$). Here, we introduced the parameter $g\!\equiv\!\rho U$, where $\rho \!=\! |a^{(0)}(0)|^2$ denotes the condensate density; the latter enters the normalization of the solution $a_n^{(0)}(t)$; the amplitudes $u_q,v_q$ of the Bogoliubov modes are defined by the expansion of the fluctuation term $\delta a_n$ in terms of Bloch waves
	\begin{equation}
		\delta a_n(t)=\sum_q u_q(t) \mathrm e^{iqn}+v^*_q(t) \mathrm e^{-iqn}.
		\label{eq:uvdef}
	\end{equation}
	Equation~\eqref{eq:BdGEk_pre} is numerically easier to integrate, and will also constitute the starting point of the analytical analysis (see Sec.~\ref{sec:ana}).
	
	\section{Numerical Simulation}
	\label{sec:Num}
	
	The general procedure outlined above can be implemented and solved numerically, regardless of the precise form of the physical band model. In more involved models, the initial state $a_n^{(0)}(0)$ can generically be determined by finding the ground state of the effective  Hamiltonian, ${H}_{\mathrm{eff}}$, i.e.~by solving the equivalent of Eq.~(\ref{eq:GPEeff}) using imaginary time propagation. The condensate wavefunction $a_n^{(0)}(t)$ is then determined by solving the time-dependent GPE with this initial condition, through real-time propagation (e.g.~using a Crank-Nicolson integration scheme). Finally, the Bogoliubov equations are also solved over one driving period by real-time propagation, yielding the operator $\Phi(T)$, which is then exactly diagonalized (e.g.~using a Lanczos algorithm).
	In the present case, this procedure can be shortcut by directly solving Eqs.~(\ref{eq:BdGEk_pre}), and we have checked that this produces the same results for all physical quantities. We now present the numerical results obtained for our model.
	
	\subsection{Stability diagrams}
	\label{sec:NumDiag}
	
	\begin{figure}[]
		
		\begin{minipage}{8cm}
			
			\includegraphics[width=8.3cm]{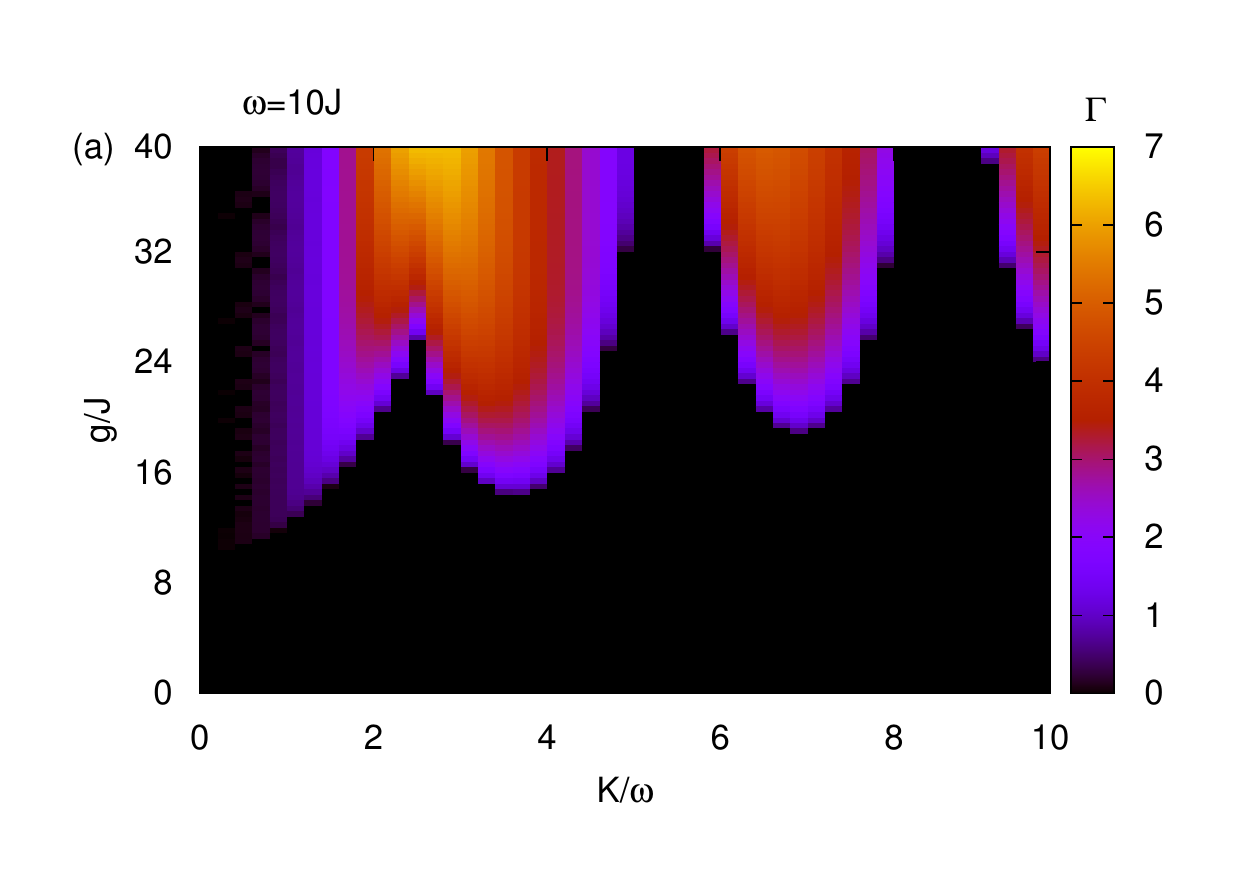}
		\end{minipage}
		\begin{minipage}{8cm}
			\centering
			\includegraphics[width=8.3cm]{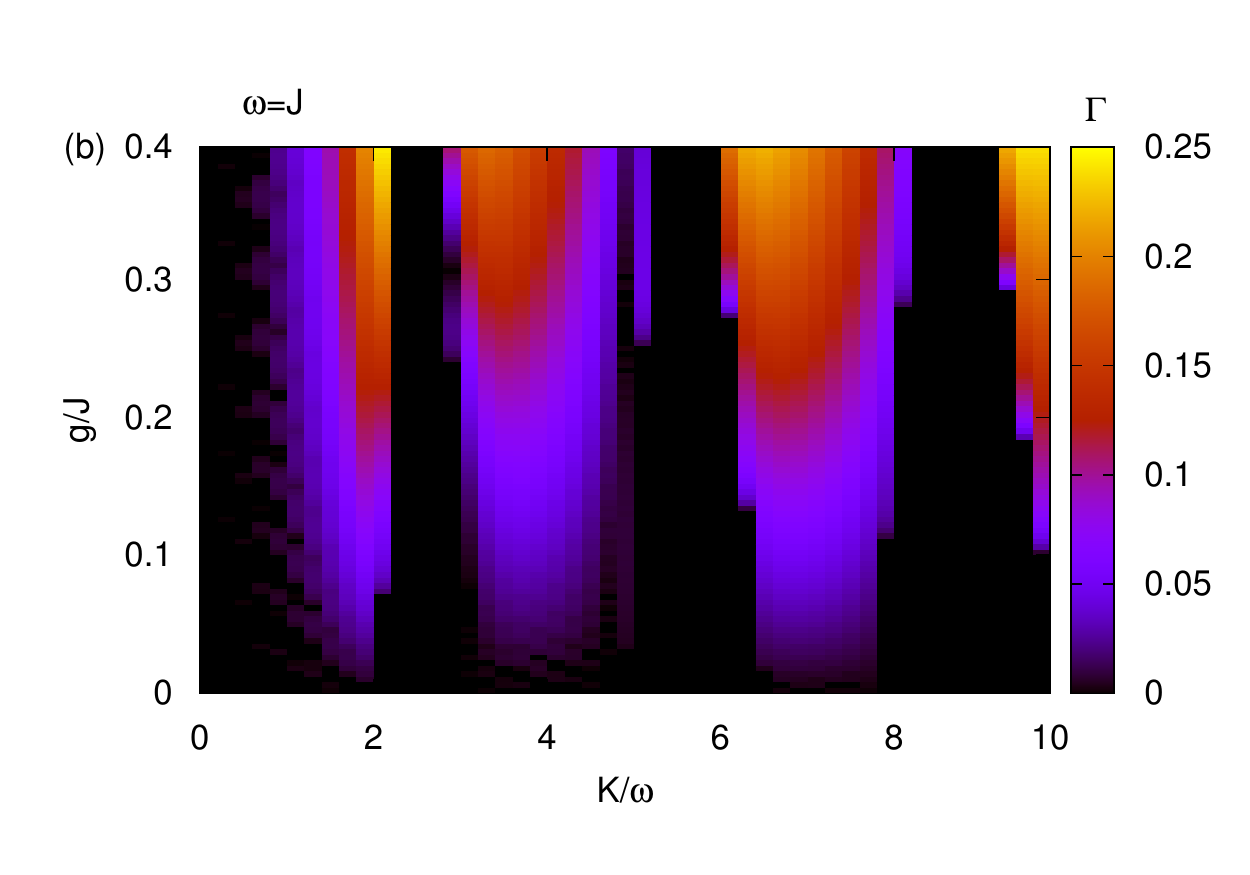}
		\end{minipage}
		\caption{Numerical instability rate $\Gamma$ as a function of interaction strength $g=U\rho$ and modulation amplitude $K/\omega$, for two values of the driving frequency ($\omega=10J$ and $\omega=J$). At large $\omega$ (\ie~outside the ``low-frequency" regime, see text), the instability rates depend mostly only on $K/\omega$ and $g/\omega^2$, while in the low frequency regime, instabilities can occur at infinitesimal interaction strength.}
		\label{fig:diagNum}
	\end{figure}
	
	The stability diagram of the model in Eq.~\eqref{eq:GPE}, which displays the behavior of the instability rate $\Gamma$ as a function of the interaction strength $g\!=\!U\rho$ and modulation amplitude $K/\omega$, is shown in Fig.~\ref{fig:2Diags}, for a reasonably large driving frequency $\omega=5J$. The stability boundary is similar to the one previously reported in~\cite{creffield2009} \footnote{At least for sufficiently high frequency, as considered in the calculations of Ref.~\cite{creffield2009}; see discussion below concerning the behavior associated with the low-frequency regime.}; in particular, the system is found to be stable in regions where $\mathcal{J}_0(K/\omega)\!=\!0$, which can be attributed to the fact that the dynamics is frozen by the cancellation of the effective tunneling~\cite{creffield2009}. At the transition to instability, the instability rate builds up continuously from zero, and then increases when going further in the unstable regime.
	
	Figure~\ref{fig:diagNum} shows similar stability diagrams for two other values of the driving frequency, $\omega\!=\!10J$ and $\omega\!=\!J$.
	In the first case, we observe that the diagram is mostly unaffected by a change of $\omega$ provided $g$ is rescaled as $g\propto\omega^2$. More generally, except in what we shall refer to as the ``low-frequency regime" (defined by the condition $\omega\!<\!4|\Jeff|$ for the present model; see the next paragraph), transitions to instability always occur at some finite $g$, and the corresponding rate is found to mostly depend  on the quantities $K/\omega$ and $g/\omega^2$\footnote{Slight deviations from this scaling appear only for small values of $K/\omega$.}. As we shall see in Sec.~\ref{sec:ana}, this can be understood from the fact that the instability rate depends on a competition between $\omega$ and the Bogoliubov dispersion associated with the linearised effective GPE [i.e.~the Bogoliubov dispersion stemming from the linear analysis of Eq.~\eqref{eq:GPEeff}], 
	\begin{equation}
		\Eav(q)=\sqrt{4|\Jeff|\sin^2(q/2)(4|\Jeff|\sin^2(q/2)+2g)},
		\label{eq:Eav}
	\end{equation}
	where the two cases $p_0\!=\!0,\pi$ are taken into account through the absolute value $\vert J_{\text{eff}}\vert$.
	As soon as the transition occurs at sufficiently large $g$, the term $4|\Jeff|\sin^2q/2$ in this dispersion becomes negligible compared to $g$, which results in $\Eav\!\propto\! \sqrt{g}$, explaining the scaling indicated above. 
	
	This is no longer true in the ``low-frequency regime" (see Fig.~\ref{fig:diagNum} for $\omega\!=\!J$), which we generically define through the criterion according to which $\omega$ is smaller than the effective free-particle bandwidth (i.e.~$\omega<4|\Jeff|$ for the model under consideration). In that case, we observe that instabilities may occur at \emph{any} finite interaction strength $g$, which is related to the fact that $\omega$ is smaller than the bandwidth of the effective Bogoliubov dispersion~(\ref{eq:Eav}) at $g=0$; see also the analytical analysis in Sec.~\ref{sec:ana} for more details.
	
Finally, in the singular case where $\Jeff\!=\!0$, the effective Bogoliubov dispersion $\Eav(q)$ in Eq.~(\ref{eq:Eav}) vanishes, and thus, the system is necessarily stable.  	
	
	\subsection{Dynamics of physical observables}
	\label{sec:NumEn}
	\begin{figure}[]
		
		\begin{minipage}{8cm}
			
			\includegraphics[width=8.3cm]{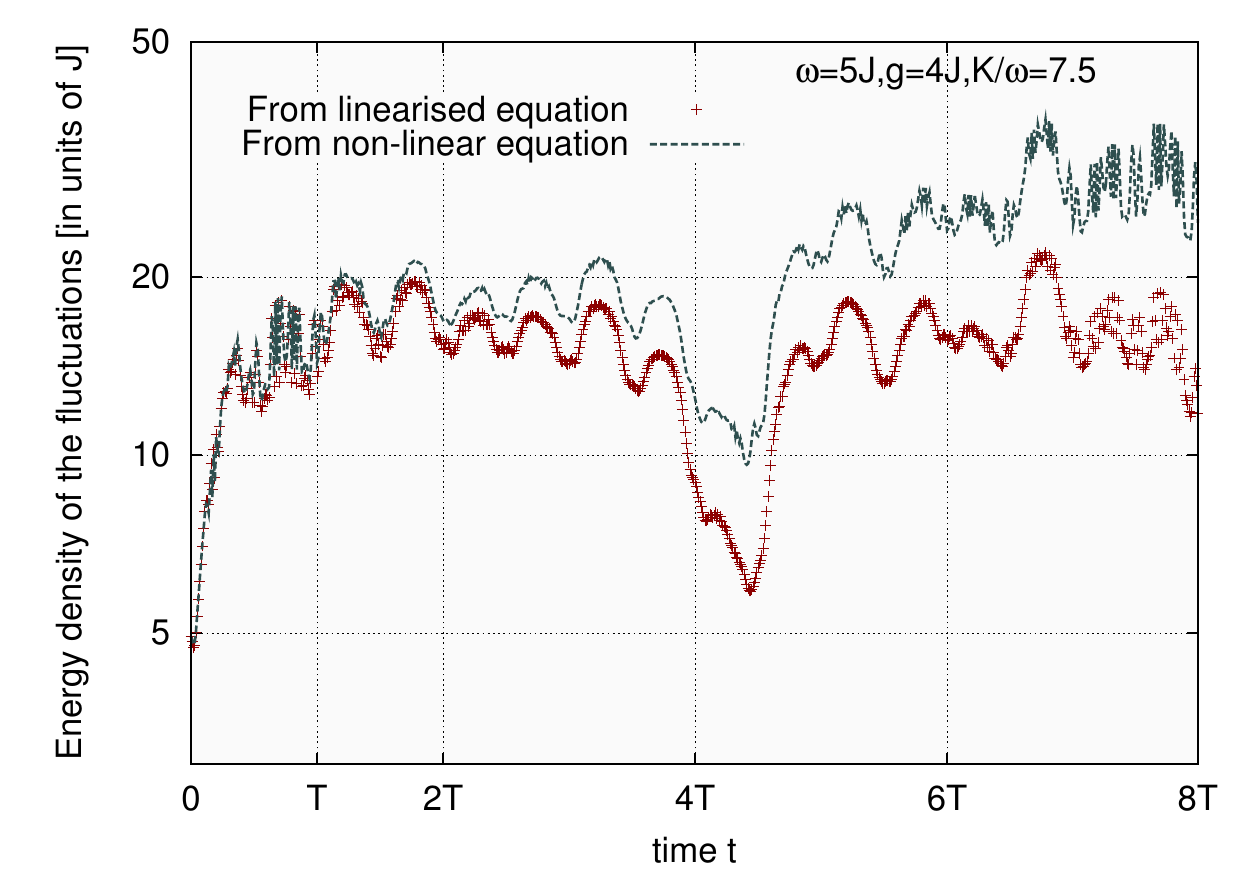}
		\end{minipage}
		\begin{minipage}{8cm}
			\centering
			\includegraphics[width=8.3cm]{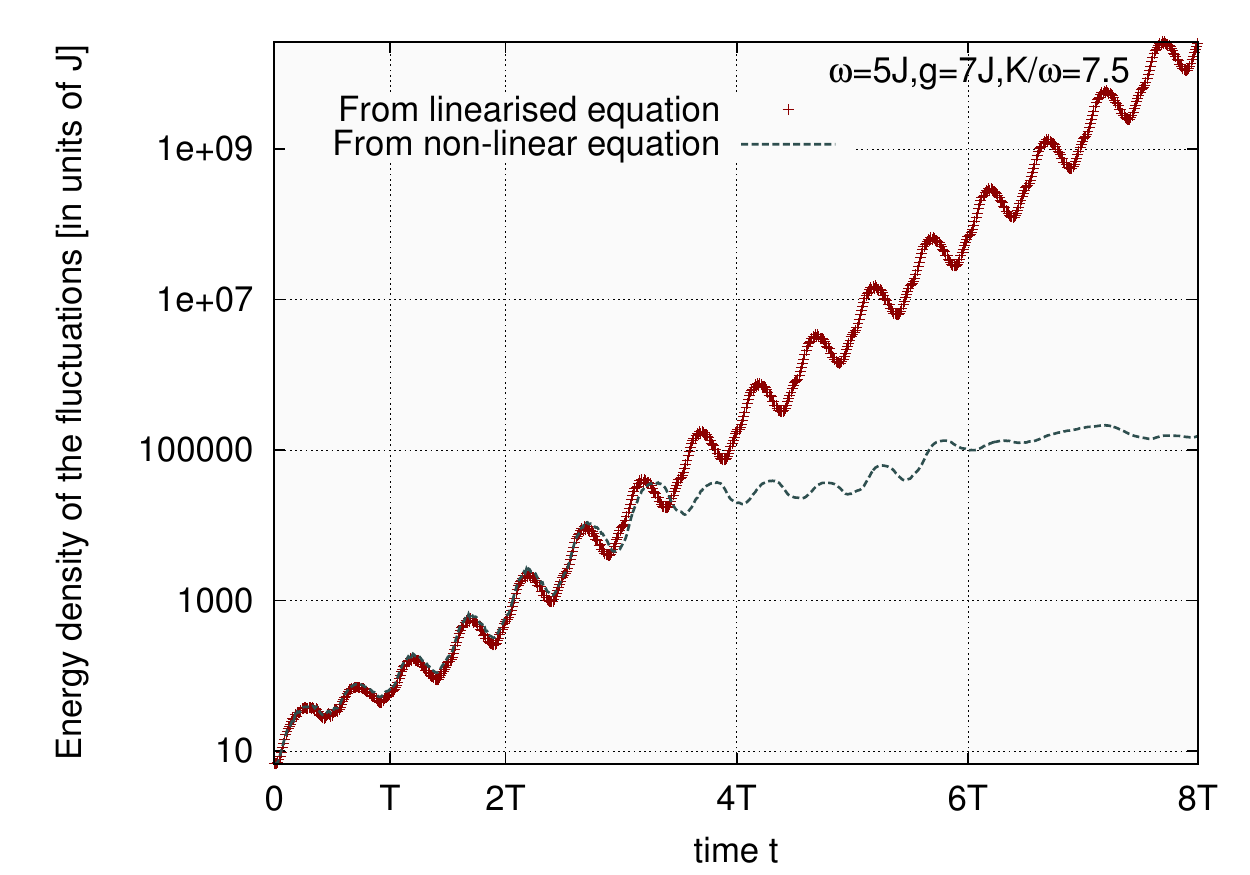}
		\end{minipage}
		\caption{ Time evolution of the energy density of the fluctuations (in units of $J$) over 8 driving periods for two values of $g$. In the stable regime (top, $g=4$), the energy displays variations due to micromotion, but no long-term growth, while in the unstable regime (bottom, $g=7$), the energy curve features exponential growth, defining a heating rate compatible with the maximal Lyapunov exponent $\Gamma$ [see Fig.~\ref{fig:DiagEner}]. The parameters are $A_q=1$ and $B_q=0$ (for the initial condition, see text), $\omega=5J$, and $K/\omega=7.5$. }
		\label{fig:Ener}
	\end{figure}
	
	Instead of solving the Bogoliubov equations [Eq.~\eqref{eq:BdGE}] over one driving period only, we can also use these to compute the full time-evolution of physical quantities, hence revealing both their long-time and micro-motion dynamics. To do so, one has to choose an initial condition for the fluctuation term, $\delta a_n(t=0)$. In this section, we will consider a generic (small) perturbation $\delta a_n(t=0)$, which has Fourier components over all Bogoliubov modes, i.e.~of the generic form $\delta a_n(0)=\alpha \sum_q A_q \mathrm e^{iqn}+ B_q \mathrm e^{-iqn}$ with $A_q, B_q$ being complex numbers and $\alpha$ a small amplitude. In Sec.~\ref{sec:BH}, we discuss a more physical initial condition, in the framework of the periodically-driven Bose-Hubbard model. Altogether, given such an initial condition, one can numerically evolve Eqs.~\eqref{eq:BdGE} using real-time propagation, which yields $\delta a_n(t)$ (and thus also $a_n(t)=a_n^{(0)}(t)[1+\delta a_n(t)]$ in the linearized approximation), hence revealing the behavior of physical quantities.
	
	\subsubsection{Energy growth and heating}
	\label{sec:Engrowth}
	As an illustration, we first evaluate the energy in the rotating frame, which is given at time $t$ by
	\begin{align}
		E(t)=&-J\sum_n \biggl [a_{n+1}(t)e^{in(K/\omega)\mathrm{sin}(\omega t)}\notag\\
		&\qquad \qquad+a_{n-1}(t)e^{-in(K/\omega)\mathrm{sin}(\omega t)} \biggr]a^*_n(t)\notag\\
		&+ \dfrac{U}{2}\sum_n |a_n(t)|^4.
		\label{eq:E}
	\end{align}
	In the linear approximation, one can explicitly recast this expression as a function of $\delta a_n$ and only keep terms of lowest order in $\delta a_n$. This yields $E(t)\!=\!E[a^{(0)}](t)+E_{\mathrm{fluct}}(t)$, where the energy of the fluctuations $E_{\mathrm{fluct}}(t)$ is given to lowest order by
	\begin{align}
		E_{\mathrm{fluct}}(t)  \approx &-J\sum_n \biggl[a^{(0)}_{n+1}\delta a_{n+1}e^{in(K/\omega)\mathrm{sin}(\omega t)} \notag\\
		&\qquad \quad +a^{(0)}_{n-1}\delta a_{n-1}e^{-in(K/\omega)\mathrm{sin}(\omega t)} \biggr]a^{*(0)}_n\delta a_{n}^* \nonumber\\
		& + \dfrac{U}{2}\sum_n |a^{(0)}_n|^4 \bigl[\delta a_n^2 + \delta a_n^{*2}+4|\delta a_n|^2 \bigr].
		\label{eq:Efluct}
	\end{align}
	
\begin{figure}[!h]
		\begin{center}
			\includegraphics[width=8.3cm]{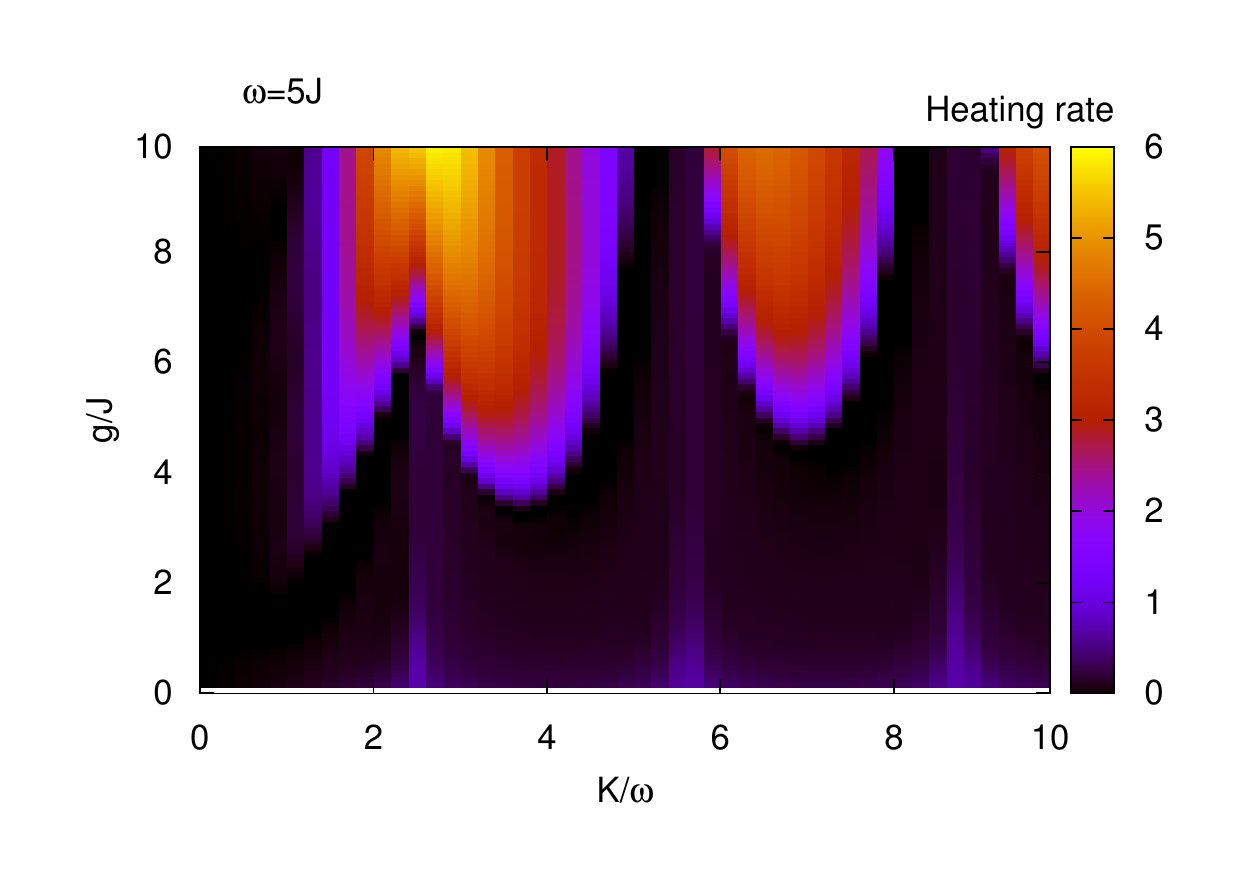}
		\end{center}
		\caption{Heating rate, obtained by fitting the exponential growth of the energy in Fig.~\ref{fig:Ener}. The resulting diagram is very similar to the instability diagram (Fig.~\ref{fig:2Diags}), up to an overall numerical factor $\approx2$ (see text).
			\label{fig:DiagEner}
		}
	\end{figure}

	The behavior of the energy of the fluctuations in Eq.~\eqref{eq:Efluct} over several driving periods is shown in Fig.~\ref{fig:Ener}, in the stable and the unstable regimes [red curves], for an initial condition corresponding to $A_q=1$ and $B_q=0$. In the stable regime, it displays modulations due to micromotion, but no long-term growth. Conversely, as anticipated for a parametric instability, the energy displays an exponential growth in the unstable regime. By fitting this growth, we find that its rate is given by $2\Gamma$ [the factor $2$ stems from the square moduli ($|\delta a_n|^2,...$) in Eq.~\eqref{eq:Efluct}; see also below, Eq.~\ref{eq:Oexp}], which expresses the fact that the maximally unstable mode dominates the growth of the energy in the system. This unstable behaviour, and the validity range of the related regime, will be further addressed in the next paragraphs. Figure~\ref{fig:DiagEner} shows the diagram obtained by extracting the heating rates from energy curves, and is found to be in excellent agreement with the stability diagram of Fig.~\ref{fig:2Diags}. Therefore, we conclude that the instability rate $\Gamma$ can be used as a quantitative estimator of heating in such systems. 

	\subsubsection{Validity regimes of the calculation}
	
	The computation of the full dynamics also highlights in which regimes our previous calculation of $\Gamma$ is expected to hold. 

	On the one hand, the calculation of the rate $\Gamma$ is based on a Floquet treatment of the Bogoliubov equations Eq.~\eqref{eq:BdGE}, and thus describes the dynamics at times $$t\gg T.$$ In turn, the dynamics within one driving period, appearent in Fig.~\ref{fig:Ener}, cannot be captured by our analysis.

	On the other hand, the analysis presented so far is based on the linear approximation of the GPE. To investigate non-linear effects, we compared the time-evolution obtained from the linearized Eq.~\eqref{eq:BdGE} (red curves in Fig.~\ref{fig:Ener}, as previously discussed) with the time-evolution of the original non-linear GPE in Eq.~(\ref{eq:GPE}) (grey dashed lines in Fig.~\ref{fig:Ener}), for the same initial condition. While the two graphs agree reasonably well in the stable regime, we find a significant deviation at longer times, within the unstable regime, due to the growth of the fluctuation term. This illustrates the intuitive fact that our linear-approximation-based treatment only holds at short times, imposing a second validity condition on our theory: $$t\ll t_{lin}.$$ At longer times, non-linear corrections to the evolution damp the exponential growth of the energy, which leads to a slowdown (saturation) of the heating dynamics. 

	Altogether, we find two natural time scales in the problem, and our approach thus requires the condition $T\ll t_{lin}$ to be relevant. Such a condition is fulfilled in a wide window of realistic system parameters.
	
	\subsubsection{Instability regimes}
	\begin{figure}[]
		
		\centering
		\includegraphics[width=8.7cm]{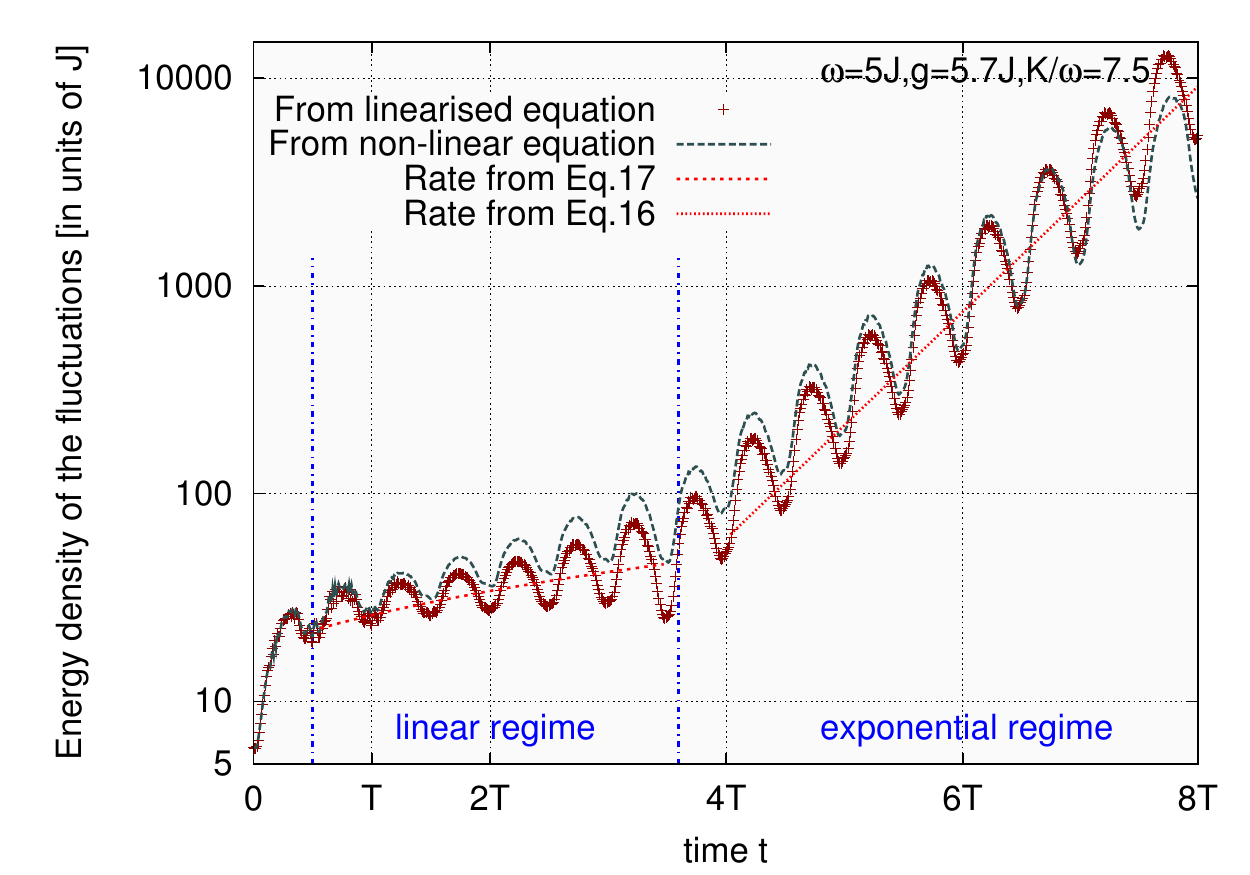}
		\caption{ Time evolution of the energy density of the fluctuations (in units of $J$) over 8 driving periods for the initial condition $A_q=1$ and $B_q=0$ (see text), for $\omega=5J$, $K/\omega=7.5$ and $g=5.7$. In this regime where $\Gamma$ is small, three regimes clearly show up: at very short times $t\ll T$, the initial growth is not described by our Floquet approach, as previously explained; at intermediate times $T \ll t \ll  \Gamma^{-1}$, the growth is linear and described by Eq.~\ref{eq:Olin} (red dotted line); at large times $t\gg \Gamma^{-1}$, the growth is exponential and described by Eq.~\ref{eq:Oexp} (red short-dotted line).}
		\label{fig:Ener2}
	\end{figure}
     A third natural time scale arises when studying the growth of physical observables. In the linear treatment of the GPE, a generic physical observable can be expressed as 
	\begin{align}
		O(t)=\sum_q \mathcal{O}[u_q(t),v_q(t)],
		\label{eq:Odef}
	\end{align}
	where $\mathcal{O}[u_q(t),v_q(t)]$ is a functional, which is quadratic in the Bogoliubov modes $u_q(t)$ and $v_q(t)$; for instance, this is straightforward for the energy when substituting Eq.~(\ref{eq:uvdef}) into Eq.~(\ref{eq:Efluct}). This also applies to the non-condensed fraction when considering a weakly-interacting Bose-Hubbard model in the meanfield interacting regime [see Eq.~(\ref{eq:fnc}) in Sec.~\ref{sec:BH}].
	Using the fact that in our Floquet treatment of the Bogoliubov equation, each mode $q$ stroboscopically evolves according to the rate $s_q$, one can rewrite Eq.~\eqref{eq:Odef} as
	\begin{align}
	O(t_N)=\sum_q \mathcal{O}[u_q(0),v_q(0)]e^{2s_q t_N}
	\label{eq:O}
	\end{align}
      where the time $t_N\equiv NT$ denotes an integer multiple of the driving period $T$. By introducing $\Gamma=\mathrm{max}_q s_q$, as considered above, two cases may arise:
	\begin{enumerate}
		\item [a)] if $\Gamma t_N\gg 1$, the growth of the maximally unstable mode dominates in Eq.~(\ref{eq:O}) and we find that the observables stroboscopically grow up exponentially with the rate $2\Gamma$, 
		\begin{align}
		O(t_N)\propto e^{2\Gamma t_N},
		\label{eq:Oexp}
		\end{align}
		as already observed for the energy in Sec.~\ref{sec:Engrowth}. Remarkably, this rate only involves the most unstable mode, and it is the same for all physical quantities. As announced above, this justifies the use of $\Gamma$ as a global instability rate in our theoretical analysis.
		\item [b)] if $\Gamma t_N \ll 1$, all exponentials in Eq.~(\ref{eq:O}) can be linearized, yielding
		\begin{align}
		O(t_N)\!=\!O(0)\!+\!t_N\sum_q 2 s_q \mathcal{O}[u_q(0),v_q(0)]+ \mathcal{O}(t_N^2).
		\label{eq:Olin}
		\end{align}
		In this regime, the energy increases linearly in time, with a rate given by the slope $2\sum_q s_q \mathcal{O}[u_q(0),v_q(0)]$. The latter now involves a summation over all modes, since no single mode contribution can be singled out at these early times ($t_N\!\ll\!1/\Gamma $). Note also that, in this regime, the rate depends on the physical observable considered (through the functional $\mathcal{O}$). 
		
		For most values of the system parameters [such as those used in Fig.~\ref{fig:Ener}], $\Gamma T\gg 1$, so that the linear regime [Eq.~(\ref{eq:Olin})] is hidden in the first oscillation (which, as stated above, is not accurately captured by our Floquet approach). Yet, very close to the stability boundary, where $\Gamma$ is small, this marginal regime can become apparent in a very narrow range of parameters, as we illustrate in Fig.~\ref{fig:Ener2}.
	\end{enumerate} 
Here, three regimes clearly show up: at very short times $t\ll T$, the initial growth is not described by our Floquet approach, as previously explained; at intermediate times $T \ll t \ll \Gamma^{-1}$, the growth is linear and accurately described by Eq.~(\ref{eq:Olin}); at longer times $t\gg\Gamma^{-1}$, the growth is exponential and well described by Eq.~(\ref{eq:Oexp}).
Note that the linear regime is expected to be hard to probe in experiments, since it appears at very short times, typically a few milliseconds (see also discussion in Sec.~\ref{sec:CCL}).\\

	\section{Analytical approach}
	\label{sec:ana}
	
	\subsection{The linearised Gross-Pitaevskii equation as a parametric oscillator}
	\label{sec:anaPO}
	
	In momentum space, the Bogoliubov-de Gennes equations~(\ref{eq:BdGEk_pre}) constitute a set of uncoupled equations, which independently govern the time evolution of excitations with a given momentum $q$: the mean-field analysis effectively reduces to a single-particle problem, and one can study how instabilities occur independently in each of those modes. The general mechanism underlying the appearance of parametric instabilities was suggested in Ref.~\cite{bukov2015}, where a weakly-interacting Bose-Hubbard model was investigated through a weak-coupling particle-number-conserving approximation; in that study, parametric resonances were shown to appear in the Bogoliubov equations, whenever the drive frequency was reduced below twice the single-particle bandwidth. A similar approach was considered in Ref.~\cite{salerno2016}, which also provides an intuitive picture of the underlying mechanisms, in terms of the so-called Krein signature associated with the Bogoliubov modes.
	
	The same phenomenon occurs in the present framework, as can be seen by performing a change of basis that recasts Eq.~(\ref{eq:BdGEk_pre}) into the following form [see Appendix~\ref{sec:A1} for details]
	\begin{align}
		& i \partial_t \left( \begin{matrix} \tilde{u}'_q  \\ \tilde{v}'_q \end{matrix} \right)=\biggl[\Eav(q)\hat{\mathbf{1}}+\hat{W}_q(t)\label{eq:BdGEPO}\\
		&\quad +\dfrac{g}{\Eav(q)}\left( \begin{matrix} 0 & h_q(t)\mathrm e^{-2i\Eav(q)t}        \\ -h_q(t)\mathrm e^{2i\Eav(q)t} & 0 \end{matrix} \right)\biggr] \left( \begin{matrix} \tilde{u}'_q  \\ \tilde{v}'_q \end{matrix} \right) .
		\notag
	\end{align}
	 Here  
	$\Eav(q)$ is the Bogoliubov dispersion associated with the effective (time-averaged) GPE, within the Bogoliubov approximation [see Eq.~(\ref{eq:Eav})]; $W_q(t)$ is a diagonal matrix of zero average over one driving period, which will play no role in the following [see Appendix~\ref{sec:A1} for its exact expression]; and $h_q(t)$ is a (real-valued) function which can be Fourier expanded as 
	\begin{align}
		h_q(t)&=\frac{J}{2}4\sin^2(q/2)\sum_{l=-\infty}^\infty [\mathcal{J}_l(K/\omega)+\mathcal{J}_{-l}(K/\omega)] \mathrm e^{il\omega t} \notag \\
		&=4J\sin^2(q/2)\sum_{l=-\infty}^\infty \mathcal{J}_{2l}(K/\omega) \mathrm e^{i2l\omega t},
		\label{eq:hq}
	\end{align}
	with $\mathcal{J}_l(z)$ the $l$-th Bessel function of the first kind. Importantly, the specific form of Eq.~\eqref{eq:BdGEPO} allows one to clearly distinguish between the contribution due to the time-averaged dynamics, as captured by the effective dispersion $\Eav(q)$, and the contribution due to the micro-motion [$W_q(t)$, $h_q(t)$]. As will be shown below, it is the micro-motion contribution that governs the existence of instabilities in the system, through the properties of the real-valued function $h_q(t)$.	
	
	Casting the equations of motion in the form~\eqref{eq:BdGEPO} makes it possible to directly identify any parametric resonance effects. To see this, we recall the reasoning of Ref.~\cite{bukov2015}, which was based on applying a rotating-wave approximation (RWA) to Eq.~(\ref{eq:BdGEPO}). If one disregards for now the expression in Eq.~\eqref{eq:hq}, and if one naively assumes that the lowest frequency appearing in $h_q(t)$ is $\omega$, then a dominant contribution to the dynamics is expected when the resonance condition $\omega\!=\!2\Eav(q)$ is fulfilled, resulting in the time-independent non-diagonal term in the matrix displayed in Eq.~(\ref{eq:BdGEPO}). Keeping  this resonant term only yields a time-independent $2\times2$ matrix which can be diagonalized, and whose eigenvalues (i.e.~the Lyapunov exponents) are found to exhibit an imaginary part, yielding a non-zero instability rate. This argument was invoked in Ref.~\cite{bukov2015} to justify the intuitive stability criterion $\omega\!>\!2W_{\mathrm{eff}}$ (with $W_{\mathrm{eff}}\!=\!\sqrt{4|\Jeff|(4|\Jeff|+2g)}$ the bandwidth of the effective Bogoliubov dispersion), which states that instability arises from the absorption of the energy $\omega$ to create a pair of Bogoliubov excitations on top of the condensate. However, a more careful examination shows that the instability rates inferred from this idea are not consistent with our numerical simulations. Moreover, to get a flavour of the additional dilemma one is faced with, we note that the expression for $h_q(t)$ in Eq.~\eqref{eq:hq} actually only contains \emph{even} harmonics of the modulation, and therefore, following the reasoning above, the resonance condition should then read $2\omega\!=\!2\Eav(q)$. Yet, such a criterion provides a wrong estimate of the stability-instability boundaries.  Hence, it appears that this simple explanation needs to be thoughtfully revised.
	
	In the following, we present a more rigorous analytical evaluation of the instability rates for our system. Similar to Ref.~\cite{bukov2015}, our derivation relies on that the Bogoliubov equations~(\ref{eq:BdGEPO}) precisely take the form of a so-called \textit{parametric oscillator}. In Appendix~\ref{sec:anaPOrapp} we briefly recall some useful results about this paradigmatic model~\cite{landau1969}, which describes a harmonic oscillator of eigenfrequency $\omega_0$ driven by a weak sinusoidal perturbation of frequency $\omega$ and amplitude $\alpha$. In brief, such a model displays a parametric instability for $\omega\!\approx\!2\omega_0$. Importantly, the instability is maximal when this resonance condition is fulfilled, but occurs in a whole range of parameters around this point. As detailed in Landau and Lifshits~\cite{landau1969}, the width of the resonance domain and the instability rates in the vicinity of the resonance can be calculated perturbatively by introducing the detuning $\delta\!=\!\omega-2\omega_0$ and solving the equations perturbatively in $\delta$ and amplitude $\alpha$ (see Appendix~\ref{sec:anaPOrapp}).
	
	More specifically, the Bogoliubov equations~(\ref{eq:BdGEPO}) exactly take the form of the parametric oscillator [Eq.~(\ref{eq:PO2})], with the frequency $\omega_0$ of the unperturbed oscillator being identified with the dispersion $\Eav(q)$. Thus, it immediately follows that the model is equivalent to a set of independent parametric oscillators, one for each mode $q$. Two points should be emphasized here:~first, all those parametric oscillators depend on $q$ in a different way, and will thus exhibit different resonance conditions; second, the function $h_q(t)$ in Eq.~(\ref{eq:BdGEPO}) is not a pure sinusoid as in Eq.~(\ref{eq:PO2}), but rather contains all (even) harmonics of the driving frequency $\omega$.  Altogether, given those two remarks, resonances are expected as soon as \textit{one} of the harmonics of the modulation, of energy $l\omega$, is close to twice the energy of \textit{any} of the (effective, time-averaged) Bogoliubov modes, $2\Eav(q)$. 
	
	Before digging into more technical details, let us acquire some intuition about the general mechanisms behind this phenomenon. To do so, consider the stability diagram on Fig.~\ref{fig:2Diags} (i.e.~outside of the ``low-frequency regime"~\footnote{The same analysis holds in the low-frequency regime unless the system is already unstable at low $g$, where a resonance can show up (since $2\Eav(q)=l\omega$ may already have a solution for some $q_0$, $l_0$). The instability rate is then no longer governed by the mode $q=\pi$ and the lowest harmonic, but by the mode $q_0$ and the harmonic $l_0$.}). In this case, in the lowest part of the stability diagram (small $g$), $\omega$ is generically large compared to the bandwidth of the effective dispersion $\Eav(q)$, so that no resonance can occur. When increasing $g$ at fixed $K/\omega$, the first mode to exhibit a resonance will be that of maximal $\Eav(q)$, i.e.~$q\!=\!\pi$. Naively, the first resonance would be due to the first harmonic $l\!=\!1$ and would occur for $2\Eav(q\!=\!\pi)\!=\!\omega$, which was also the argument in~\cite{bukov2015}. However, as already discussed, the function $h_q(t)$ only contains  even harmonics, and the first resonance to occur is, therefore, the one corresponding to $l=2$. At first sight, this might seem to be in contradiction with the intuitive stability criterion $2\Eav(\pi)\!=\!\omega$, since the $l\!=\!2$ resonance is centered around $\Eav(\pi)=\omega$; however, as we shall discuss in more detail below, the key point is that the resonance domain has a finite width:~although \textit{centered} around $\Eav(\pi)=\omega$, its \textit{boundaries} are in fact close to the point where $2\Eav(\pi)=\omega$. Resonances due to higher harmonics (fourth, sixth, etc.) become important only for higher values of $g$, and we can, to first approximation, restrict the analysis to the second harmonic only. 
	
\begin{figure}[!t]
		
		\begin{minipage}{8cm}
			
			\includegraphics[width=8.3cm]{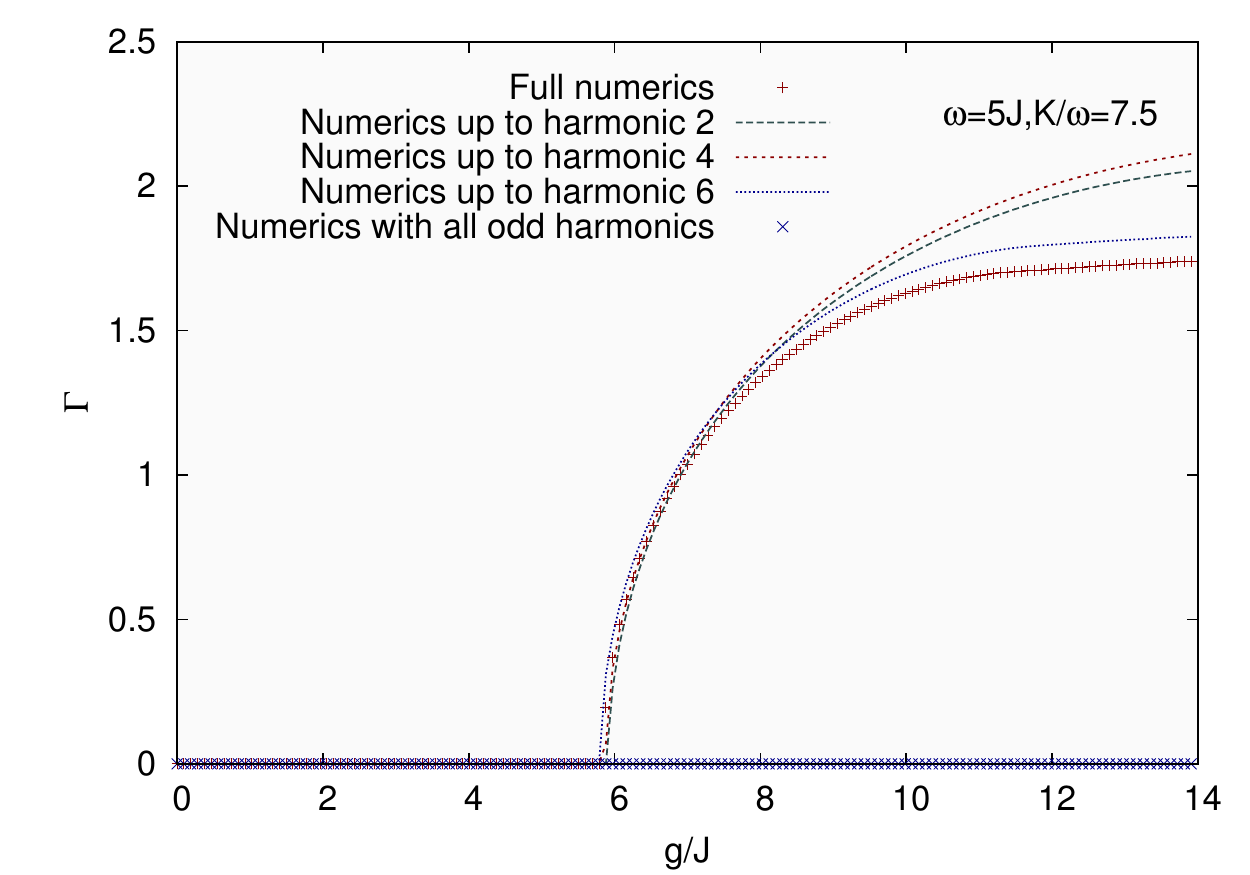}
		\end{minipage}
		\begin{minipage}{8cm}
			\centering
			\includegraphics[width=8.3cm]{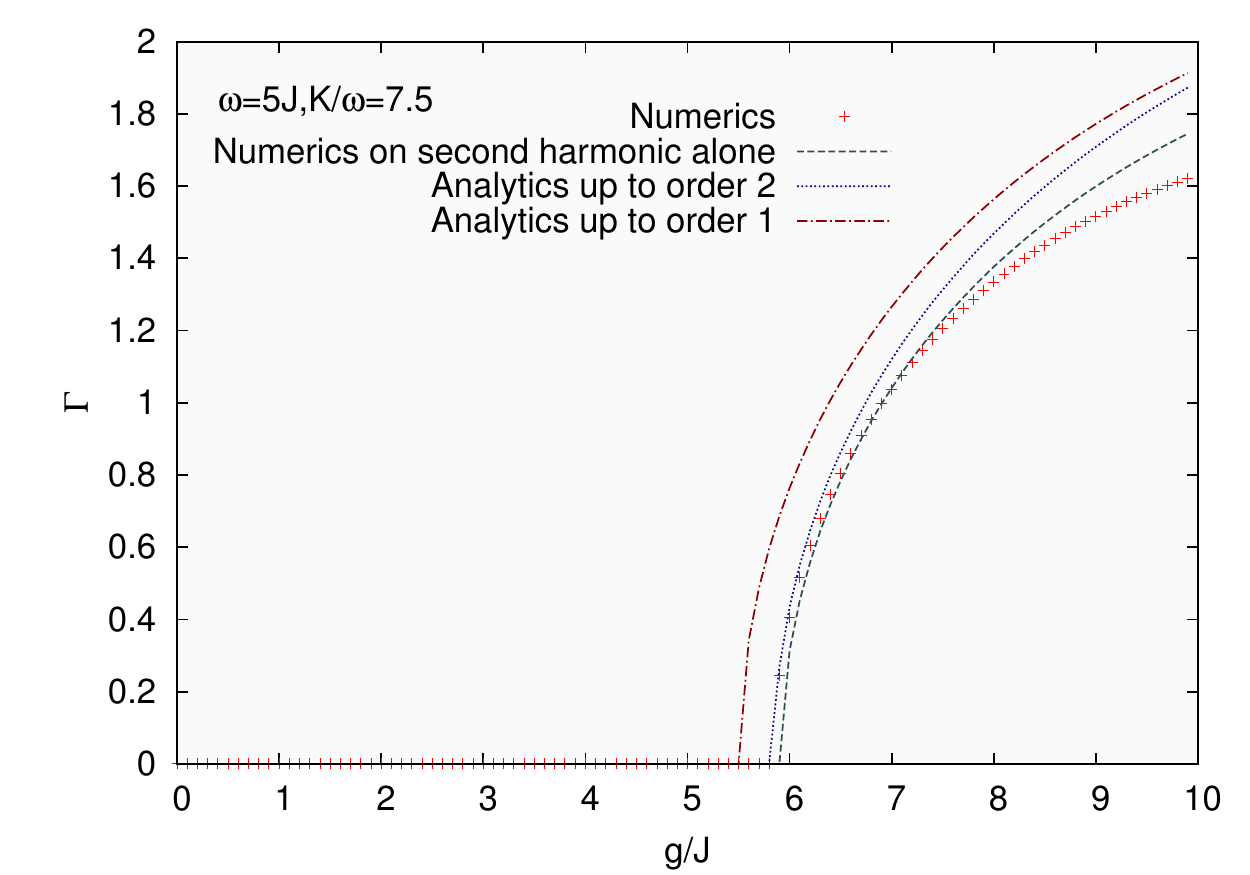}
		\end{minipage}
		\caption{ Top: Instability rate $\Gamma$ as a function of the interaction strength $g$ for $\omega=5J$ and $K/\omega=7.5$, numerically computed from the Bogoliubov equations [cf.~Sec.~\ref{sec:Imodel}] taking into account only a few harmonics of the modulation in the function $h_q(t)$. As theoretically predicted from the expression of $h_q(t)$ [Eq.~(\ref{eq:hq})], the odd harmonics do not contribute, while the first even ones successfully describe the full numerical calculation. Bottom : Comparing the analytical and numerical results for the instability rate $\Gamma (g)$, using the same parameters as above. The red crosses correspond to the full numerical solution, while the blue dashed lines are numerically obtained by only taking the second harmonic of the drive  into account. As generically observed in the A zones of the stability diagram [Fig.~\ref{fig:2Diags}], the numerical instability rates are well captured by the perturbative analytical approach, implemented here up to order 1 (red dotted-dashed line) and 2 (blue dotted line).}
		\label{fig:HarmFavo}
	\end{figure}

	To confirm this intuition, we have verified that restricting our analysis to the second harmonic only is generally sufficient to estimate the stability boundary, and to recover the instability rate in its vicinity~\footnote{See below for a precise discussion about the validity of this approximation.}. Deeper in the unstable region, the fourth, and eventually the sixth harmonics are progressively required to recover the agreement with the numerical results [see Fig.~\ref{fig:HarmFavo}]
	
	\subsection{Effective model} 
	\label{sec:anaEff}
	
	To simplify the calculations, we restrict our analysis to the second harmonic (see discussion above); we stress that higher-order resonances can be treated along the same lines. The equations of motion [Eq.~\eqref{eq:BdGEPO}] then read
\begin{widetext}
	\begin{align}
		&i \partial_t \! \left( \begin{matrix} \tilde{u}'_q  \\ \tilde{v}'_q \end{matrix} \right)=\biggl[ \Eav(q)\hat{\mathbf{1}}
		+\hat{W}_q(t) 
		+\!\dfrac{\alpha_q\Eav(q)}{2}\left( \begin{matrix} 0 & \cos(2\omega t)\mathrm e^{-2i\Eav(q)t}        \\ -\cos(2\omega t)\mathrm e^{2i\Eav(q)t} & 0 \end{matrix} \right)\biggr]\! \left( \begin{matrix} \tilde{u}'_q  \\ \tilde{v}'_q \end{matrix} \right),
		\label{eq:BdGEO2}
	\end{align}
	with $$\alpha_q=16J \mathcal{J}_2(K/\omega)\sin^2(q/2) \frac{g}{[\Eav(q)]^2}.$$ 
	\end{widetext}
	
\begin{figure}[]
		\begin{center}
			\includegraphics[width=8.3cm]{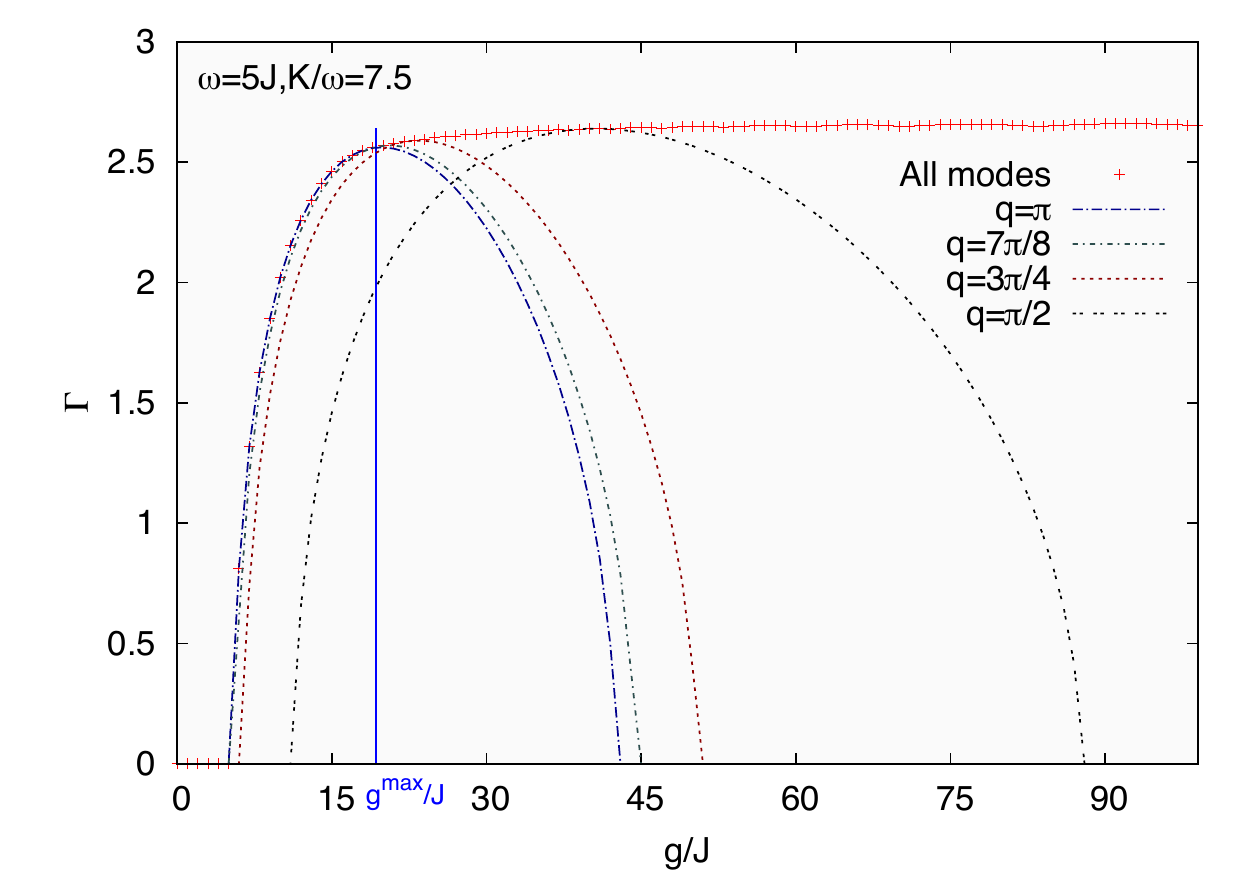}
		\end{center}
		\vspace{0.1cm}
		\caption{Total instability rate $\Gamma$ (red crosses) and instability rates $s_q$ of a few individual modes (dashed lines), numerically computed as a function of interaction strength $g$ for $\omega=5J$ and $K/\omega=7.5$. The instability rate associated with an individual mode $q$ is maximal for $\Eav(q)=\omega$ and decreases around it. The total instability rate is given by the envelope of all those curves. 
			\label{fig:Bumps}
		}
	\end{figure}

	Therefore, for a fixed $q$, these equations are now strictly equivalent to the equations of motion for the parametric oscillator of Eq.~(\ref{eq:PO2}), with the following identifications: 
	\begin{eqnarray}
		\omega_0 & \rightarrow & \Eav(q)
		\nonumber \\
		\omega & \rightarrow & 2 \omega
		\nonumber \\
		\alpha & \rightarrow & \alpha_q.
		\label{eq:Subst}
	\end{eqnarray}
	As a result, each mode $q$ exhibits a resonance domain centred around $\Eav(q)=\omega$. The instability rate is generically maximal around the resonance point (defining the maximal rate $s_q^{\mathrm{max}}$) and it decreases  until it cancels at the edges of the resonance domain. In order to obtain analytical estimates of the instability rate and the width of the resonance domain, one can apply the procedure detailed in Appendix~\ref{sec:anaPOrapp}, which amounts to solving the problem perturbatively in the detuning $\delta=2\omega-2\Eav(q)$ and amplitude $\alpha_q$. 
	To zeroth order, the instability only arises when the resonance condition is precisely fulfilled, which defines the rate ``on resonance" $s_q^{*}$, which is given by
	\begin{equation}
		s_q^{*}=  g\dfrac{4J \mathcal{J}_2(K/\omega)\sin^2(q/2)}{\Eav(q)}.   
		\label{eq:sstar}
	\end{equation}
	To first order, the instability rate reads [see Eq.~(\ref{eq:s1}) with the substitutions~(\ref{eq:Subst})]
	\begin{equation}
		s_q=   s_q^{*} \sqrt{1-\left(\dfrac{[\omega-\Eav(q)]\Eav(q)}{4J \mathcal{J}_2(K/\omega)\sin^2(q/2) g}\right)^2}  ,
		\label{eq:s1ana}
	\end{equation}
	whenever the argument in the square root is positive, and $s_q=0$ otherwise. The associated instability rate $s_q$ is thus maximal on resonance (hence, $s_q^{\mathrm{max}}\!=\!s_q^{*}$) and it decreases when going towards the boundaries of the resonance domain.
	At second order, $s_q$ is given by the solution of the implicit equation~(\ref{eq:det}) with the substitutions~(\ref{eq:Subst}), which can be found numerically; this can be performed to improve the accuracy of the analytical rate's value. In particular, at this order, we note that the actual maximal instability point is then slightly shifted from the resonance point (i.e.~$s_q^{\mathrm{max}}\!\neq\! s_q^{*}$).
	Altogether, the total instability rate is given by [Eq.~(\ref{eq:GammaDef})]
	\begin{equation}
		\Gamma=\max_q s_q.
		\label{eq:sqMaxq}
	\end{equation}
	As a function of $g$ [which enters $\Eav(q)$, see Eq.~(\ref{eq:Eav})], the instability rate of a given mode $q$ forms a ``bumped curve"  [see Fig.~\ref{fig:Bumps}], which to first approximation simply follows from Eq.~\eqref{eq:s1ana}. Since the resonance domain for each mode is centered around a different energy [and therefore a different $g$], the curves corresponding to various modes are slightly shifted, and the total instability rate is given by the envelope of all those curves [see Fig.~\ref{fig:Bumps}]. 
	
		In particular, the curve centered around the smallest values of $g$ (i.e.~the curve most located to the left in Fig.~\ref{fig:Bumps}, near the transition point) is the one associated with the mode of highest $\Eav$, namely $q\!=\!\pi$. Let us denote by $g^\mathrm{max}$ the value of $g$ where this curve is maximal: at first order, this corresponds to the solution of the resonance condition $\omega=\Eav(\pi)=\sqrt{4|\Jeff|(4|\Jeff|+2g)}$. Then, we identify two cases:
		
		\begin{itemize}

		\item[(a)] For $g>g^\mathrm{max}$, there always exists, in the thermodynamic limit, a single mode $q$ which is maximally unstable, and the total instability rate, Eq.~(\ref{eq:sqMaxq}) is thus given by the rate of this particular mode. At first order [see discussion above], this mode is the one fulfilling the resonance condition $\omega\!=\!\Eav(q)$, and the instability rate is given by the rate \textit{on resonance} of this particular mode $q_\mathrm{res}$, so that : 
		\begin{equation}
			\Gamma=s_{q_\mathrm{res}}^*=\left|\dfrac{\mathcal{J}_2(K/\omega)}{\mathcal{J}_0(K/\omega)}\right|\dfrac{g}{\omega}(\sqrt{g^2+\omega^2}-g),
			\label{eq:sresR}
		\end{equation}
where we have explicitly re-expressed $q_\mathrm{res}$ using the expression $\omega\!=\!\Eav(q_\mathrm{res})$.
	
		\item[(b)] Conversely, for $g<g^\mathrm{max}$, the instability rate is only governed  by the mode $q=\pi$ [see Fig.~\ref{fig:Bumps}], and it cannot be estimated from the knowledge of the associated rate \textit{on resonance}. 
		Therefore, to capture the behaviour of the instability rate near the transition point, as well as the stability boundary itself, one can indeed restrict our analysis to the mode $q\!=\!\pi$, but one has in turn to resort to the perturbative expansion around the resonance point discussed in Appendix~\ref{sec:anaPOrapp}. 
		In other words, in that case, one has 
		\begin{equation}
			\Gamma=s_{\pi},
			\label{eq:spi}
		\end{equation}
		where $s_{\pi}$ is given at lowest order by Eq.~(\ref{eq:s1ana}) with $q\!=\!\pi$. The stability boundary can also be computed at any required order following the procedure indicated in Appendix~\ref{sec:anaPOrapp}, which yields, up to second order, 
		\begin{equation}
			\omega=\Eav(\pi)+\dfrac{4J \mathcal{J}_2(K/\omega) g}{\Eav(\pi)}-\dfrac{4J^2 \mathcal{J}_2^2(K/\omega) g^2}{\Eav(\pi)^3} + \mathcal{O}(g^3). 
			\label{eq:bound2}
		\end{equation}
		Interestingly,  in the high-frequency regime, since the transition occurs at sufficiently large $g$, so that $\Eav\propto \sqrt{g}$, one recovers that the boundary is a function of the combination $g/\omega^2$, as previously observed numerically in Sec.~\ref{sec:Num}.
	
	\end{itemize}	
	
	When decreasing the frequency, the transition occurs at lower and lower values of $g$:~since all corrections of any order tend to vanish at small $g$ in Eq.~\eqref{eq:bound2}, the transition point at vanishing $g$ is simply obtained for $\omega=4\Jeff$, recovering the criterion for entering the ``low-frequency" regime, previously discussed. In this regime, which in some sense corresponds to a negative $g^\mathrm{max}$, one is always in case (a) and the rate $\Gamma$ is given by Eq.~(\ref{eq:sresR}).\\
	
	Altogether, to capture the behaviour of the instability rate in the whole range of parameters, one should use Eqs.~\eqref{eq:s1ana}-\eqref{eq:sqMaxq}, which include all modes as well as the first correction due to the finite detuning. 
	Note that the perturbative expansion yielding those expressions is expected to hold provided $\alpha_q$ is not too large, i.e~$\Eav(q)$ is not too small [see below for a discussion of the breaking points of the approach]. 
	
	\subsection{Results based on the analytical approach}
	\label{sec:AnaRes}
	
	We now show the results obtained from this analytical approach for the instability rate, Eq.~(\ref{eq:sqMaxq}), with $s_q$ being generically computed to second order in the detuning (unless explicitly specified).
	The stability diagram is very similar to the one obtained numerically [see Fig.~\ref{fig:2Diags}].  
	Let us investigate in more detail this agreement and comment on the small differences between the numerical and analytical diagrams. \\
	
	\subsubsection{Agreement Regions}For values of $K/\omega$ which are not too close to the zeros of the Bessel functions $\mathcal{J}_0$ and $\mathcal{J}_2$~\footnote{We recall that the zeros of $\mathcal{J}_2$ are very close to the zeros of $\mathcal{J}_0$, but slightly smaller. Moreover, $\mathcal{J}_0(x)$ has an additional zero around $x=2.4$ where $\mathcal{J}_2(x)$ is not small compared to unity.} (i.e.~not too close to the edges of the lobes in the stability diagram, zones A on Fig.~\ref{fig:2Diags}), the analytical calculation gives a very good estimate of both the transition point and the instability rate, up to moderate interactions $g$ [Fig.~\ref{fig:HarmFavo}]. At higher $g$, more harmonics become important (presumably, in a complex and  ``coupled" way: we verified that building independent effective models for separate harmonics does not reproduce the numerics accurately). Zones where $\mathcal{J}_4$  vanishes are ``favorable" since the first correction to the effective model [Eq.~(\ref{eq:BdGEO2})] is absent, and the agreement between analytics and numerics survives for larger values of $g$.
	
	In the vicinity of the ``common" zeroes of $\mathcal{J}_2(K/\omega)$ and $\mathcal{J}_0(K/\omega)$ (i.e.~``between" the lobes of the stability diagram, zones B on Fig.~\ref{fig:2Diags}), both the analytics and numerics agree and predict a stable regime.\\
	
 \subsubsection{Disagreement Regions}A detailed description of the small disagreement regions is provided in Appendix~\ref{sec:A2}. In brief, the analytics breaks down in two main regions: (i) around the first zero of $\mathcal{J}_0$ (i.e.~the only zero of $\mathcal{J}_0$ where $\mathcal{J}_2$ is significant; zones C on Fig.~\ref{fig:2Diags}), the perturbative approach breaks down due to the vanishing of the effective dispersion $\Eav$, and the analytical approach fails. (ii) Near the common zeros of $\mathcal{J}_0$ and $\mathcal{J}_2$ (\ie~``between" the lobes of the stability diagram), we find that the numerics and the analytics disagree in the way the stable zones close at large $g$. We identified two main causes for this effect: the breakdown of the perturbative approach due to the vanishing of $\Eav$, and the vanishing of the second harmonics near the zeros of $\mathcal{J}_2$. The consequences of these factors taken together are very involved, since they can both compensate for each other and compete. In brief, at the left border of each lobe (zones D on Fig.~\ref{fig:2Diags}), the numerics displays a transition to instability (although showing up only at large $g$), which is not captured by the analytics. At the right border of the lobes (zones E), the instability predicted by the full numerics is most likely due to a joint effect of the second and higher-order harmonics, and is therefore not accurately captured by the analytical approach [see Fig.~\ref{fig:defavo}(b)]. 
	
	Nevertheless, the disagreement regions constitute very narrow zones in the stability diagram. Figure~\ref{fig:2Diags} summarizes the zones where the analytical and numerical calculations agree (in green) and the ones where they disagree qualitatively (in red) or just quantitatively (in orange).\\
	
\subsubsection{Scaling in limiting cases}The analytical solutions provide a general scaling behaviour for the instability rates as a function of the model parameters, which we analyze for three limiting cases. 
		
	\paragraph{Weak interactions:} If $\omega$ is larger than the effective free-particle bandwidth of the model  (``high-frequency" regime), the system is stable at $g\approx 0$, and features a transition to an unstable phase at some finite $g_c$, with a scaling [see Eq.~(\ref{eq:s1ana})] $$\Gamma \propto \sqrt{(g-g_c)}.$$ Conversely, if $\omega$ is smaller than the effective free-particle bandwidth (``low-frequency" regime), the system is typically unstable at very small $g$, and [see Eq.~(\ref{eq:sresR})] $$\Gamma \propto g.$$ 
	
	 \paragraph{Weak driving amplitudes:} The analysis of the transition to the unstable phase from $K\!=\!0$ (where the system is stable) to finite $K$ is subtle, since this limit does not commute with the thermodynamic limit: indeed, one has to note that the resonance domain of each mode has a vanishing width when $K\to 0$; therefore, for any finite-size system, instabilities at vanishing $K$ occur only for discrete values of the parameters (fulfilling one of the resonance conditions associated with a specific mode); it is only when increasing $K$ that those instability regions grow in parameter space, eventually merging and forming the first lobe of the stability diagram. In the thermodynamic limit, one nevertheless finds the general scaling $$\Gamma \propto K^2,$$ which stems from the second order Bessel function which dominates at low $K$ in Eq.~(\ref{eq:sresR}).
	 
	 \paragraph{Low driving frequency:} In this limit, one finds [see Eq.~(\ref{eq:sresR})] $$\Gamma \propto \omega \times f(K/\omega),$$ where $f$ is a generic function of $K/\omega$ only.

\section{Finite-Size Systems: from a Double Well to the Entire Lattice}  
	\label{sec:FiniteSize}

	\begin{figure}[]
		
		\begin{minipage}{8cm}
			
			\includegraphics[width=8cm]{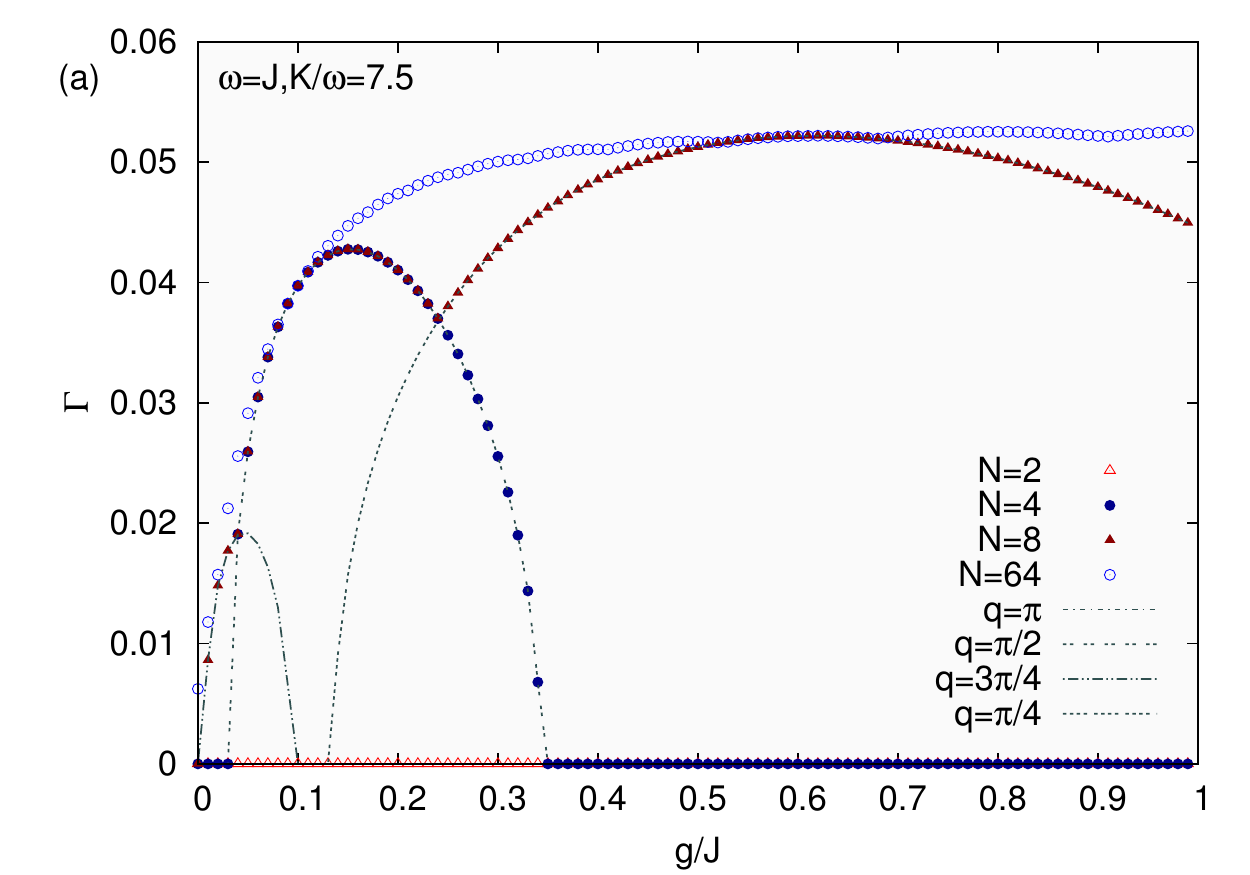}
		\end{minipage}
		\begin{minipage}{8cm}
			\centering
			\includegraphics[width=8cm]{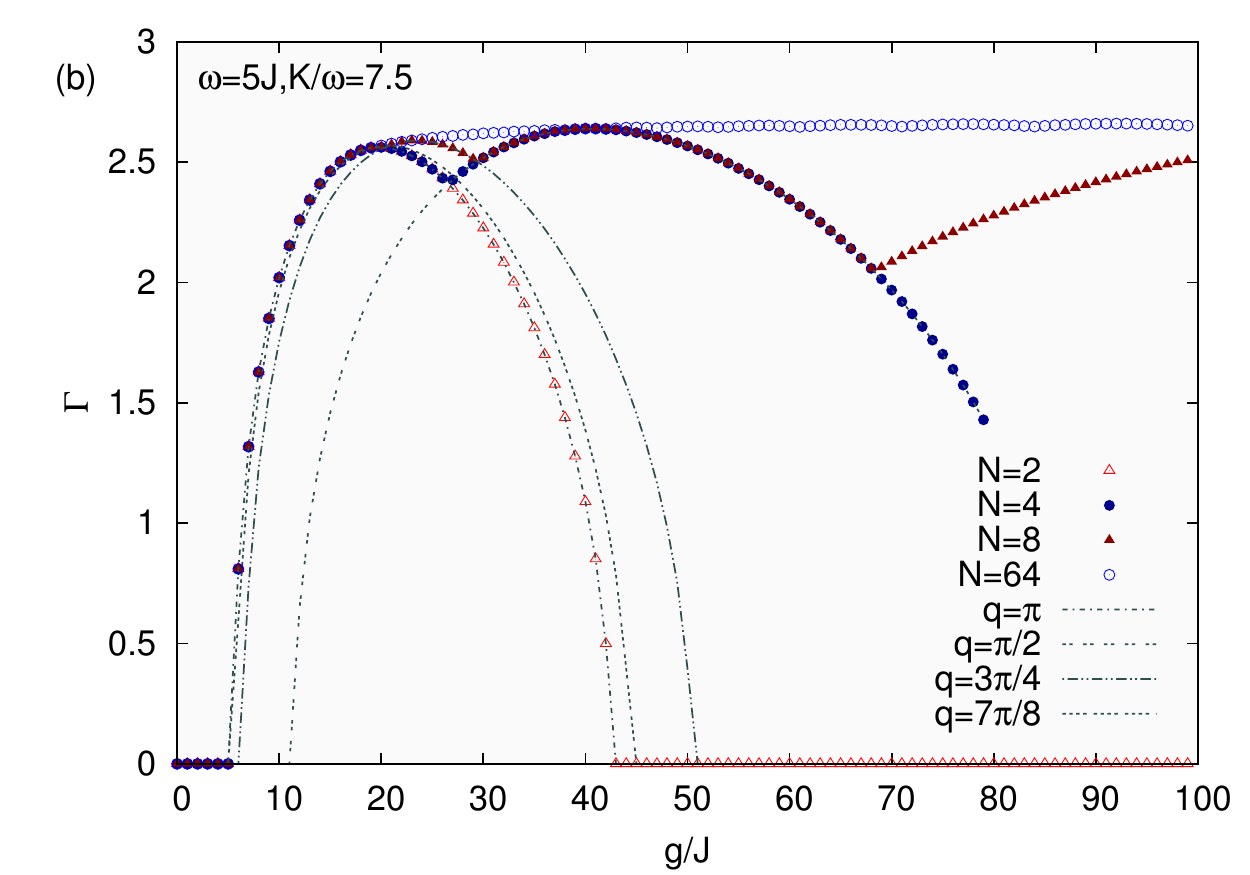}
		\end{minipage}
		\caption{Total instability rate and contributions of a few modes as a function of the interaction strength $g$, for $\omega=J$ (``low-frequency" regime) and $\omega=5J$ (outside the ``low-frequency" regime), and for an increasing number of lattice sites $N=2,4,6,64$. The total instability rate is the envelope of the curves associated with ``available" modes in the system (for $N=2$, only $q=0,\pi$; for $N=4$, only $q=0,\pi/2,\pi,3\pi/2$...). The curve gets smoother when increasing the number of sites. Outside the low-frequency regime, the transition to instability is governed by the mode $\pi$ and already captured by a two-site system. In the low-frequency regime, the instability rate at low $g$, which is governed by a certain mode $q_\mathrm{res}$, is probed with more and more accuracy when gradually increasing $N$: here, the 2-site system is always stable since $q_\mathrm{res}>\pi$, the 4-site system displays a transition since $q_\mathrm{res}>\pi/2$, and then this transition lowers when probing $q_\mathrm{res}$.
			\label{fig:FSS1}}
	\end{figure}
	The analytical approach, which treats different modes independently, allows one to readily predict the behaviour of finite-size systems under periodic modulation, close to a parametric resonance. In this case, the ``continuous" dispersion $\Eav(q)$ is replaced by discrete energy modes. For a small number of lattice sites (and thus of modes), the total instability rate $\Gamma\!=\!\max_q s_q$, which is given by the envelope of the curves associated with individual modes (see Fig.~\ref{fig:Bumps}), may not be a monotonic curve as a function of $g$. Instead, it is rather composed of disjointed bumps if the resonance domains of two consecutive modes are disconnected. Figure~\ref{fig:FSS1} shows how individual modes contribute to the total growth rate when increasing the number of sites, both in and outside the low-frequency regime. Interestingly, the situation is slightly different in these two cases. 
	Outside the low-frequency regime, i.e.~for $\omega\gtrsim \Jeff$, the onset of instability is governed by the mode $q\!=\!\pi$ [see Sec.~\ref{sec:ana}], and therefore, the critical frequency, below which the system becomes unstable, happens to be the same  for both an infinite and a two-site system. Since the mode $q=\pi$ remains the ``most unstable" one, all the way up to large values of $g$ [see Fig.~\ref{fig:Bumps}], the stability diagram of a two-site system is in fact very similar to the one associated with an infinite system (whereas a one-site system is trivially stable) \footnote{Only very small systems with an odd number of sites (thus without featuring a mode $q\!=\!\pi$) are expected to display differences on the stability boundary.}. Conversely, in the low-frequency regime ($\omega\lesssim \Jeff$), instability occurs already at vanishingly small $g$, induced by a certain resonant mode $q_{\mathrm{res}}$ fulfilling the resonance condition [see Sec.~\ref{sec:ana}]. 
	Increasing the number of sites therefore lowers the instability boundary as a function of $g$, since increasing the number of points in momentum space allows for the states to get closer to the maximally unstable mode. Moreover, if the number of sites $N$ is too small, the discretisation in momentum space is so rough that the resonance domains of consecutive modes may be disconnected, resulting in ``islands" of instability in the phase  diagram. These islands begin to merge with increasing $N$, as the resonance domains of adjacent modes overlap, only gradually approaching the phase diagram of an infinite system. Both our analytical and numerical methods capture this behaviour, and agree for any number of lattice sites. Interestingly, since the instability rate is determined by its maximal value over all modes (i.e.~the envelope of all curves in Fig.~\ref{fig:FSS1}), its estimate obtained for finite-size systems is already very decent, and further increasing the number of sites just smoothens the curve of the instability rate as a function of $g$. 
	
	We point out that for very small systems, there might be minor corrections due to edge states: indeed, the previous treatment, formulated in momentum space, tacitly implies periodic boundary conditions. However, in the lab frame, the drive actually breaks translational invariance and the boundary conditions are intrinsically open, possibly yielding different selection rules for states near the edges. Although such a difference is expected to play a negligible role in large systems, it may lead to minor but noticeable deviations in small systems.

	\section{The Effect of Transverse Directions: 2D lattices vs tubes}
	\label{sec:2D}
	
	So far, our study focused on the origin of parametric instabilities that occur in a driven system, which satisfies the 1D GPE on a lattice~\eqref{eq:GPE}. As further discussed in Sec.~\ref{sec:BH}, this model can be used to describe the physics of weakly-interacting bosonic atoms, trapped in a 1D optical lattice. However, from an experimental point of view, it is intriguing to determine the effects of transverse directions on the stability diagram and instability rates. Indeed, optical-lattice experiments involving weakly-interacting bosons~\cite{aidelsburger2014,kennedy2015} typically feature continuous transverse degrees of freedom, commonly referred to as ``tubes" or ``pancakes". Besides, experiments realizing artificial magnetic fields~\cite{dalibard2011,goldman2014a} involve two-dimensional optical lattices, and hence, it is relevant to study the fate of instabilities as one transforms a 1D optical lattice into a full 2D lattice, by adding sites (and allowing for hopping processes) along a transverse direction.


In this section, we extend our previous analysis to study the effects of (i) a secondary tight-binding-lattice direction (resulting in a ladder or a full 2D lattice), and (ii) an additional continuous (``tube'") degree of freedom. In the following, we assume that the periodic modulation remains exclusively aligned along the original tight-binding lattice dimension.


	\subsection{Two-dimensional lattice geometry}
	\label{sec:2Dlat}
	
	In this Section, we consider the addition of a transverse lattice direction, aligned along the $y$-axis; by doing so, we keep the same time-dependent modulation as before, which is hence exclusively aligned along the $x$-axis. Our aim is to study how an increase in the number of sites along the transverse direction affects the instability rates, previously evaluated for the 1D configuration. Theoretical studies~\cite{bilitewski2015,choudhury2015a} have reported that heating and collisional processes are enhanced by the presence of a transverse direction, which provides crucial information for current experimental studies. 
	
	For this extended model, the time-dependent GPE  reads 
	\begin{align}
		i \partial_t a_{m,n}=&-J(a_{m,n+1}+a_{m,n-1} + a_{m+1,n}+a_{m-1,n}) \notag \\
		&+K\cos(\omega t)ma_{m,n}+U|a_{m,n}|^2 a_{m,n},
		\label{eq:GPE2D}
	\end{align}
	where each site of the underlying (2D) square lattice is now labelled by two integers $(m,n)$. As for the 1D case [Sec.~\ref{sec:Imodel}], the condensate wavefunction at time $t\!=\!0$ is again given by a Bloch state $\mathrm e^{i(p_x m+p_y n)}$, with momentum $p_x=0$ if $\mathcal{J}_0(K/\omega)>0$ and $p_x=\pi$ if $\mathcal{J}_0(K/\omega)<0$; for this choice of the time-modulation, the momentum along the $y$ direction is necessarily $p_y=0$. The time-dependent Bogoliubov-de Gennes equations, which govern the time evolution of the mode ${\bf q}=(q_x,q_y)$, take the form
	\begin{equation}
		i \partial_t \left( \begin{matrix} u_q  \\ v_q \end{matrix} \right)=\left( \begin{matrix}  \varepsilon(\mathbf{q},t)+g & g        \\ -g & -\varepsilon(-\mathbf{q},t)-g \end{matrix} \right)\left( \begin{matrix} u_q  \\ v_q \end{matrix} \right),
		\label{eq:BdGEk2D}
	\end{equation}
	where $$\varepsilon(\mathbf{q},t)=4J\sin\dfrac{q_x}{2}\sin\left(\dfrac{q_x}{2}+p_x-\dfrac{K}{\omega}\sin\omega t\right) + 4J\sin^2\dfrac{q_y}{2},$$ which is a straightforward generalization of Eq.~(\ref{eq:BdGEk_pre}). Therefore, both the numerical and the analytical methods presented in the previous sections can be readily applied to Eq.~(\ref{eq:BdGEk2D}).
	
	\begin{figure}[h!]
		
		\begin{minipage}{8cm}
			
			\includegraphics[width=8.3cm]{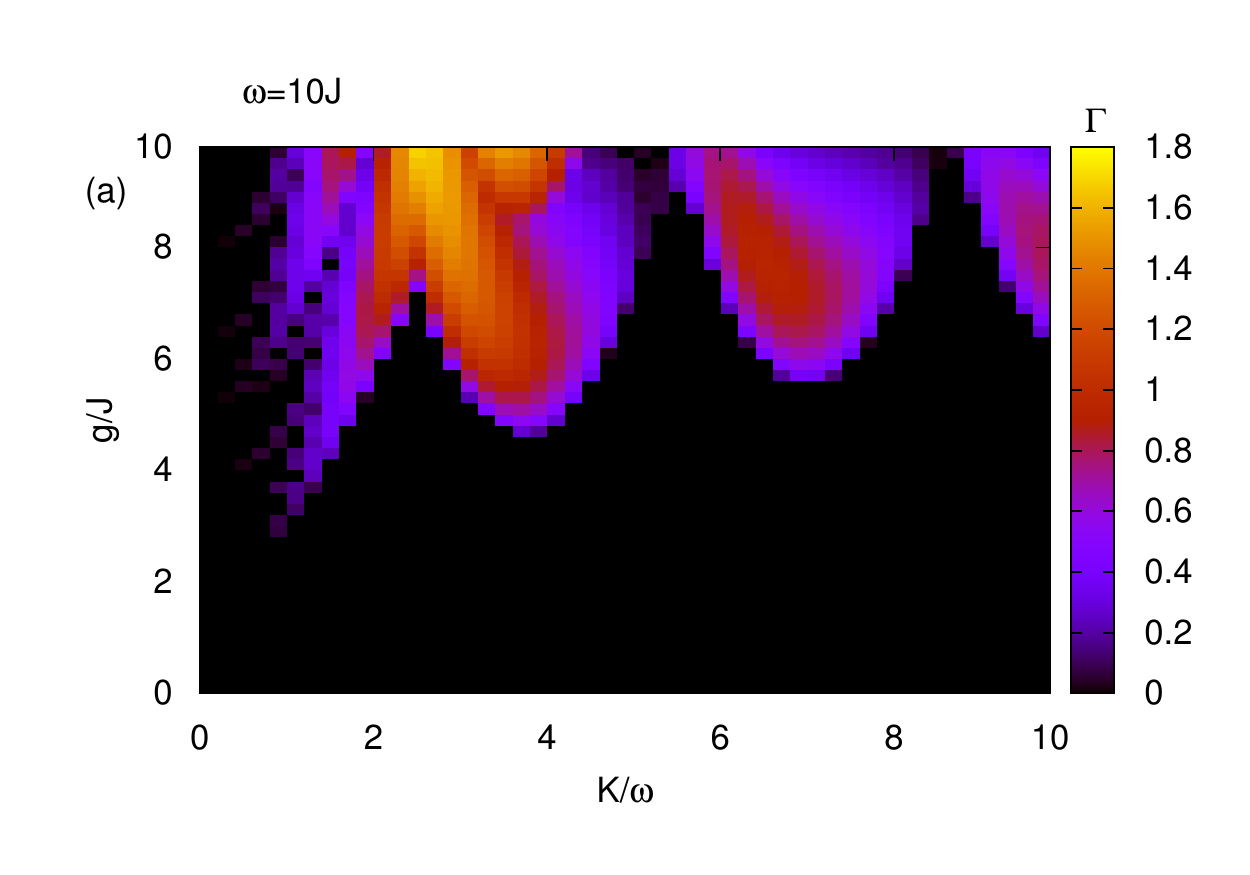}
		\end{minipage}
		\begin{minipage}{8cm}
			\centering
			\includegraphics[width=8.3cm]{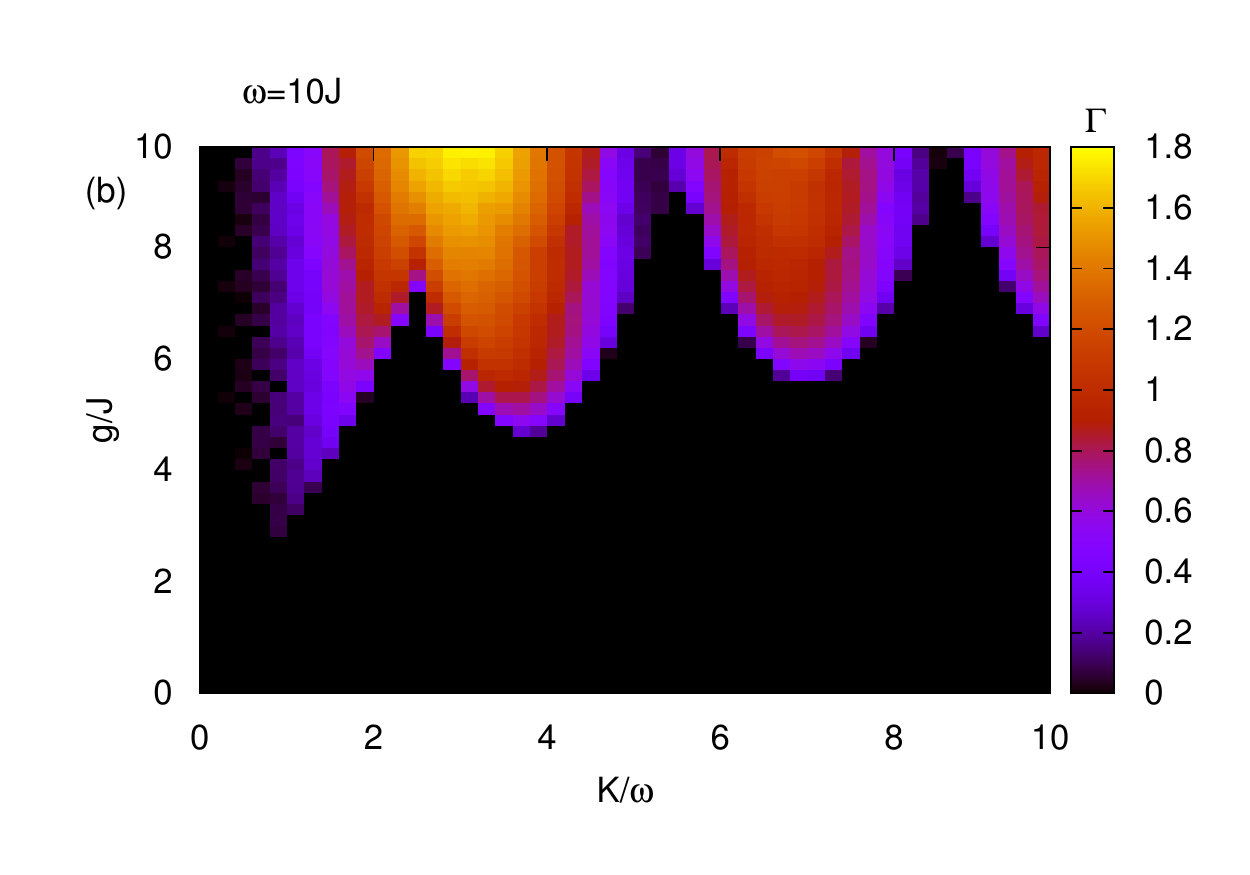}
		\end{minipage}
		\caption{Numerical instability rate as a function of interaction strength $g=U\rho$ and modulation amplitude $K/\omega$, for $\omega=10J$ (i.e. outside the ``low-frequency" regime) and for two different number of lattice sites in the transverse direction: $N_y=4$ (top) and $N_y=16$ (bottom). 
			\label{fig:2DNum10}}
	\end{figure}
	\begin{figure}[h!]
		
		\begin{minipage}{8cm}
			
			\includegraphics[width=8cm]{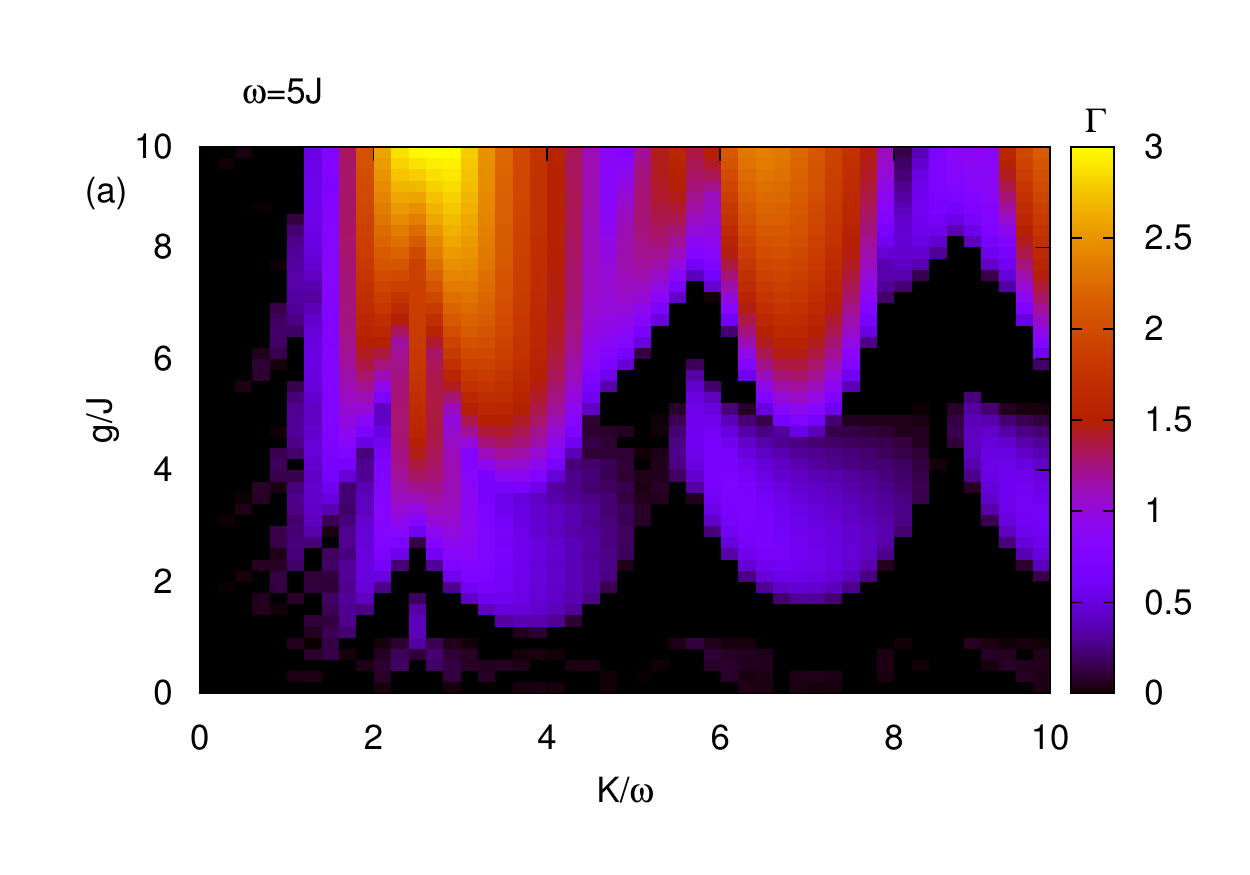}
		\end{minipage}
		\begin{minipage}{8cm}
			\centering
			\includegraphics[width=8cm]{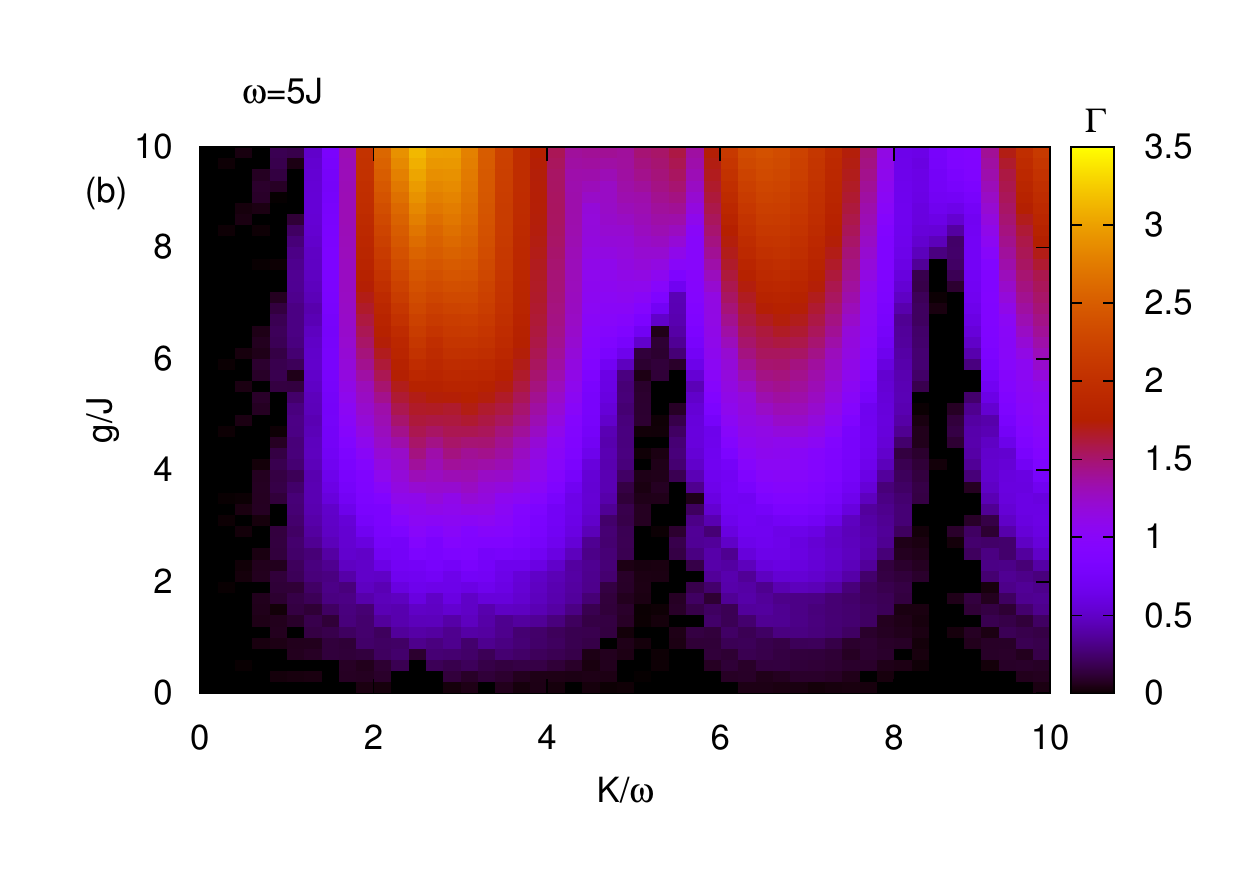}
		\end{minipage}
		\caption{Numerical instability rate as a function of interaction strength $g=U\rho$ and modulation amplitude $K/\omega$, for $\omega=5J$ (i.e. in the ``low-frequency" regime) and for two different number of lattice sites in the transverse direction: $N_y=4$ (top) and $N_y=16$ (bottom). 
			\label{fig:2DNum5}}
	\end{figure}
	Figures~\ref{fig:2DNum10} and~\ref{fig:2DNum5} show the stability diagrams, obtained numerically by increasing the number $N_y$ of transverse lattice sites (from $4$ to $16$); they correspond to the drive frequency $\omega=10J$ (i.e.~``high-frequency" regime, where $\omega$ is greater than the effective bandwidth) and $\omega=5J$ (i.e.~``low-frequency" regime), respectively. Since the motional degrees of freedom along the $x$ and $y$ directions are independent of each other, the problem being separable, the situation is analogous to that previously discussed in the case of finite-size systems. 
	
	Let us summarize the results: away from the ``low-frequency" regime [Fig.~\ref{fig:2DNum10}], namely for $\omega\!>\!4(J+\Jeff)$ [the drive frequency is again compared to the modified (effective) free-particle bandwidth], the system is stable for small $g$ and the onset of instability is governed by the mode of maximal energy (i.e.~the first mode to exhibit a resonance), which now corresponds to $\mathbf{q}\!=\!(\pi,\pi)$. Therefore, as soon as two sites are present in the transverse direction, the transition point to the unstable regime is well-captured. Moreover, both the instability boundary and the instability rates in its vicinity are expected to remain unaffected when increasing $N_y$, since they are dominated by the mode $(\pi,\pi)$ [see Eq.~(\ref{eq:spi})]. Conversely, in the low-frequency regime, there is an instability at vanishingly small $g$, which is driven by a certain mode $(q_x,q_y)$: in this case, adding the number of transverse sites lowers the instability boundary, as increasing the resolution in momentum space allows one to get closer to this maximally unstable mode. 
	
\begin{figure}[h!]
		
		\begin{minipage}{8cm}
			\vspace*{-0.8cm}
			\includegraphics[width=8.3cm]{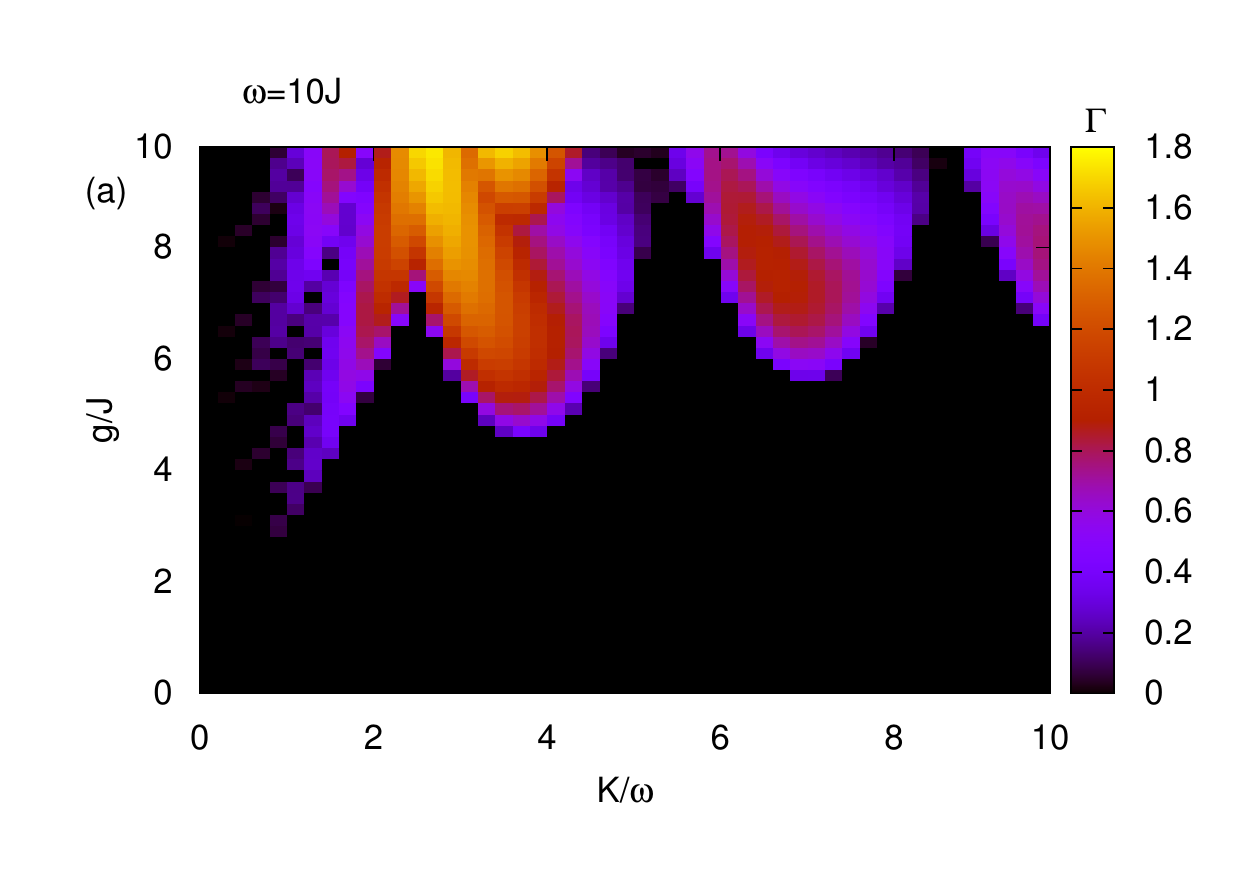}
		\end{minipage}
		\begin{minipage}{8cm}
			\centering
			\vspace*{-0.8cm}
			\includegraphics[width=8.3cm]{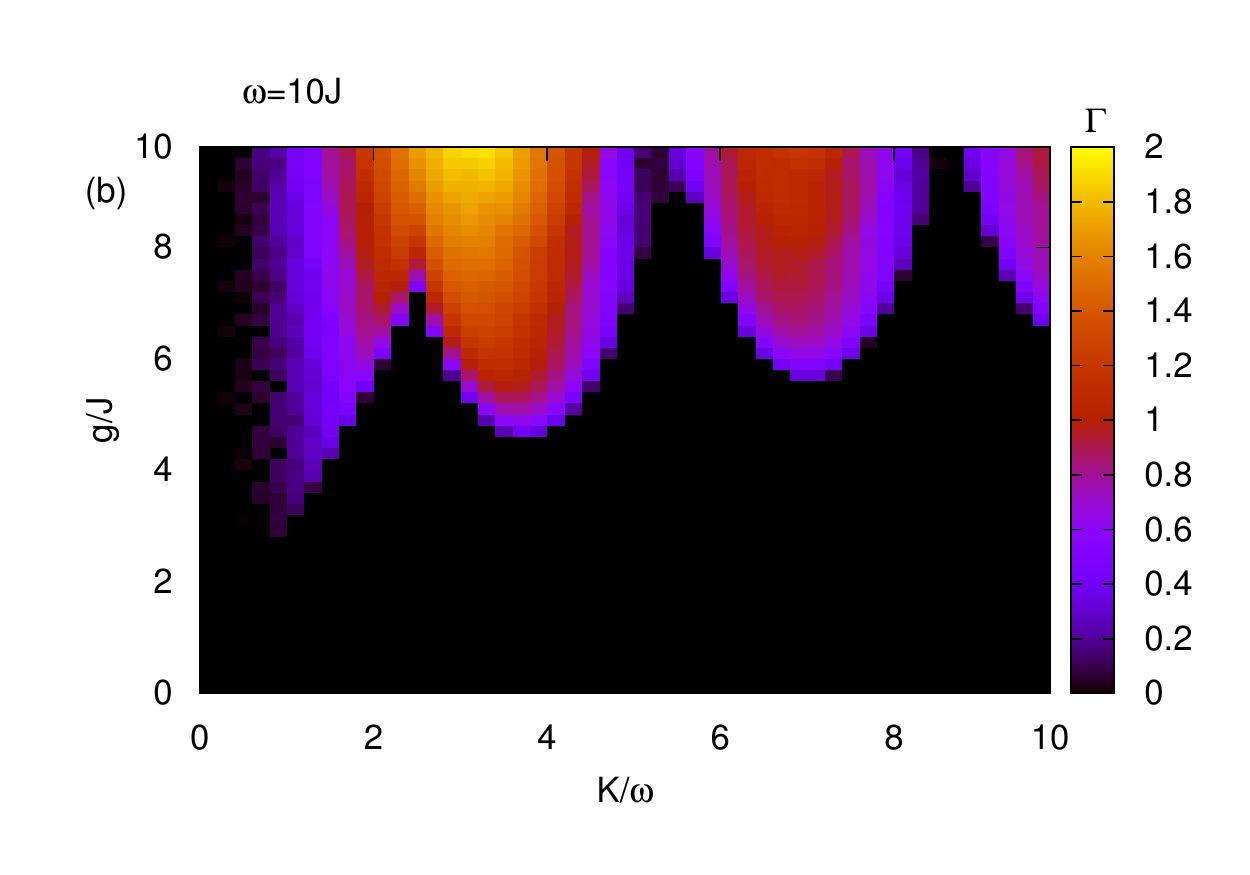}
		\end{minipage}
		\caption{Analytical instability rate as a function of interaction strength $g=U\rho$ and modulation amplitude $K/\omega$, for $\omega=10J$ (i.e. outside the ``low-frequency" regime) and for two different number of lattice sites in the transverse direction: $N_y=4$ (top) and $N_y=16$ (bottom). To be compared with Fig.~\ref{fig:2DNum10}.  
			\label{fig:2DAna10}}
	\end{figure}

	Using Eq.~(\ref{eq:BdGEk2D}), one can readily extend the analytical approach of Sec.~\ref{sec:ana}. Indeed, all the results of Sec.~\ref{sec:ana} remain applicable  in the present case, provided that the dispersion $\Eav(\mathbf{q})$ is now replaced by
	\begin{align}
		\Eav(\mathbf{q})=&\sqrt{\left(4|\Jeff|\sin^2 q_x/2 + 4J\sin^2 q_y/2 \right)} \label{eq:Eav2D} \\
		&\times \sqrt{\left(4|\Jeff|\sin^2 q_x/2+4J\sin^2 q_y/2 +2g\right)}.\notag		
	\end{align}
	The resulting stability diagrams are shown in Fig.~\ref{fig:2DAna10} for the same parameters as in Fig.~\ref{fig:2DNum10}, and they reveal an excellent agreement with the latter. Interestingly, the agreement between analytics and numerics is even better than for the 1D configuration, since the presence of transverse modes in the effective dispersion Eq.~\eqref{eq:Eav2D} prevents $\Eav(\mathbf{q})$ from vanishing when $\mathcal{J}_0(K/\omega)\!=\!0$~\footnote{we recall that this singular vanishing of the Bessel function is responsible for a breakdown of the perturbative approach in the 1D case, which resulted in small disagreement zones [C and D] between the numerical and analytical stability diagrams [Fig.~\ref{fig:2Diags}]}.
	
	%
	
	\subsection{Continuous transverse degrees of freedom: the case of ``tubes" }
	\label{sec:2DCont}
	
	We now consider the case where the transverse degree of freedom corresponds to free-motion along a continuum, in one or more transverse directions. In this case, the Bogoliubov-de Gennes equations still take the form of Eq.~(\ref{eq:BdGEk2D}), now with 
	$$\varepsilon(\mathbf{q},t)=4J\sin\dfrac{q_x}{2}\sin\left(\dfrac{q_x}{2}+p_x-\dfrac{K}{\omega}\sin\omega t\right) + \dfrac{\qp^2}{2m},$$ so that the previous analysis still holds, provided that the time-averaged Bogoliubov dispersion $\Eav(\mathbf{q})$ is now replaced by
	\begin{align}
		\Eav(\mathbf{q})=&\sqrt{\left(4|\Jeff|\sin^2 q_x/2 + \qp^2/2m \right)} \label{eq:Eav2DC0}\\
		&\times \sqrt{\left(4|\Jeff|\sin^2 q_x/2+\qp^2/2m +2g\right)}.\notag		
	\end{align}
	Interestingly, since this dispersion is unbounded, $\omega$ will always be smaller than the total bandwidth, and hence, there will always be (at least) one mode that is precisely set on resonance:~in other words, the system necessarily falls into the  ``low-frequency regime" previously introduced, where it is unstable at any finite $g$. As explained in Sec.~\ref{sec:anaEff}, the instability rate can be evaluated, at lowest order, by calculating the rate on resonance associated with the mode(s) satisfying the condition $\omega\!=\!\Eav(\qbf^\mathrm{res})$ [see Eq.~(\ref{eq:sresR})]: 
	\begin{align}
	s^*_{\qbf^\mathrm{res}}&=4g \mathcal{J}_2(K/\omega)\sin^2(q_x^\mathrm{res}/2) J/\Eav(\qbf^\mathrm{res}), \notag \\
	&=4g \mathcal{J}_2(K/\omega)\sin^2(q_x^\mathrm{res}/2) J/\omega .\label{gamma_res}
	\end{align}
Importantly, although there are generically several resonant modes $\qbf^\mathrm{res}$ (corresponding to different $q_x$ and $\qp$), they do not have the same instability rate $s^*_{\qbf^\mathrm{res}}$:~the total instability rate $\Gamma$ is then attributed to the most unstable mode,   namely 
	\begin{equation}
	\Gamma\!=\! \max_{\qbf^\mathrm{res}} \, \left (s^*_{\qbf^\mathrm{res}} \right ) = s^*_{\qbf^\mathrm{mum}},
	\end{equation} 
	where we introduced the notation $\qbf^\mathrm{mum}$ to denote the momentum of the \emph{most unstable mode}.
	
	We then identify two cases:
	\begin{itemize}
		\item[(i)] If $\omega\!>\!\sqrt{4\vert \Jeff \vert(4\vert\Jeff\vert+2g)}$, which is often the case in realistic configurations, there exists a resonant mode at $q_x^\mathrm{mum}\!=\!\pi$, which is thus the most unstable one; $\qp^\mathrm{mum}$ is then adjusted so as to respect the resonance condition. The maximally unstable mode thus reads
		\begin{equation}
		q_x^\mathrm{mum}=\pi; \quad (\qp^\mathrm{mum})^2/2m = \sqrt{g^2+\omega^2}-g-4|\Jeff|.
		\label{eq:qres2DC0}
		\end{equation}
		In this case, the total instability rate is given by the simple analytical formula 
		\begin{equation}
		\Gamma= 4J\left|\mathcal{J}_2(K/\omega)\right|\dfrac{g}{\omega}.
		\label{eq:tauxAna2DC0}
		\end{equation}
		
		\item[(ii)] If $\omega\!<\!\sqrt{4\vert\Jeff\vert(4\vert\Jeff\vert+2g)}$, the most unstable mode, which obeys the resonance condition $\omega\!=\!\Eav(\qbf)$, is necessarily reached at $\qp^\mathrm{mum}\!=\!0$, yielding 
		\begin{equation}
		q_x^\mathrm{mum}= 2 \arcsin \sqrt{\dfrac{\sqrt{g^2+\omega^2}-g}{4|\Jeff|}} ; \quad \qp^\mathrm{mum}=0.
		\label{eq:qres2DC0bis}
		\end{equation}
		In this case, the total instability rate is given by 
		\begin{align}
		\Gamma=(\sqrt{g^2+\omega^2}-g) \left|\dfrac{\mathcal{J}_2(K/\omega)}{\mathcal{J}_0(K/\omega)}\right|\dfrac{g}{\omega}.
		\label{eq:tauxAna2DC0bis}
		\end{align}
	\end{itemize}

\begin{figure}[!t]
	
	\begin{minipage}{8cm}
		\includegraphics[width=8.3cm]{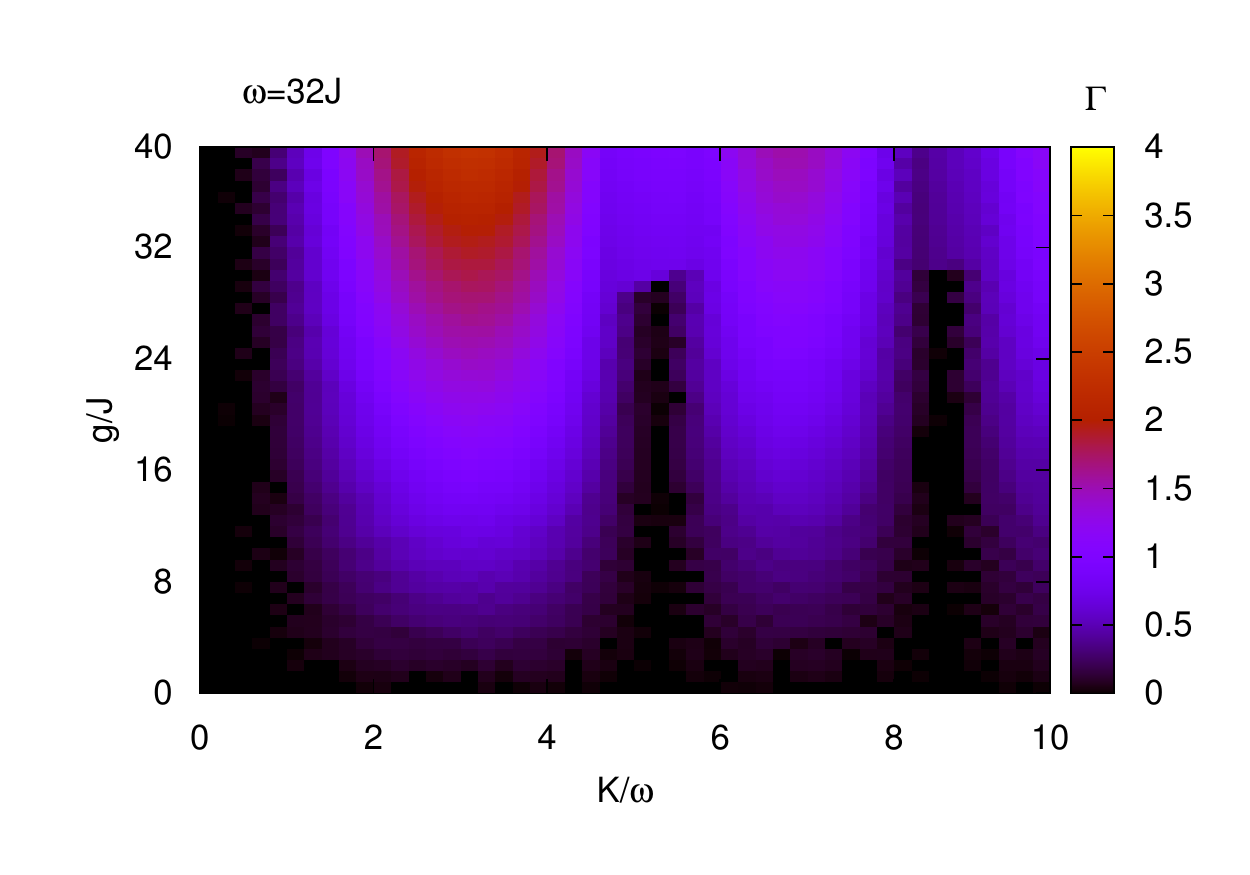}
	\end{minipage}
	\begin{minipage}{8cm}
		\centering
		\includegraphics[width=8.3cm]{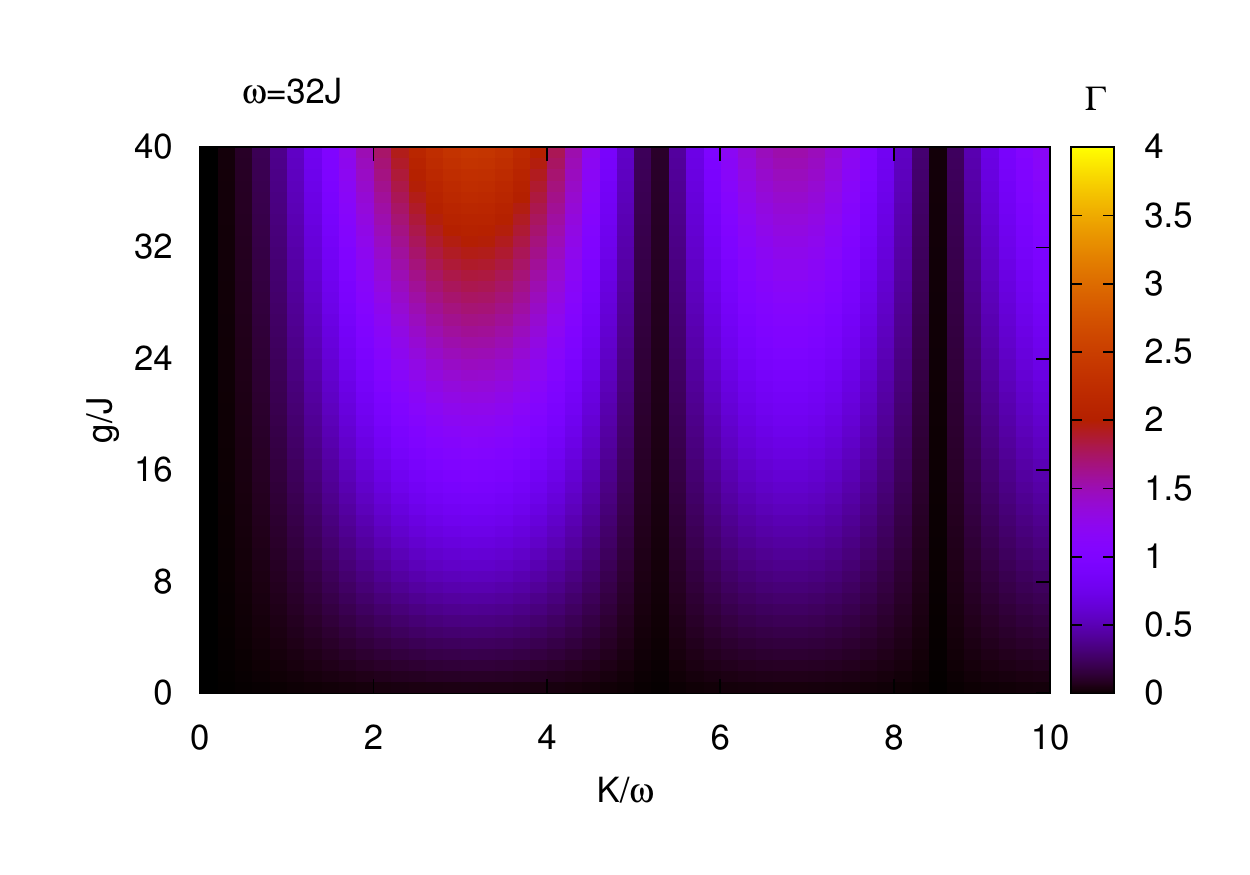}
	\end{minipage}
\vspace{0.5cm}
	\caption{Numerical (top) and analytical (bottom) instability rates $\Gamma$ as a function of the modulation amplitude $K/\omega$ and interaction strength $g$, for a high drive frequency $\omega\!=\!32J$.
      \label{fig:FSS}}
\end{figure}
\begin{figure}[!t]
	
	\begin{minipage}{8cm}
		
		\includegraphics[width=8.3cm]{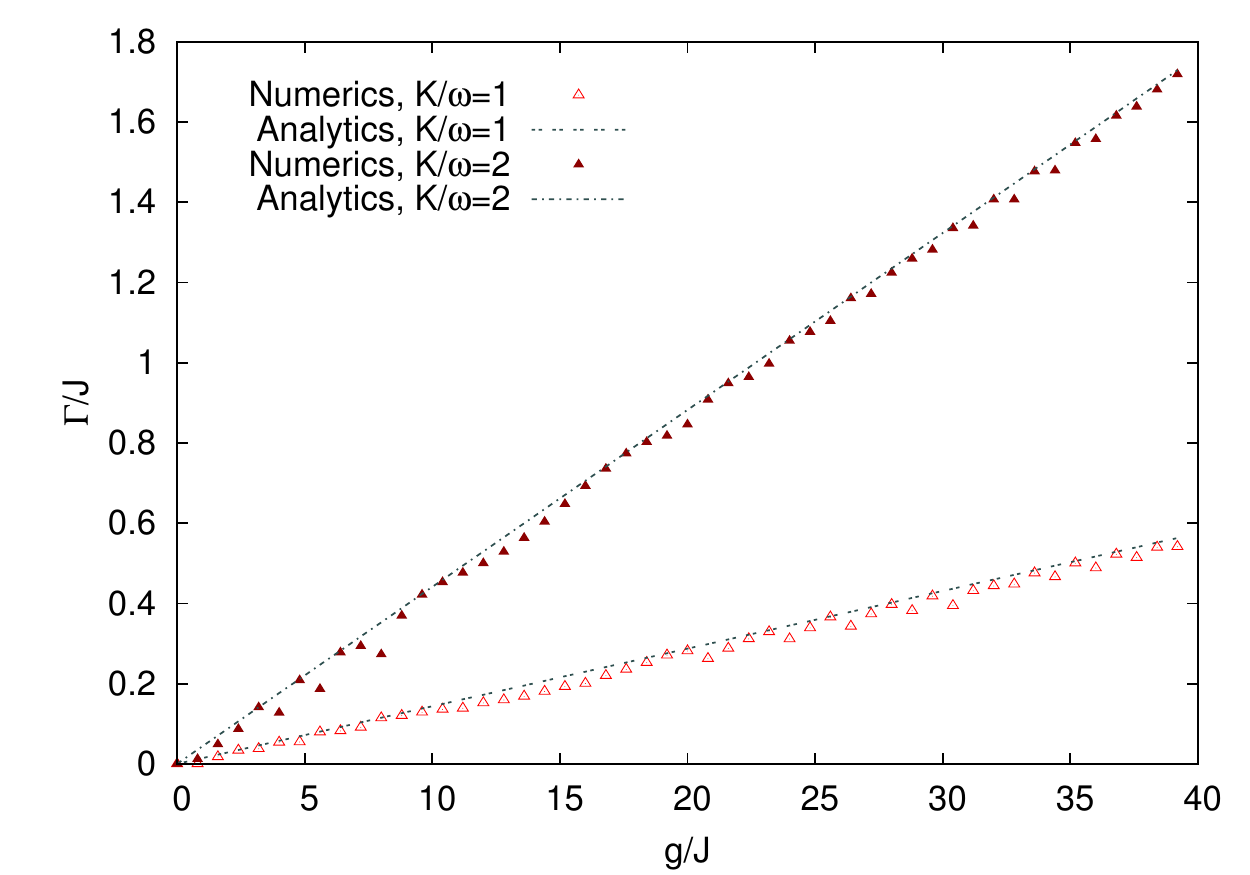}
	\end{minipage}
	\begin{minipage}{8cm}
		\centering
		\includegraphics[width=8.3cm]{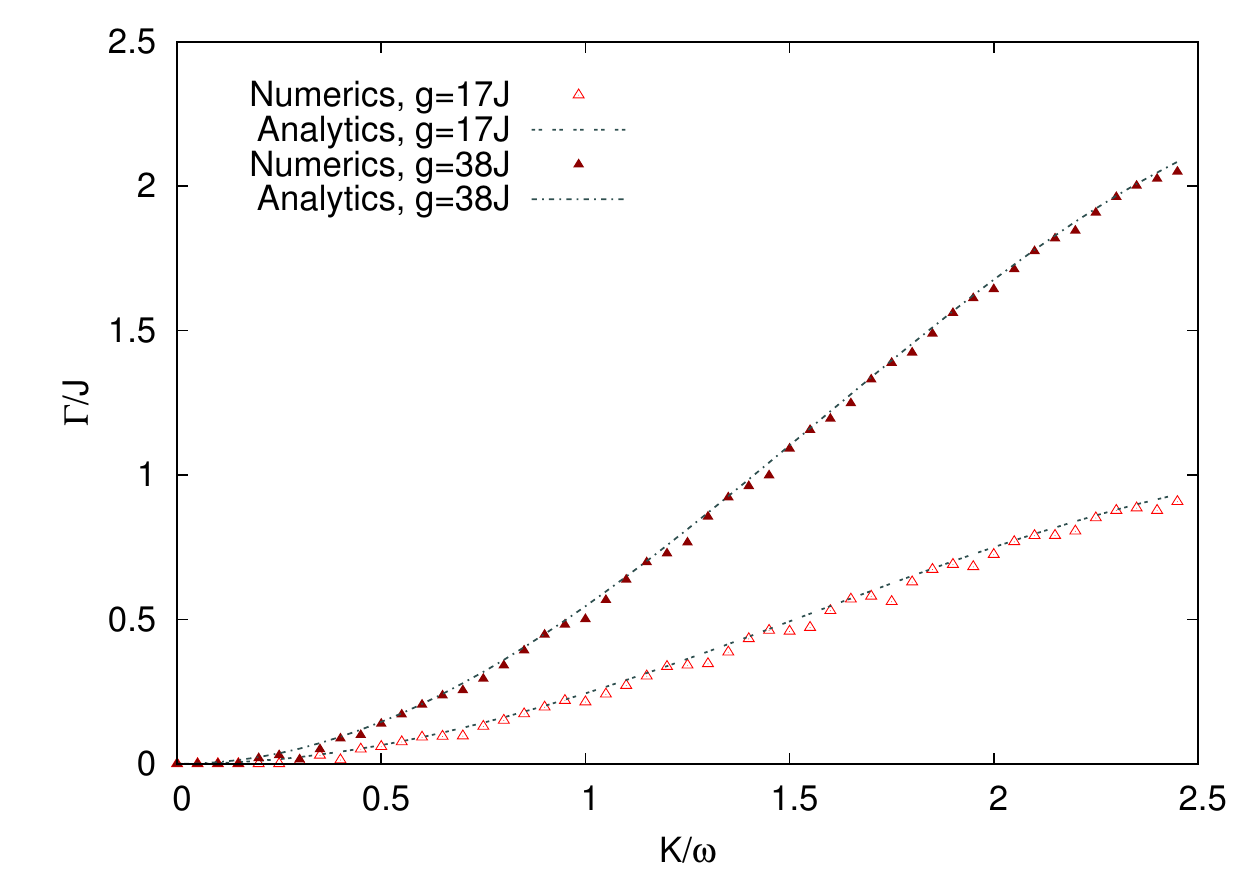}
	\end{minipage}
\vspace{1cm}
	\caption{Top: Instability rate as a function of the interaction strength $g$ for $\omega\!=\!32J$ and for two values of the modulation amplitude. As captured by the analytical expression Eq.~(\ref{eq:tauxAna2DC0}), the instability rate increases linearly with $g$. 
		 Bottom: Instability rate as a function of the modulation amplitude $K/\omega$ for $\omega\!=\!32J$ and for two values of the interaction parameter $g$. As captured by the analytical expression Eq.~(\ref{eq:tauxAna2DC0}), the instability rate increases quadratically with $K/\omega$ at low amplitudes. 
    \label{fig:FSS2}}
\end{figure}
Interestingly, these results do not depend on the number of transverse dimensions, since all that matters is the existence of an unbounded continuum.
Figure~\ref{fig:FSS} compares the numerical stability diagram to the analytical formula Eq.~(\ref{eq:tauxAna2DC0}), for realistic (experimental) parameters. As visible on Fig.~\ref{fig:FSS2}, which shows cuts through the diagrams of Fig.~\ref{fig:FSS}, the agreement is excellent at low interaction/modulation amplitude, and the scaling of the instability rate is remarkably simple: $\Gamma$ increases linearly with $g$ and quadratically with $K$. Such simple results could promisingly be compared with experiments. 
Note that in experiments, one typically has $J/\hbar\approx \mathrm{100 Hz-1kHz}$, so that the inverse instability rates are typically of the order of $\mathrm{1-10 ms}$.

	At this stage, it is worth comparing those results with the more traditional Fermi-Golden-Rule (FGR) approach: first, in the present case, and contrary to what is expected for a FGR regime~\cite{bilitewski_14}, the final rate can be inferred from the contribution of a single (well-identified) mode, and not from a sum over all modes (which would then involve the density of states in the description). Second, the form of the final formula in Eq.~\eqref{eq:tauxAna2DC0} is significantly different from that resulting from a FGR argument, where the rate would typically scale as $g^2$ (as dictated by the squared matrix element of the perturbation~\cite{bilitewski_14}). When considering a driven weakly-interacting degenerate Bose gas, one expects the short-time dynamics to be dominated by the parametric instability identified in this work; at longer times, as the condensate significantly depletes, heating rates are expected to be dominated by a FGR behavior~\cite{bilitewski_14,bilitewski2015}, not captured in the present analysis. We believe that those two distinct mechanisms and instability regimes could be probed by current ultracold-atom experiments.\\
	
	As a practical remark, we emphasize that the results presented in this section suggest that working with a lattice seems to be more favorable than with tubes/pancakes, in view of reducing parametric instabilities in periodically-driven Bose systems.
	
	\section{The Periodically-Driven Bose-Hubbard model}
	\label{sec:BH}

		In this section, we discuss the relation between the Gross-Pitaevskii equation introduced in Eq.~\eqref{eq:GPE} and the (full quantum) driven Bose-Hubbard model~\cite{Eckardt:2016Review}. Our goal is to clarify the validity of our analysis in view of describing parametric instabilities and heating in the context of this quantum model.
		
		\subsection{Effective Hamiltonian and micromotion operator}	
		
		For the sake of concreteness, here we focus  on the 1D model of Eq.~\eqref{eq:GPE}, but the discussion applies to all models investigated in the paper (in particular the two-dimensional models of Section~\ref{sec:2D}).
		Consider a system of weakly-interacting bosons, trapped in a shaken 1D lattice, as described by the periodically-driven Bose-Hubbard Hamiltonian~\cite{Eckardt:2016Review}
		\begin{align}
			\hat{H}(t)=&-J\sum_n (\ahd_{n+1}\ah_n+ \mathrm{h.c.})+K\cos(\omega t)\sum_n n \ahd_n\ah_n \notag \\
			&+ \frac{U}{2} \sum_n \ahd_n\ahd_n\ah_n\ah_n,
			\label{eq:H}
		\end{align}
		where $J>0$ denotes the tunnelling amplitude of nearest-neighbour hopping, and $U>0$ is the on-site interaction strength. The on-site potential term describes a time-periodic modulation of amplitude $K$ and frequency $\omega=2\pi/T$.  

		As we stated already in the introduction, the dynamics is stroboscopically governed by the effective Floquet Hamiltonian, see Eq.~\eqref{eq:Floquet_thm}.
		In the absence of interactions, it is exactly~\footnote{Equation~\eqref{eq:Heff} was derived in the thermodynamic limit. For finite-size systems with open boundary conditions, there are additional corrections that become relevant for $\omega\lesssim 2J_\mathrm{eff}$.} given by~\cite{Eckardt:2016Review,goldman2014b,bukov2014}
		\begin{align}
		\label{eq:Heff}
			&\hat{H}_{\mathrm{eff}}(U=0)=-\Jeff\sum_n (\ahd_{n+1}\ah_n+\mathrm{h.c.}) , \\
			& \Jeff\!=\!J\mathcal{J}_0(K/\omega) .\notag			
		\end{align}
		Besides, the kick operator [Eq.~\eqref{eq:Floquet_thm}] reads~\cite{goldman2015a}
		\begin{equation}
			\hat{K}_\mathrm{kick}(t)= i\log\left[ \hat{R}(t)\mathrm{e}^{-i\hat{K}^\mathrm{rot}_\mathrm{kick}(t)}  \right],
			\label{eq:Klab}
		\end{equation}
		where $\hat{R}(t)\!=\!\e^{-iK/\omega\sin(\omega t)\hat{X}}$ is the unitary transformation to the rotating frame (with $\hat{X}\equiv\Sigma_n n \ahd_n\ah_n$ the position operator on the lattice). Here $\hat{K}_\mathrm{kick}^\mathrm{rot}$ denotes the kick operator in the rotating frame~\cite{goldman2015a} , which is explicitly given by 
		\begin{align}
			\hat{K}_\mathrm{kick}^\mathrm{rot}(t) = -2J\sum_{q\in\mathrm{BZ}} \biggl[& \int_0^t \cos(q - K/\omega\sin\omega t')\mathrm{d} t' \notag \\
			& - \mathcal{J}_0(K/\omega)\cos(q)  \biggr]   a^\dagger_ka_k.
			\label{eq:Krot}
		\end{align}
		
		Whenever interactions are present, Eq.~\eqref{eq:Heff} is no longer exact, and the effective Hamiltonian is a much more complicated (possibly nonlocal) object~\cite{bukov_14,anisimovas2015}. In the high-frequency regime, it can be approximated  using the inverse-frequency expansion~\cite{goldman2014b}, which to lowest order yields
		\begin{align}
			\hat{H}_{\mathrm{eff}} =& -\Jeff\sum_n (\ahd_{n+1}\ah_n+\mathrm{h.c.}) \notag \\
			&+\dfrac{U}{2}\sum_n \ahd_n\ahd_n\ah_n\ah_n + \mathcal{O}(\omega^{-1}).
			\label{eq:HeffG}
		\end{align}
		Hence, in the infinite-frequency limit, the periodic drive merely renormalizes the hopping matrix element $J\to J_\mathrm{eff}$. Due to the Bessel function taking both positive and negative values, it is possible to tune the amplitude-to-frequency ratio such that $\mathcal{J}_0(K/\omega)\!=\!0$, in which case tunneling is completely suppressed~\cite{dunlap_86,dunlap_88,eckardt2005}. Therefore, even a weakly-interacting lattice system can effectively behave as a strongly interacting one under periodic driving, in the sense that the interaction strength $U$ potentially dominates over the tunneling when $\Jeff\!\approx\!0$. We recall that the lattice dispersion flips sign when  $\mathcal{J}_0(K/\omega)\!<\!0$, in which case the stable minimum appears at quasi-momentum $q\!=\!\pi$:~namely, at equilibrium, bosons are expected to condense in a finite-momentum state in this regime. We point out that corrections to the Hamiltonian~\eqref{eq:HeffG} become non-negligible away from the high-frequency limit, and that these are expected to manifest themselves over longer time scales.\\ 
		
		\subsection{Relation to the mean-field approach and observables}
		
		The mean-field approach presented in this paper [based on Eq.~\eqref{eq:GPE}] is expected to provide a good description for the shaken 1D Bose-Hubbard model [Eq.~\eqref{eq:H}] at short times, as far as the weakly-interacting regime is concerned.
		More specifically, if one assumes that the initial state is that of fully condensed bosons, the system in Eq.~\eqref{eq:H} can be treated within mean-field theory~\cite{popov1983}. In this approximation, the annihilation and creation operators $\ah_n,\ahd_n$ are merely replaced by classical fields $a_n,a^*_n$, and the Heisenberg equations of motion for $\ah_n,\ahd_n$ lead to the time-dependent GPE in Eq.~\eqref{eq:GPE}. 
		
		Throughout this work, our choice for the initial state corresponds to the Bogoliubov ground-state of the effective Hamiltonian \eqref{eq:HeffG} [see Section~\ref{sec:Imodel}].  In order to compute heating rates for such a model, following the procedure detailed in Sec.~\ref{sec:NumEn}, we also had  to fix the initial condition for the fluctuation term $\delta a$, which in our GPE-based approach, could be taken to be an arbitrary small perturbation on top of the condensate. \\
		
		However, our choice for the initial condensate wave function (i.e.~ground state of $\hat{H}_{\mathrm{eff}}$) suggests a natural choice for the initial fluctuation term $\delta a(t\!=\!0)$: indeed, one could assume that at $t\!=\!0$, this fluctuation term is precisely given by the Bogoliubov wave function of a condensate in the ground-state of $\hat{H}_{\mathrm{eff}}$, as obtained from a numerical diagonalization of the corresponding (effective)  Bogoliubov equations, namely, $$\delta a_n(t=0)=\sum_q u_q \mathrm e^{iqn}+v^*_q \mathrm e^{-iqn},$$ where $u_q, v_q$ correspond to the solution of
		\begin{equation}
			\left( \begin{matrix}  \varepsilon_{\text{av}}(q)+g & g        \\ -g & -\varepsilon_{\text{av}}(-q)-g \end{matrix} \right)\left( \begin{matrix} u_q  \\ v_q \end{matrix} \right)=\Eav(q)\left( \begin{matrix} u_q  \\ v_q \end{matrix} \right),
			\label{eq:BdGEk}
		\end{equation}
		 where $\varepsilon_{\text{av}}(q)$ and $\Eav(q)$ depend on the model under consideration. Here, $\varepsilon_{\text{av}}(q)$ denotes the time-average of the instantaneous dispersion $\varepsilon(q,t)$ [see Eq.\eqref{epsilon_av} for the 1D model, and Eq.\eqref{eq:BdGEk2D} in the 2D case], and $\Eav(q)$ is the time-averaged Bogoliubov dispersion [see Eq.~\eqref{eq:Eav} for the 1D model, and Eq.\eqref{eq:Eav2D} in the 2D case].
		
\begin{figure}[!h]
			\begin{center}
				\includegraphics[width=8cm]{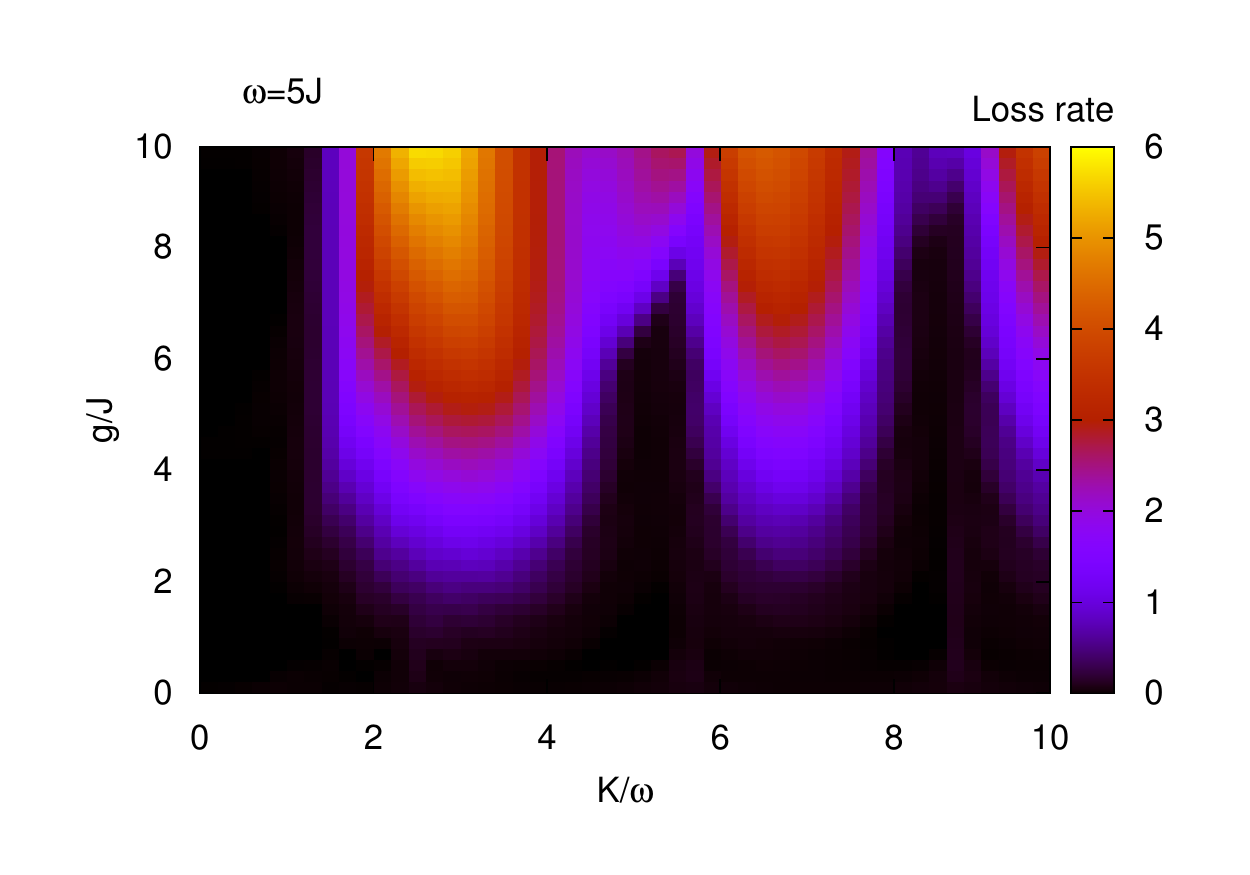}
			\end{center}
			\caption{Loss rate (exponential growth rate of the non-condensed density) for the 2D lattice model of Sec.~\ref{sec:2Dlat}, for $\omega=5J$ and $N_y=16$. This rate is obtained by fitting the exponential growth of the non-condensed fraction, similarly to what was done in Fig.~\ref{fig:DiagEner} for the energy.
				\label{fig:DiagEner2D}
			}
		\end{figure}

		A relevant observable to quantify heating and losses, which can be probed in experiments, is the non-condensed density, which, in the Bogoliubov approximation, reads~\cite{castin2001}
		\begin{equation}
		\rho_{nc}(t)  = \dfrac{1}{\mathrm{Vol}}\sum_q |v_q(t)|^2,
		\label{eq:fnc}
		\end{equation}
		with $\mathrm{Vol}$ the volume of the system. Given the above initial condition, it is straightforward to apply the numerical tools developed in Sec.~\ref{sec:NumEn}~\footnote{at least in dimension $d>1$, since the 1D case is problematic due to the divergence of the quantum depletion in the thermodynamic limit (which is independent of the drive, and hence, exists even at time $t\!=\!0$).}, so as to describe the time evolution of the non-condensed density Eq.\eqref{eq:fnc} in the system. 
		For instance, Fig.~\ref{fig:DiagEner2D} shows the exponential growth rate of the non-condensed fraction (referred to as the loss rate) obtained for the 2D lattice model of Sec.~\ref{sec:2Dlat}, which is therefore expected to describe particle losses in the associated 2D driven Bose-Hubbard model. Note that, unsurprisingly, this is very similar to the stability diagram Fig.~\ref{fig:2DNum5}(b), which corresponds to the same parameters [see Sec.~\ref{sec:NumEn}].

		The aforementioned aspects are intrinsic to the mean-field Bogoliubov approximation in the frame of the model under consideration. More generally, there are a few important points one has to keep in mind, when adopting a mean-field approach to study out-of-equilibrium ergodic (non-integrable) bosonic systems subject to periodic driving: 
		\begin{itemize}
			\item As already alluded to above, the Floquet Hamiltonian becomes an increasingly nonlocal operator, the further the drive frequency deviates from the infinite-frequency limit. An intriguing and interesting part of this intrinsic nonlocality is due to Floquet many-body resonances~\cite{bukov_15_erg} appearing because of hybridisation of many-body states induced by the drive. These resonances appear as a result of energy absorption and provide shortcuts between states in energy space separated by multiples of the drive frequency. Hence, in the unstable phase, they are expected to affect the heating rates at times $t\sim U^{-1}$, when interaction effects between quasiparticles become important. 
			\item  In fact, one can anticipate (using the inverse-frequency expansion in the rotating frame) that the Floquet Hamiltonian of the driven Bose-Hubbard model contains complicated three- and higher-body interaction terms, starting from order $\omega^{-3}$. It is currently an open problem how these terms modify and limit the application of Bogoliubov's mean-field theory associated with the effective (Floquet) Hamiltonian.	
		\end{itemize}	
		Despite the open character of these potential issues related to interacting bosonic systems, we expect that our GPE-based analysis captures the dominant contribution to the short-time evolution of the periodically-driven Bose-Hubbard model in Eq.~\eqref{eq:H}, in particular the onset of instability.
		
		\subsection{Time evolution beyond the linearised regime: the conserving approximation}

		Since the mean-field Bogoliubov approximation assumes a macroscopic occupation of the condensate mode, it remains valid provided the number of non-condensed atoms remains small compared to the total number of atoms throughout the entire subsequent evolution. As expected, this assumption fails in the unstable regions, where the depletion of the condensate grows exponentially and the mean-field approach typically holds at short times. Thus, the time scale for reaching a sizeable dynamical depletion sets a natural upper bound on the validity of mean-field approaches.
				 
		One way to avoid the problems associated with particle conservation, is to apply the weak-coupling conserving approximation (WCCA)~\cite{bukov2015}. Based on a Keldysh field theory formalism, the WCCA is the minimal extension of Bogoliubov theory, which includes the proper effective interactions between the Bogoliubov quasiparticles and the condensate, order by order in the original interaction strength $U$, while ensuring particle conservation at all times. Truncated to linear order in $U$, the WCCA is equivalent to the Bogoliubov-Hartree-Fock approximation~\cite{griffin_95}, and in the following we shall restrict to this case. 
The unknown variables in the WCCA formalism are~\footnote{We implicitly assume here that the system is translational invariant and condensation occurs in the $q=0$ mode. For the general case, please consult the Supplemental material to Ref.~\cite{bukov2015}.}
		\begin{align}
		&\phi(t) = e^{-i\int_0^t \mu(t')\mathrm{d} t'}\langle a_{q=0} \rangle, \notag \\
		&F_{11}(t;q) = \frac{1}{2}\langle\{a_q(t),a_q^\dagger(t)\}\rangle_c, \notag \\
		&F_{12}(t;q) = \frac{e^{-2i\int_0^t \mu(t')\mathrm{d} t'}}{2}\langle\{a_q(t),a_{-q}(t)\}\rangle_c,
		\label{eq:WCCA_vars}
		\end{align}
		where $\phi(t) \!=\!a^0_n(t) e^{-i\int_0^t \mu(t')\mathrm{d} t'}\sqrt{\mathrm{Vol}}$ is the $q\!=\!0$ mode of the spatially-homogeneous (i.e.~$n$-independent) condensate wave function, and where $\mu(t)\!=\!-zJ\cos(K/\omega\sin\omega t)+g$ is the chemical potential, which is irrelevant for U(1)-conserving dynamics. The subscript $_c$ stands for \emph{connected} correlators: $\langle A(t)B(t)\rangle_c \!=\! \langle A(t)B(t)\rangle \!-\! \langle A(t)\rangle\langle B(t)\rangle$. The equal-time correlator $F_{11}(t;q)$ is, apart from an additive constant, the phonon density. In momentum space, it is related to the momentum distribution function of the quasiparticles $n_q(t)$ by $n_q(t) \!=\! F_{11}(t;q) - 1/2$. Note that $n_q(t)$ does \emph{not} include the condensate delta function peak at $q=0$. The WCCA EOM for the equal-time correlators represent a simple system of non-linear, non-local in space equations
	\begin{widetext}
		\begin{align}
		\label{eq:WCCA_EOM}
		i\partial_t\phi(t) &= \varepsilon_{q=0}(t)\phi(t)
	    + \frac{U}{\mathrm{Vol}}\bigg[ \left[\phi(t)\right]^*\left[\phi(t)\right]^2
	     + 2\phi(t) \int_{q'} F_{11}(t;q') + \left[\phi(t)\right]^*\int_{q'} F_{12}(t;q') \bigg],\\
		\partial_t F_{11}(t;q) &= 2\text{Im}\left\{  \frac{U}{\mathrm{Vol}}\left( \left[\phi(t)\right]^2 + \int_{q'} F_{12}(t;q')\right)\left[F_{12}(t;q)\right]^* \right\},\nonumber\\
		i\partial_tF_{12}(t;q) &\!=\! \bigg\{ [\varepsilon_q(t)\!+\!\varepsilon_{-q}(t)\!]F_{12}(t;q)
		\!+\! 2\frac{U}{\mathrm{Vol}}\bigg[ 2\left( |\phi(t)|^2 \!+\! \int_{q'} F_{11}(t;q') \right)F_{12}(t;q)		
		\!+\! \left( \left[\phi(t)\right]^2 \!+\! \int_{q'} F_{12}(t;q')\right)\! F_{11}(t;q)  \bigg]\bigg\},\nonumber
		\end{align}
		where $^*$ denotes complex conjugation. 
		\end{widetext}
		
		The conserved quantity itself is the total number of particles (condensed and excited):
		\begin{align}
		N &= |\phi(t)|^2 + \int_q n_q(t) \notag \\
		&= |\phi(t)|^2 + \int_q \left( F_{11}(t;q) - \frac{1}{2}\right) = \mathrm{const.}
		\end{align}  
		One recognises the GPE equation for the spatially homogeneous condensate density $\phi(t)$, and identifies the additional phonon feedback terms. Quite generally, if one eliminates all terms containing integrals over $q$, the WCCA EOM decouple into the familiar GPE for the condensate, see Eq.~\eqref{eq:GPE}, while the second two equations are equivalent to the BdG equations Eq.~\eqref{eq:BdGEk}
	\footnote{
	The Bogoliubov equations [see~Eq.(8)] can indeed be exactly rewritten under the form
	\begin{align*}
	\partial_t F_{11}(t;q) &= 2g\text{Im}\left ( \left[F_{12}(t;q)\right]^* \right ), \\ 
	i\partial_tF_{12}(t;q) &=  [\varepsilon_q(t)+\varepsilon_{-q}(t)-2\mu + 4g]F_{12}(t;q)+ 2g F_{11}(t;q) ,
	\label{eq:BdF_F}
	\end{align*}
	where we have defined the correlators in Eq.~\eqref{eq:WCCA_vars}.
	}.  
		
		\begin{figure}[t!]
			\centering
				\includegraphics[width=8.3cm]{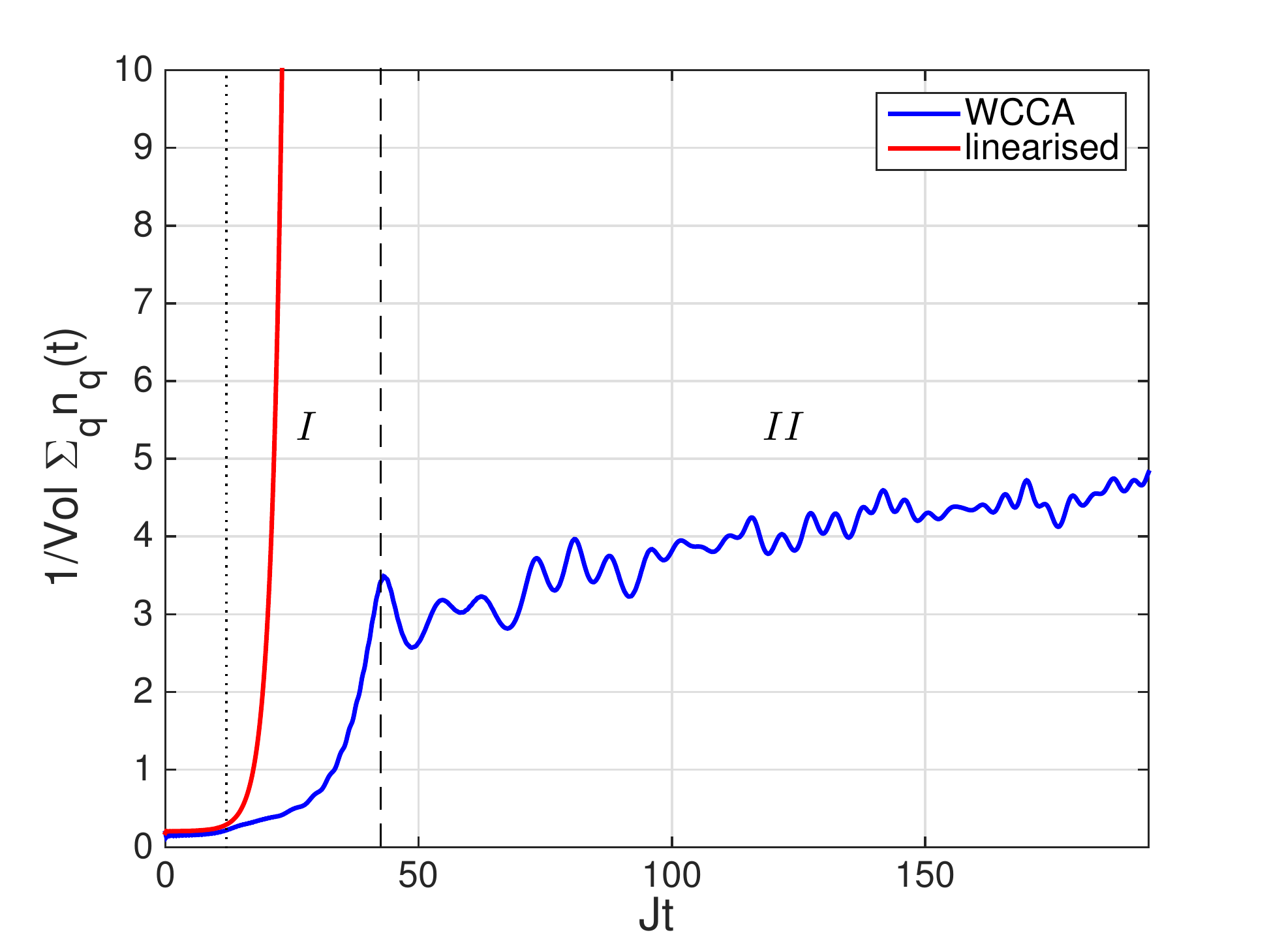}
				\includegraphics[width=8.3cm]{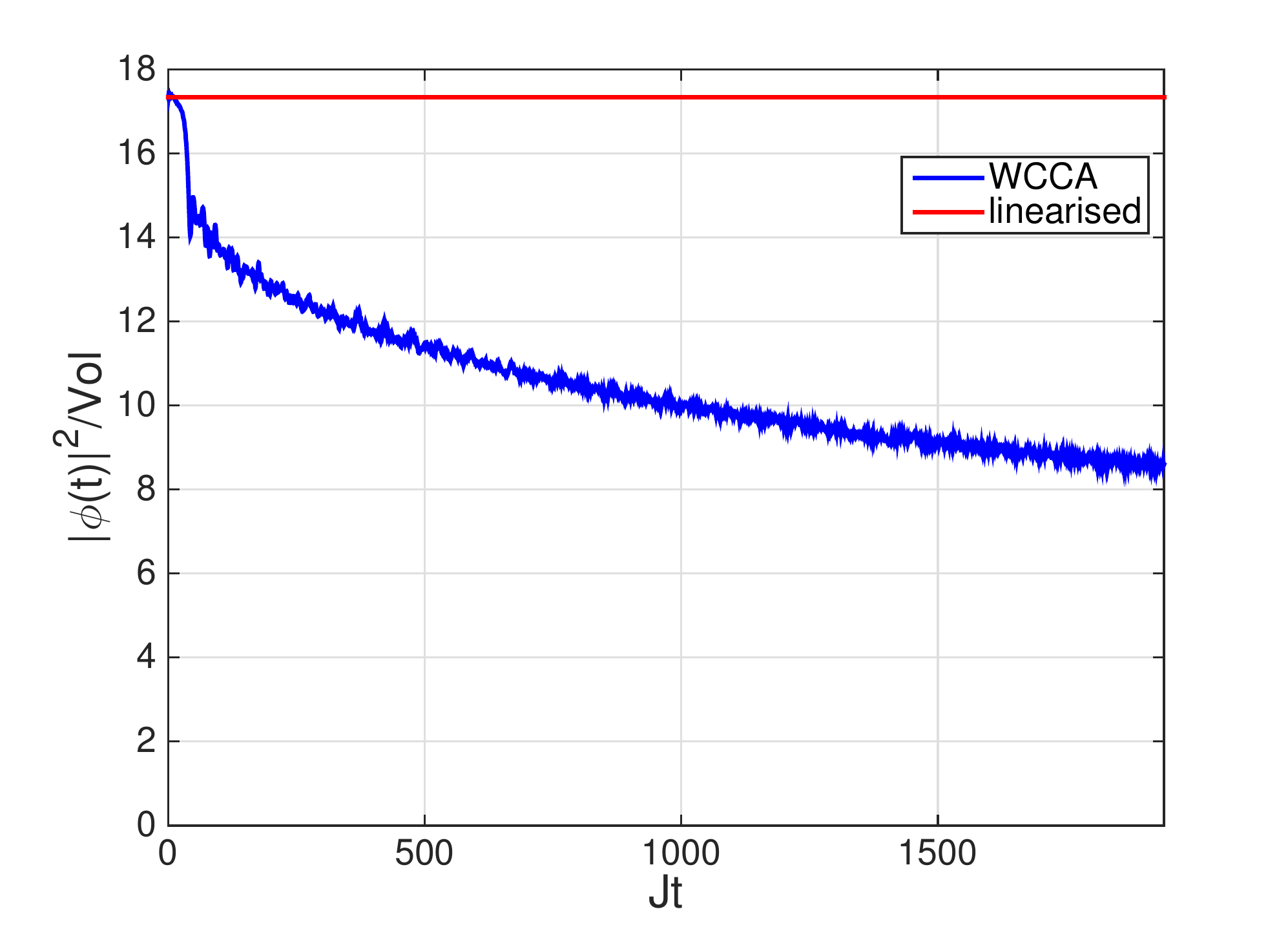}
			\caption{\label{fig:WCCA_vs_lin}  Time dependence of the quasiparticle (phonon) density (top) and the condensate density (bottom) using the linearised (red) and the conserving (blue) approximation, for the 2D band model from Sec.~\ref{sec:2DCont} (with a lattice along the $x$-direction and continuum along the $y$-direction). There are two discernible regimes: $I$ and $II$ (see text), delineated by a vertical dashed line. The vertical dotted line marks the time up to which the linearised mean-field dynamics and the WCCA agree well. The model parameters are $\rho/\mathrm{Vol}=17.34$, $U/J=1.011$ ($g/J=17.53$), $m=1/(20J)$, healing length to $x$-lattice constant ratio $\xi/a_x=1.47$, $\omega/J=32.37$, and $K/\omega=1.0$. We use a total of $N=N_xN_y=5\times 10^4$ momentum points with $N_x=50$ and $N_y=1000$ to reach the thermodynamic limit. As an initial state we choose the Bogoliubov ground state.}  
		\end{figure}
		
		Opening up a channel between the quasiparticles and the condensate, we find that the phonon build-up rate decreases (and consequently, the condensate depletion slows down) compared to the linearised BdG equations, in agreement with intuitive expectations. In other words, the WCCA deviates from the exponentially growing linearised solution, which sets a natural scale on the validity of the mean-field approximation. 
		
\begin{figure}[!t]
				\centering
				\includegraphics[width=0.9\columnwidth]{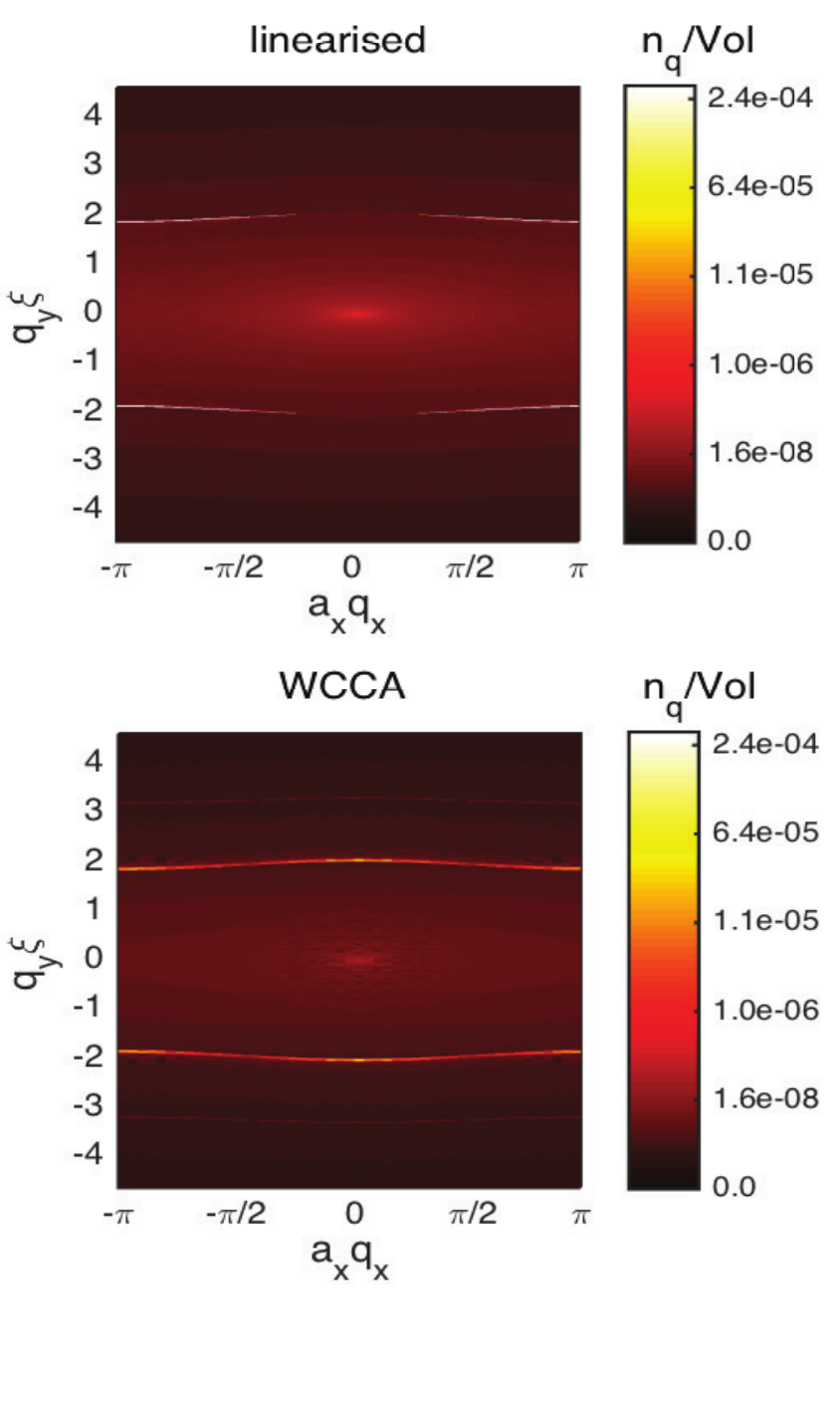}
				\caption{\label{fig:nk_WCCA_vs_lin} Quasiparticle momentum distribution function $n_q$ after a duration of $t\!=\!1000T$, for the 2D band model from Sec.~\ref{sec:2DCont} (with a lattice along the $x$-direction and continuum along the $y$-direction):~(top) linearised mean-field model,  and (bottom) the conserving approximation (WCCA) to order $U$. The model parameters are the same as in Fig.~\ref{fig:WCCA_vs_lin}. Here, we set the lattice constant $a_x\!=\!1$ to unity, while $\xi\!=\!1.47a_x$ denotes the BEC healing length. 
				}  
		\end{figure}		
		
		To justify the applicability of the mean-field approach at short times, we compare the dynamical build-up of quasi-particle excitations in the parametrically unstable regime, predicted by the linearised EOM [see~Eq.~\eqref{eq:BdGEk_pre}], and the WCCA (to order $U$). Here, we initialise the system in the Bogoliubov ground state corresponding to the non-driven Hamiltonian $H(K\!=\!0)$. Figure~\ref{fig:WCCA_vs_lin} shows the time evolution of the quasiparticle excitations (top) and the condensate density (bottom) for the 2D model of Sec.~\ref{sec:2DCont}. One can clearly identify two stages of evolution: in stage $I$, which continues for about $t\lesssim 220T$ driving periods, the dynamics is governed by the single-particle mean-field physics. The dotted vertical line at $t^\ast\approx 60T$ marks the time up to which the dynamics predicted by the linearised mean-field equations agrees with the WCCA. Thus, for $t\leq t^\ast$, this initial regime features an exponential build-up of ``phonons" at the momenta satisfying the resonant condition $\omega=E_\mathrm{av}({\bf q})$. For $60T\lesssim t\lesssim 220T$, the time evolution is characterised by a different growth rate than the one predicted by the mean-field equations for this system configuration. By the time $t\approx t^\ast$, the back-action effect of the quasiparticles onto the condensate becomes sizeable and the dynamics no longer follows the initial linearised exponential growth. Based on the available data, it is not possible to determine whether the growth remains exponential or follows another law. In stage $II$, $t\gtrsim 220T$, the population of the resonant $q$-modes slows down significantly, although it never really saturates, as can be seen from the condensate depletion at longer times, see Fig.~\ref{fig:WCCA_vs_lin} (bottom panel). Due to the unbounded character of the dispersion along the $q_y$-direction, the presence of resonant modes set by the condition $\omega\!=\!E_\mathrm{av}({\bf q})$ does not allow for the formation of a Floquet steady state. Nevertheless, the growth rate diminishes significantly and the dynamics slows down, as expected for a pre-thermal regime. Interestingly, the condensate evolution curve [Fig.~\ref{fig:WCCA_vs_lin} (bottom panel)] changes curvature in stage $II$, compared to stage $I$.
		Curiously, we note that a very similar change of behaviour was reported in a cosmological context~\cite{berges2014}.
		
		Important features of the quasiparticles dynamics are conveniently reflected in their momentum distribution function $n_q(t)$\footnote{The condensate delta peak at $q=0$ is not part of $n_q(t)$ since this mode is treated additionally via the field $\phi(t)$.}. Figure~\ref{fig:nk_WCCA_vs_lin} shows a comparison between the momentum distributions, as obtained by time-evolving the linearised and WCCA EOM over a long duration ($t\!=\!1000T$), for the same 2D model. 
	The visible lines, which materialize largely-populated modes, correspond to on-resonance modes [$\Eav(\bs q)\!=\!\omega$], which have indeed been shown to be unstable. In the linearised dynamics, the largest peaks are visible around the maximally unstable modes, as predicted by our theory [see Eqs.~\eqref{eq:qres2DC0} and~\eqref{eq:qres2DC0bis}]; for the considered parameters, the most unstable mode indeed corresponds to $q_x^{\text{mum}}\!=\!\pm\pi$ and $q_y^{\text{mum}}$ is adjusted so that the resonance condition $\omega\!=\!E_\mathrm{av}({\bf q})$ is fulfilled [see Sec.~\ref{sec:2DCont}]. 
	We stress that the large width of the peaks around $q_x\!=\!\pm\pi$, in the top panel of Fig.~\ref{fig:nk_WCCA_vs_lin}, is an artifact due to the significant population of momentum modes at long times, which yields saturation on this plot; we refer to Fig.~\ref{fig:nk_intro} for a more striking illustration of these peaks, as calculated at shorter times (for the same model parameters).
	In contrast, the WCCA  couples all momenta through the $q'$-integrals in Eq.~\eqref{eq:WCCA_EOM}, allowing the population to further spread along the $x$-axis. All modes fulfilling the resonance condition now exhibit peaks of similar magnitudes, precisely materializing the curve of equation $\omega\!=\!E_\mathrm{av}({\bf q})$; note though that the populations of the modes along this curve is not completely homogeneous, presumably due to the non-linear character of the WCCA EOM and/or the differences in the instability rates of resonant modes. Moreover, as we discussed in Sec.~\ref{sec:AnaRes}, modes which are not precisely on resonance, but sufficiently close to it, are also affected by the instability. 
	
	A summary of the various behavior types of the momentum distribution, which are expected for the different parameters regimes, are presented in the concluding Section~\ref{sec:CCL}, see Figs.~\ref{fig:CCL_nq_integrated}-\ref{fig:CCL_nq_integrated} and Table~\ref{fig:table2}.
	

		We conclude this Section by noticing that while the WCCA to leading order in $U$ features a pre-thermal Floquet regime at high driving frequencies~\cite{bukov2015}, the absence of quasiparticle collisions prevents the onset of thermalising dynamics at later times. Including the next-to-leading order in $U$ has been done for the periodically-driven $O(N)$ model~\cite{Weidinger:2016,Chandran:2016}, where the crossover from a  pre-thermal steady state to a thermal regime becomes clearly visible.

\begin{figure}[t!]
				\centering
				\includegraphics[width=8.9cm]{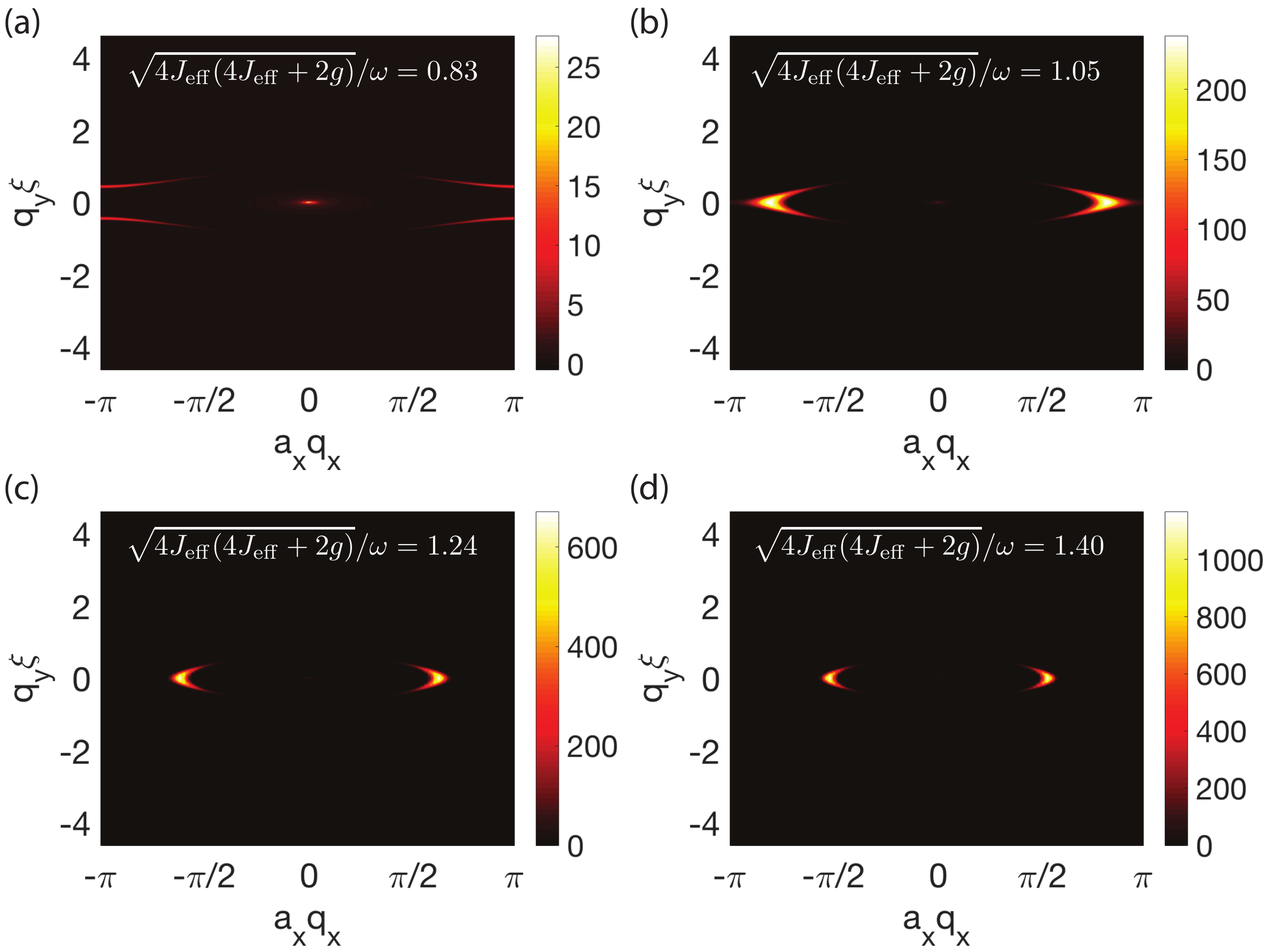}
				\caption{\label{fig:CCL_n_vs_qx_qy_lin} Quasiparticle momentum distribution function $n_q$ after a duration of $t\!=\!10T$, for the 2D band model from Sec.~\ref{sec:2DCont} (with a lattice along the $x$-direction and continuum along the $y$-direction), computed using the linearised dynamics, for different values of the relevant parameter $\sqrt{4\vert\Jeff\vert(4\vert\Jeff\vert+2g)}/\omega$; see Table~\ref{fig:table2}. The suggested ellipses correspond to on-resonance modes [$\Eav(q)\!=\!\pi$], which are unstable, while the most unstable modes, which are revealed by the main peaks in the momentum distribution, are accurately described by our theoretical predictions [see Table~\ref{fig:table2}]. }  
		\end{figure}

\begin{figure*}[t!]
				\centering
				\includegraphics[width=\textwidth]{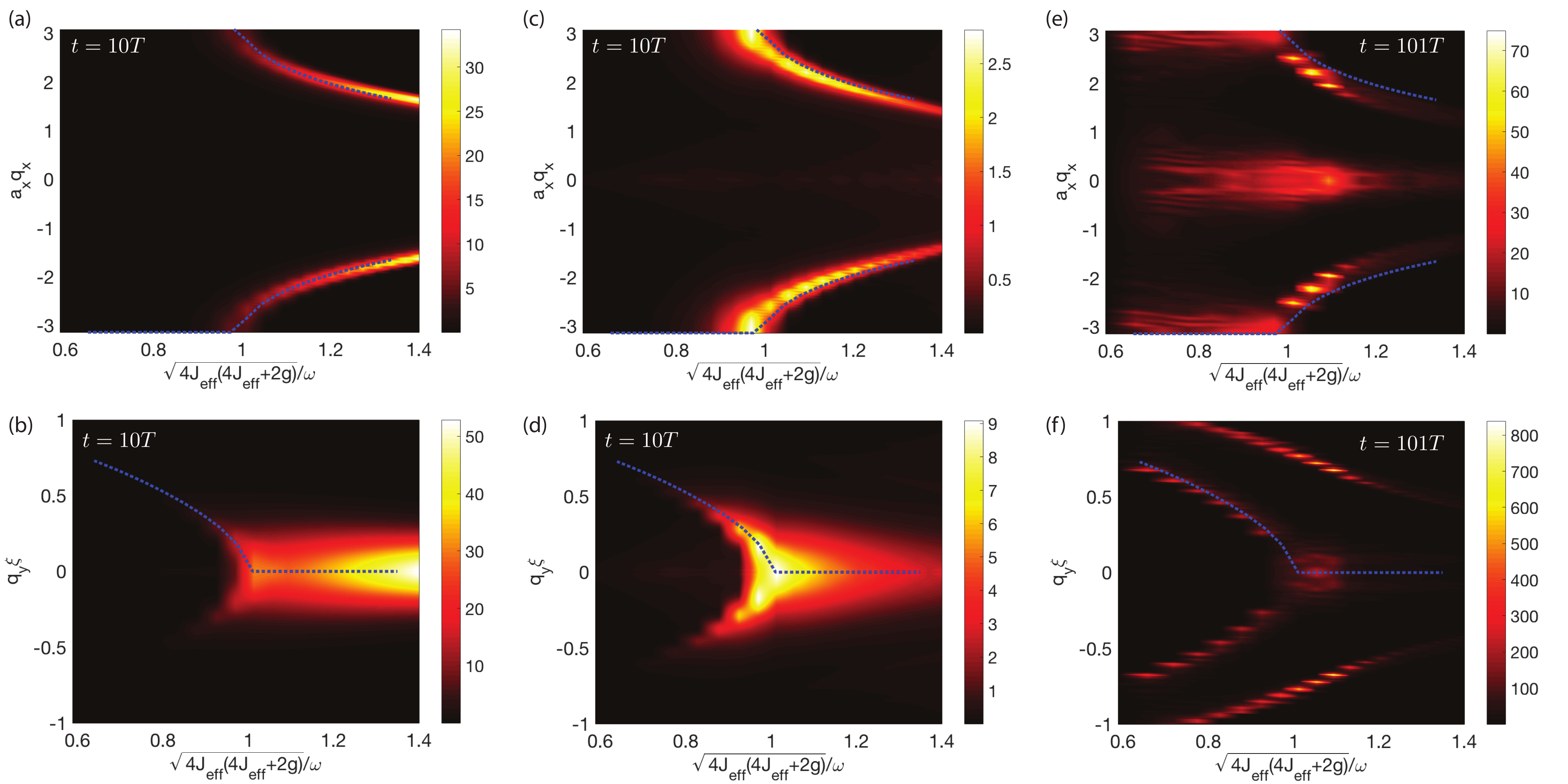}
				\caption{\label{fig:CCL_nq_integrated} Quasiparticle momentum distribution function $n_q$ for the 2D band model from Sec.~\ref{sec:2DCont} (with a lattice along the $x$-direction and continuum along the $y$-direction), as a function of the parameter $\sqrt{4\vert\Jeff\vert(4\vert\Jeff\vert+2g)}/\omega$ and of $q_x$ (upper line)[resp.~$q_\perp$, (lower line)], after integration over $q_\perp$ (resp.~$q_x$), and computed : (left) after a time $t\!=\!10T$ using the linearised dynamics; (middle) same time but using the WCCA; (right) after a longer time $t\!=\!101T$ using the WCCA. The dotted blue line indicates our theoretical prediction for the most unstable mode [see Table~\ref{fig:table2}], which is verified in all regimes:~for $\sqrt{4\vert\Jeff\vert(4\vert\Jeff\vert+2g)}/\omega\!>\!1$, we find $q_\perp^\mathrm{mum}\! =0\! $ and $q_x^\mathrm{mum}<\pi$, as predicted by Eq.~(\ref{eq:qres2DC0bis}); for $\sqrt{4\vert\Jeff\vert(4\vert\Jeff\vert+2g)}/\omega\!<\!1$, we find $q_x^\mathrm{mum}\!=\!\pi$ and $q_\perp^\mathrm{mum}>0$, as given by Eq.~(\ref{eq:qres2DC0}); let us note that, in the latter regime, the peaks are only apparent after a long time (here $101T$, right panel) due to the small values of the corresponding instability rates; see also Fig.~\ref{fig:CCL_n_vs_qx_qy_lin}(a), where these peaks are made visible. As already observed in Fig.~\ref{fig:WCCA_vs_lin}, both linearized and WCCA agree very well on the position of the peaks, although the WCCA predicts a broadening of the instability zone due to non-linear couplings between the modes. Note that at longer times (right panel), higher order parametric resonances are also visible. The apparent ``quantised" character of the resonant peaks for $t\!=\!101T$ is a numerical artifact due to the discretization of the variable $\sqrt{4\vert\Jeff\vert(4\vert\Jeff\vert\!+\!2g)}/\omega$ used in the numerical simulation. }  
\end{figure*}

\begin{widetext}
	
	\section{Summary of the main results and concluding remarks}
	\label{sec:CCL}
	
	The analysis presented in this work reveals the crucial role played by parametric instabilities in the short-time dynamics of periodically-driven band models. It provides a quantitative description and a qualitative understanding of the stability diagrams of these periodically-driven systems, capturing both the stability boundaries and the instability rates in its vicinity. This work also identifies clear signatures of these early-stage instabilities, which could be detected in current cold-atom experiments.
	
	On the one hand, the quantitative description of the instability rates will prove helpful to guide experimentalists in finding stable regimes of operation. A first indicator is the instability rates themselves. Table~\ref{fig:table} summarizes the scaling behaviour found for instability rates in the thermodynamic limit as a function of the model parameters, as derived in Sec.~\ref{sec:AnaRes}. Remarkably, these scaling laws hold in all cases investigated in this work, i.e~1D/2D lattices and high/low-frequency regime.

\begin{table}[h!]
\begin{center}
\begin{tabular}{|c|c||c|c|}
\hline
\multicolumn{2}{|c||}{low $g$} & low $K$ & low $\omega$ \\
\cline{1-2}
high-frequency regime & low-frequency regime & & \\ 
\hline
$\Gamma \propto \sqrt{(g-g_c)}$ & $\Gamma \propto g$ &$\Gamma \propto K^2$  & $\Gamma \propto \omega \times f(K/\omega)$ \\
\hline
\end{tabular}
\end{center}
\caption{Limiting scalings of instability rates in the thermodynamic limit, in terms of the interaction strength $g$, the modulation amplitude $K$ and frequency $\omega$. The ``high" and ``low" frequency regimes are defined through a comparison between $\omega$ and the free-particle effective bandwidth, see Sec.~\ref{sec:NumDiag}. \label{fig:table}}
\end{table}		
		
	A second important information concerns the identification of the most unstable mode $\qbf^\mathrm{mum}$. Table~\ref{fig:table2} recalls the characteristics of this most unstable mode for the various relevant regimes discussed in this work. In the ``high-frequency" regime, where $\omega$ is larger than the free-particle effective bandwidth, the onset of instability is necessarily driven by the mode $q\!=\!\pi$ (or $q_{x,y,z}\!=\!\pi$ if several lattice degrees of freedom are present). In the ``low-frequency" regime, the most unstable mode $\qbf^\mathrm{mum}$ obeys the resonance condition $\omega\!=\!\Eav(\qbf)$. Yet, in dimensions higher than one, we found that this condition is not sufficient to unambiguously determine $\qbf^\mathrm{mum}$, since a whole set of resonant modes (associated with different rates) typically satisfy it. Our analysis showed that two situations can occur, and that these could be distinguished by comparing the frequency $\omega$ to the effective Bogoliubov bandwidth associated with the shaken-lattice direction alone: if $\omega\!>\!\sqrt{4\vert \Jeff \vert(4\vert\Jeff\vert+2g)}$, then $q_x^\mathrm{mum}\!=\!\pi$ and $q_\perp^\mathrm{mum}\! \neq 0$ [see Eq.~(\ref{eq:qres2DC0})]; conversely, if  $\omega<\sqrt{4\vert\Jeff\vert(4\vert\Jeff\vert+2g)}$, we find $q_x^\mathrm{mum}\!<\!\pi$ and $q_\perp^\mathrm{mum}\! =\! 0$ [see Eq.~(\ref{eq:qres2DC0bis})]. Let us emphasize that, in principle, the so-called ``low-frequency" regime can be reached at arbitrarily large frequencies; for instance, this typically occurs whenever the free-particle effective dispersion is unbounded, as in the case where a continuous direction is present. Figures~\ref{fig:CCL_n_vs_qx_qy_lin} and~\ref{fig:CCL_nq_integrated} show the momentum distribution as calculated for the hybrid-2D band model of Sec.~\ref{sec:2DCont} (i.e.~a shaken lattice along the $x$-direction and a continuum along the $y$-direction); by exploring the different parameter regimes, these results illustrate all the behaviors described above [Table~\ref{fig:table2}].

\begin{table}[h!]
\begin{center}
\begin{tabular}{|c|c||c|}
\hline
  \multicolumn{2}{|c||}{low-frequency regime}  & high-frequency regime \\
\cline{1-2}
$\omega<\sqrt{4\vert\Jeff\vert(4\vert\Jeff\vert+2g)}$ & $\omega>\sqrt{4\vert\Jeff\vert(4\vert\Jeff\vert+2g)}$ &  \\ 
\hline
$q_x^\mathrm{mum}\!<\!\pi$ & $q_x^\mathrm{mum}\!=\!\pi$ & $q_{x,y,z}^{\text{mum}}=\pi$\\
$q_\perp^\mathrm{mum}\! =\! $0 & $q_\perp^\mathrm{mum}\! > 0$ & (at the onset of instability )  \\
 see Eq.~(\ref{eq:qres2DC0bis}) & see Eq.~(\ref{eq:qres2DC0})&   \\
\hline
\end{tabular}
\end{center}
\caption{ Characterization of the most unstable mode (mum) for various relevant regimes.  The ``low-frequency" regime is generically defined through the criterion according to which the drive frequency $\omega$ is smaller than the effective free-particle bandwidth (e.g.~$\omega\!<\!4|\Jeff|$ for the 1D shaken-lattice model of Section~\ref{sec:NumDiag}). We stress that the effective free-particle dispersion is unbounded for models involving a continuous spatial dimension, in which case the system necessarily corresponds to the ``low-frequency" regime (for all $\omega$). \label{fig:table2}}
\end{table}		

\end{widetext}

Those predictions on the instability rates can readily be tested in present-day experiments. For typical experimental parameters, the ratio $\omega/J$ is generally set in the range $10\!-\!100$; therefore, in the case of pure-lattice systems, experiments would typically lie in the ``high-frequency" regime defined in Section~\ref{sec:NumDiag}, and the most unstable mode is thus expected to be $q_{x,y,z}^{\text{mum}}\!=\!\pi$. In turn, experiments with a continuous degree of freedom fall into the second column of Table~\ref{fig:table2}, and we thus expect the maximal instability to occur through $q_x^\mathrm{mum}\!=\!\pi$ and finite $q_\perp^\mathrm{mum}$. We point out that experiments could also investigate the ``low-frequency" regime of driven-lattice systems (first column of Table~\ref{fig:table2}), by working with lower drive frequencies $\omega\!\sim\!J$ (see, e.g., Ref.~\cite{sias2008}).

The different observables that we introduced are accessible in current experiments and probing them should permit to have clear experimental signatures of the predicted instability behaviour. In particular, measuring the momentum distribution of the gas, which is routinely performed through time-of-flight techniques, and identifying the positions of the peaks, would give access to the momentum of the most unstable modes, which is one of the most recognizable characteristics of parametric instabilities.

	On the other hand, several important conceptual points arise from the detailed analysis presented in this paper: in particular, the intuitive stability criterion $\omega>2W_\mathrm{eff}$ is not sufficient to properly estimate the instability boundary and rates in the system, as becomes clear from the rigorous derivation we provided. More precisely, the parametric resonance condition for the instability is fulfilled for $\omega\!=\!\Eav(q)$, but in addition, one has to take the (possibly) large width of the resonance window into account in order to properly estimate the position of the  boundary; we note that similar effects were found in the context of the many-body Kapitza pendulum~\cite{citro2015}.
	
	As a result of our study, the following general guidelines deserve to be highlighted:~in lattice systems, working at high frequency (much larger than the effective bandwidth of Bogoliubov excitations) is favourable, since no parametric resonance can then occur, ensuring a stable evolution. We also analyzed the effect of transverse directions, be it a deep lattice or tubes; although the instability is generically increased in both cases by the fact that more and more modes are available in the system (offering more possibilities to have resonances), we showed that working with a lattice should be much more favorable to find stable regimes and have less heating. We stress that our analysis results from a tight-binding approximation, namely, the restriction of the study to a single Bloch band. This simplification is reasonable for situations where the drive frequency is set away from any inter-band resonances, which is indeed possible in experiment. Interestingly, we note that such inter-band excitations have been generated and studied experimentally in Ref.~\cite{weinberg2015}. The interplay between inter-band transitions, as generated by setting the drive frequency close to resonance with inter-band spacing, and the existence of parametric instabilities, should show rich behaviors, which could be explored through the theoretical approach presented in this work.

		Finally, we emphasize that the instability rates derived from parametric resonances (see, e.g., the scalings in Table~\ref{fig:table}) differ quantitatively from the ones computed within a Fermi Golden Rule (FGR) approach~\cite{bilitewski2015}, and which generically scale as $g^2$ for low interactions.  In particular, we note that contrary to the case of parametric resonances, the rates emanating from the FGR approach strongly depend on the dimensionality of the system, through its density of states, which indicates that the two phenomena are indeed very distinct in essence. 
Since the system is bosonic, we expect that the parametric instability should dominate the short-time dynamics, due to its exponential growth rate. It is only at later times that Fermi's Golden Rule comes to rule the behaviour of the system, when the dynamics will be dominated by collision processes leading to thermalisation at a higher temperature, a feature which is generically known for quantum Bose fluids~\cite{berges2014}. Hence, both phenomena take place in different regimes of the time evolution.
More precisely, for typical experimental parameters, the instability rates derived from parametric resonances are found to be of the order of $\mathrm{1-10 ms}$, while the Fermi Golden Rule decay usually spreads over times of the order of $\mathrm{100ms}$~\cite{bilitewski2015}. It is within the scope of present-day experiments to observe those two distinct regimes.\\

	\section*{Acknowledgments}
	The authors are pleased to acknowledge J.~Berges, I.~Bloch, J.~C.~Budich, J.~Dalibard, M.~Dalmonte, P.~Gaspard, M.~Lacki, C.~Laflamme, J.~Lutsko, J.~Nager, S.~Nascimbene, H.~Pichler, A.~Polkovnikov, M.~Reitter, U.~Schneider, K.~Wintersperger and P.~Zoller for insightful discussions. 
	This research was financially supported by the Belgian Science Policy Office under the Interuniversity Attraction Pole project P7/18 ``DYGEST".
	N.~G.~is supported by the Universit\'e Libre de Bruxelles (ULB) and the FRS-FNRS (Belgium). M.~B.~acknowledges support by NSF DMR-1506340 and ARO W911NF1410540. E.~D.~was supported by Harvard-MIT CUA, NSF Grant No. DMR-1308435, AFOSR Quantum Simulation MURI, AFOSR MURI Photonic Quantum Matter.
	The authors are pleased to acknowledge that part of the computational work reported on in this paper was performed on the Shared Computing Cluster which is administered by Boston University's Research Computing Services. The authors also acknowledge the Research Computing Services group for providing consulting support which has contributed to the results reported within this paper.
	\\



%

	
	
\newpage

\onecolumngrid

\appendix	

		\section{Different parametrizations of the Bogoliubov expansion in the time-dependent case}
		\label{sec:appBogoparam}
		
	In this Appendix, we recall the different parametrizations that are commonly used for the Bogoliubov expansion in time-dependent systems. 
	Starting from the time-dependent Gross-Pitaevskii equation (GPE) on the lattice, 
		\begin{equation}
			i \partial_t a_n=-J \left (a_{n+1}+a_{n-1} \right)+K\cos(\omega t)na_n+U|a_n|^2 a_n,
			\label{eq:GPEapp}
		\end{equation}
		and given the time-dependent solution for the condensate wave function $a_n^{(0)}(t)$, the basic idea of the Bogoliubov expansion is to consider a small perturbation on top of $a_n^{(0)}(t)$.\\
		
		A natural approach consists in writing $a_n(t)\!=\!a_n^{(0)}(t)\!+\!\delta a_n(t)$, and to linearize the Gross-Piteavskii equation in terms of $\delta a_n$. This yields the time-dependent Bogoliubov-de Gennes equations (BdGEs), which take the general form
		\begin{equation}
			i \left( \begin{matrix} \dot{\delta a_n}  \\ \dot{\delta a}_n^{*} \end{matrix} \right) =\left( \begin{matrix} \mathcal{M}_n(t) & U {a_n^{(0)}}^{2} \\ -U {a_n^{(0)}}^{*2} & -\mathcal{M}_n(t)^{*} \end{matrix} \right)\left( \begin{matrix} \delta a_n  \\ \delta a_n^{*} \end{matrix} \right),
			\label{eq:BdGEapp1}
		\end{equation}
		where we introduced the notation 
		\begin{align}
			\mathcal{M}_n(t)\delta a_n\equiv &-J(\delta a _{n+1}+\delta a_{n-1}) + 2U|a_n^{(0)}|^2+K\cos(\omega t)n\delta a_n .
			\label{eq:Fntapp1}
		\end{align}
		The resulting form of the Bogoliubov equations in \eqref{eq:BdGEapp1}, which features spurious complex time-dependent off-diagonal terms, is quite unusual, and in fact, not very convenient. First of all, we find that those terms can lead to numerical instabilities, when solving this equation using a finite-difference scheme. Then, we note that the form~\eqref{eq:BdGEapp1} does not allow for a simple comparison with the standard static case:~indeed, setting $K\!=\!0$ and writing $a_n^{(0)}(t)\!=\!\sqrt{\rho}e^{-i\mu t}$, with $\rho$ the condensate density and $\mu\!=\!g-2J$ the chemical potential (and $g\equiv U\rho$), we find that Eq.~\eqref{eq:BdGEapp1} yields the non-standard form
		\begin{equation}
			i \left( \begin{matrix} \dot{\delta a_n}  \\ \dot{\delta a}_n^{*} \end{matrix} \right) =\left( \begin{matrix} -J\hat{L}+2g & g e^{-2i\mu t} \\ g e^{2i\mu t} & J\hat{L}-2g \end{matrix} \right)\left( \begin{matrix} \delta a_n  \\ \delta a_n^{*} \end{matrix} \right),
			\label{eq:BdGEapp1bis}
		\end{equation}
		where the discrete operator $\hat{L}$ is defined by $\hat{L}(\cdot)_n\equiv (\cdot) _{n+1}+(\cdot)_{n-1}$. The equations of motion~\eqref{eq:BdGEapp1bis} are still explicitly time-dependent, and hence, they do not allow for a direct identification of the quasiparticle spectrum.\\
		
		In the static case, a common way of solving this issue is to use a slightly different parametrization, namely $a_n(t)=e^{-i\mu t}[a_n^{(0)}(t)\!+\!\delta a_n(t)]$, where we have factored out the dynamical phase $\mu$ associated with the free evolution. Using this parametrization, we obtain the Bogoliubov equations in the form
		 	\begin{equation}
			i \left( \begin{matrix} \dot{\delta a_n}  \\ \dot{\delta a}_n^{*} \end{matrix} \right) =\left( \begin{matrix} -J\hat{\Delta}+g & g  \\ g  & J\hat{\Delta}-g \end{matrix} \right)\left( \begin{matrix} \delta a_n  \\ \delta a_n^{*} \end{matrix} \right),
			\label{eq:BdGEapp2bis}
		\end{equation}
	where $\hat{\Delta}\delta a_n\equiv \delta a_{n+1}-2\delta a_{n}+\delta a_{n-1}$ denotes the discrete Laplacian. The resulting equations of motion correspond to the standard form of the BdGEs, which are indeed \textit{time-independent} (the spurious complex off-diagonal terms have disappeared), allowing for a direct extraction of the Bogoliubov spectrum.\\
	
	In the driven-system situation, the time-dependent phase of the BEC wavefunction $a_n^{(0)}(t)$ does not reduce to a trivial phase $\mu$, as it also contains an additional dynamical phase induced by the periodic driving. Therefore, factorizing $e^{-i\mu t}$ alone in the Bogoliubov parametrization would not be sufficient to get rid of all the spurious terms. In order to solve that issue, we generalize the method discussed above by factorizing the entire time-dependent phase of $a_n^{(0)}(t)$:~this is achieved by writing
\begin{equation}
a_n(t)\!=\!e^{i\phi_n(t)}[\rho_n(t)\!+\!\delta a_n(t)],\label{para_app_full}
\end{equation}
where we have introduced the modulus and phase of the BEC wavefunction, $a_n^{(0)}(t)=\rho_n(t)e^{i\phi_n(t)}$. 

For the model considered in this work, translational invariance is recovered in the rotating frame, which allows one to further simplify the parametrization:~indeed, taking into account the fact that $\rho_n(t)$ is uniform (it does not depend on $n$), we can equivalently write Eq.~\eqref{para_app_full} in the form [Eq.~\eqref{para_text}]
\begin{equation}
a_n(t)\!=\!a_n^{(0)}(t)[1\!+\!\delta a_n(t)],\label{para_app}
\end{equation}
where we have now factorized the full condensate wavefunction. Linearizing the Gross-Piteavskii equation in $\delta a_n$ then yields the time-dependent Bogoliubov-de Gennes equations, which take the general form considered in the main text
		\begin{equation}
			i \left( \begin{matrix} \dot{\delta a_n}  \\ \dot{\delta a}_n^{*} \end{matrix} \right) =\left( \begin{matrix} \mathcal{F}_n(t) & U |{a_n^{(0)}}|^{2} \\ -U |{a_n^{(0)}}|^{2} & -\mathcal{F}_n(t)^{*} \end{matrix} \right)\left( \begin{matrix} \delta a_n  \\ \delta a_n^{*} \end{matrix} \right),
			\label{eq:BdGEapp2}
		\end{equation}
		where we introduced the operator $\mathcal{F}_n(t)$, whose action on $\delta a_n$ is defined by
		\begin{align}
			\mathcal{F}_n(t){\delta a_n}\equiv &-J\dfrac{\hat{L}(a^{(0)}{\delta a_n})_n}{a_n^{(0)}} + 2U|a_n^{(0)}|^2{\delta a_n}+K\cos(\omega t)n{\delta a_n}-i\dfrac{\dot{a}_n^{(0)}}{a_n^{(0)}}{\delta a_n}.
			\label{eq:Fntapp2}
		\end{align}
		Compared to Eq.~(\ref{eq:BdGEapp1}), the differences are twofold. First of all, the off-diagonal terms are real in this approach, which is numerically more practical. Then, we note that the last term in Eq.~(\ref{eq:Fntapp2}), which involves the logarithmic derivative of the BEC wavefunction $a_n^{(0)}(t)$, features a time-dependent chemical potential, which takes into account the entire dynamical phase induced by the periodic driving; in particular, we recover the chemical potential term $-\mu=-g+2J$ in the static limit $K\!\rightarrow\!0$. \\
		
		For the reasons detailed above, we have chosen to consider the convenient parametrization defined in Eq.~\eqref{para_app}, both in our numerical and analytical extraction of the instability rates; see also Eqs.~\eqref{eq:WCCA_EOM} for the extension to WCCA. We stress that the parametrization in Eq.~\eqref{para_app} should be revised for more elaborate models (multiband, multicell...), where $\rho_n(t)$ is not uniform, using the more general form~\eqref{para_app_full}. Finally, we emphasize that all parametrizations lead to the same physical results, and that choosing one or the other is just a matter of technical convenience.

		\section{Useful results on the parametric oscillator}
		\label{sec:anaPOrapp}
		
		The simplest parametric oscillator is the system described by the equation of motion
		\begin{equation}
			\ddot{x}(t)+\omega_0^2[1+\alpha \cos (\omega t)]x(t)=0,
			\label{eq:PO}
		\end{equation}
		describing a harmonic oscillator of pulsation $\omega_0$ perturbed by a sinusoidal modulation of pulsation $\omega$ and amplitude $\alpha\ll 1$. 
		Equivalently, one can introduce the standard operators $\gamma,\gamma^\dagger$ defined by $x=(\gamma+\gamma^\dagger)/\sqrt{2\omega_0}$ and $p=i\sqrt{\omega_0/2}(\gamma^\dagger-\gamma)$, and use the parametrization $\gamma(t)=\tilde{u}'(t)\gamma(t=0)-\tilde{v}'^*(t)\gamma^\dagger(t=0)$, to rewrite the equation of motion~(\ref{eq:PO}) in the following way (see~\cite{bukov2015} for details)
		\begin{equation}
			i \partial_t \left( \begin{matrix} \tilde{u}'(t)  \\ \tilde{v}'(t) \end{matrix} \right)=\left[ \omega_0 \hat{\mathbf{1}}+W(t)\hat{\sigma}_z+\dfrac{\alpha\omega_0}{2}\left( \begin{matrix} 0 & \cos(\omega t)e^{-2i\omega_0 t}        \\ -\cos(\omega t)e^{2i\omega_0 t} & 0 \end{matrix} \right)\right] \left( \begin{matrix} \tilde{u}'(t)  \\ \tilde{v}'(t) \end{matrix} \right),
			\label{eq:PO2}
		\end{equation}
		where $W(t)=\alpha\omega_0/2 \cos(\omega t)$.
		As mentioned above, the simplest way to identify the appearance of parametric resonance in this model is to perform a RWA treatment of Eq.~(\ref{eq:PO2}), which will yield a pronounced contribution if $\omega=2\omega_0$. Keeping only the resonant terms yields a time-independent 2x2 matrix in Eq.~(\ref{eq:PO2}) which can straightforwardly be diagonalized. Its eigenvalues exhibit then an imaginary part responsible for an exponential growth of the solution. The instability rate (here on resonance) $s^{(R)}$ is thus given by the imaginary part of the eigenvalues of this 2x2 matrix, yielding $s^{(R)}= \alpha\omega_0/4$. This corresponds to the rate \textit{on resonance}, i.e. precisely when $\omega=2\omega_0$ \footnote{There are also other resonances for $\omega=2\omega_0/n$ but they are of higher order, in the sense that both the magnitude of the instability rates and the width of the resonance domain scale as $\alpha^n$)}. 
		
		However, a key feature that will have a crucial importance in the following is that instability in fact arises not only when the resonance condition is precisely satisfied, but in a full range of parameters around the resonance point.
		To investigate more precisely how the rate behaves in the vicinity of the resonance, and at which point it will eventually drop to zero (which will give the limits of the instability domain), one can follow the general procedure indicated in Landau and Lifshits'~\cite{landau1969}.
		The idea is to write $\omega=2\omega_0 +\delta$ and to solve the problem perturbatively in the detuning $\delta$ and $\alpha$ \footnote{The idea is indeed that the n-th order correction in $\delta$ will scale as $\alpha^n$.}.
		At lowest order, we seek for a solution of Eq.~(\ref{eq:PO}) of the form $x(t)=A(t)\cos[(\omega_0+\delta/2)t]+B(t)\sin[(\omega_0+\delta/2)t]$, where $A(t)$ and $B(t)$ vary slowly compared to the oscillating terms. Inserting it in Eq.~(\ref{eq:PO}) and solving it to first order in $\delta$ eventually yields that $A(t)$ and $B(t)$ behave as $e^{st}$, with a rate $s$ given by
		\begin{equation}
			s=s^{(R)}\sqrt{1-(2\delta/\alpha\omega_0)^2}; \quad s^{(R)}= \alpha\omega_0/4.
			\label{eq:s1}
		\end{equation}
		At first order, the instability rate is thus maximal on resonance and decreases around it, and instability arises for all values $$\omega\in[2\omega_0-\alpha\omega_0/2;2\omega_0+\alpha\omega_0/2].$$
		This calculation can be extended to higher orders, by adding in the ansatz for the solution harmonics that will be of higher order in the detuning. For instance, at second order, the idea is to seek for a solution of the form  $$x(t)=A_0(t)\cos[(\omega_0+\delta/2)t]+B_0(t)\sin[(\omega_0+\delta/2)t]+A_1(t)\cos[3(\omega_0+\delta/2)t]+B_1(t)\sin[3(\omega_0+\delta/2)t]$$ and to solve Eq.~(\ref{eq:PO}) to second order in $\delta$ and $\alpha$. We find an instable regime for $$\omega\in[2\omega_0-\alpha\omega_0/2-\alpha^2\omega_0/32;2\omega_0+\alpha\omega_0/2-\alpha^2\omega_0/32],$$ while the instability rate $s$ is solution of the implicit equation 
		\begin{equation}
			\begin{vmatrix} s^2+\omega_0^2-\omega^2/4+\alpha\omega_0^2/2 & \omega s & \alpha\omega_0^2/2 & 0      \\ -\omega s & s^2+\omega_0^2-\omega^2/4-\alpha\omega_0^2/2 & 0 & \alpha\omega_0^2/2 \\ \alpha\omega_0^2/2 & 0& s^2-8\omega_0^2 & 3\omega s \\ 0 & \alpha\omega_0^2/2 & -3\omega s  & s^2-8\omega_0^2 \end{vmatrix}=0,
			\label{eq:det}
		\end{equation}
		which can be numerically solved. Similarily to the first order result, the instability rate continuously decreases from its maximal value to zero at the edges of the instability domain; yet, the latter is no longer centered on the resonance point but slightly shifted; therefore, the maximal instability rate, which is reached at the center of the instability domain, does not coincide with the resonance point (although it is not far from it).

		\section{The change of basis leading to Eq.~\eqref{eq:BdGEPO}}
		\label{sec:A1}
		
		In this Appendix, we describe the change of basis introduced in Eq.~\eqref{eq:BdGEPO}. Starting from the Bogoliubov equations written in Fourier space [Eq.~(\ref{eq:BdGEk_pre})] 
		\begin{equation}
			i \partial_t \left( \begin{matrix} u_q  \\ v_q \end{matrix} \right)=\left( \begin{matrix}  \varepsilon(q,t)+g & g        \\ -g & -\varepsilon(-q,t)-g \end{matrix} \right)\left( \begin{matrix} u_q  \\ v_q \end{matrix} \right),
			\label{eq:BdGEappen}
		\end{equation}
		where $$\varepsilon(q,t)=4J\sin\left(\dfrac{q}{2}\right)\sin\left(\dfrac{q}{2}+p_0-\dfrac{K}{\omega}\sin(\omega t)\right),$$
		one first introduces the change of basis which diagonalizes the time-averaged part of the equation,
		\begin{equation}
			\left( \begin{matrix} u_q  \\ v_q \end{matrix} \right)=\left( \begin{matrix} \mathrm{cosh}(\theta_q) & \mathrm{sinh}(\theta_q)        \\ \mathrm{sinh}(\theta_q) & \mathrm{cosh}(\theta_q) \end{matrix} \right)\left( \begin{matrix} u'_q  \\ v'_q \end{matrix} \right),
			\label{eq:change1}
		\end{equation}
		where $\mathrm{cosh}(2\theta_q)\equiv (4|\Jeff|\sin^2(q/2)+g)/\Eav(q)$ and $\mathrm{sinh}(2\theta_q)\equiv g/\Eav(q)$ with $$\Eav(q)=\sqrt{4|\Jeff|\sin^2(q/2)(4|\Jeff|\sin^2(q/2)+2g)}$$ the time-averaged Bogoliubov dispersion. Defining then a second transformation
		\begin{equation}
			\left( \begin{matrix} \tilde{u}'_q  \\ \tilde{v}'_q \end{matrix} \right)=\left( \begin{matrix} e^{2i\Eav(q)t} & 0       \\ 0 & 1 \end{matrix} \right)\left( \begin{matrix} u'_q  \\ v'_q \end{matrix} \right),
			\label{eq:change2}
		\end{equation}
		the Bogoliubov equations take the final form
		\begin{equation}
			i\partial_t \left( \begin{matrix} \tilde{u}'_q  \\ \tilde{v}'_q \end{matrix} \right)=\left[ \Eav(q)\hat{\mathbf{1}}+\hat{W}_q(t)+\mathrm{sinh}(2\theta_q)\left( \begin{matrix} 0 & h_q(t)e^{-2i\Eav(q)t}        \\ -h_q(t)e^{2i\Eav(q)t} & 0 \end{matrix} \right)\right] \left( \begin{matrix} \tilde{u}'_q  \\ \tilde{v}'_q \end{matrix} \right).
			\label{eq:BdGEPOappen}
		\end{equation}
where $\mathrm{sinh}(2\theta_q)= g/\Eav(q)$, $\Eav(q)$ is the Bogoliubov dispersion associated with the effective GPE, within the Bogoliubov approximation [see Eq.~(\ref{eq:Eav})], $W_q(t)$ is a diagonal matrix of zero average over one driving period,
		\begin{equation}
			W_q(t)=4J\mathrm{sin}(q/2)\sum_{l\neq 0}\mathcal{J}_l(K/\omega)\left( \begin{matrix} \mathrm{cosh}^2\theta_q\mathrm{sin}(\dfrac{q}{2}-l\omega t)+\mathrm{sinh}^2\theta_q\mathrm{sin}(\dfrac{q}{2}+l\omega t) & 0       \\ 0 & -\mathrm{sinh}^2\theta_q\mathrm{sin}(\dfrac{q}{2}-l\omega t)+\mathrm{cosh}^2\theta_q\mathrm{sin}(\dfrac{q}{2}+l\omega t) \end{matrix} \right),
			\label{eq:W}
		\end{equation}
 and $h_q(t)$ is a (real-valued) function which can be Fourier expanded as 
	\begin{align}
		h_q(t)&=\frac{J}{2}4\sin^2(q/2)\sum_{l=-\infty}^\infty [\mathcal{J}_l(K/\omega)+\mathcal{J}_{-l}(K/\omega)] \mathrm e^{il\omega t} \notag \\
		&=4J\sin^2(q/2)\sum_{l=-\infty}^\infty \mathcal{J}_{2l}(K/\omega) \mathrm e^{i2l\omega t},
		\label{eq:hq_appendix}
	\end{align}
	with $\mathcal{J}_l(z)$ the $l$-th Bessel function of the first kind.

		\section{Disagreement regions }
		\label{sec:A2}
		\begin{figure}[]
			\begin{center}
				\includegraphics[width=10cm]{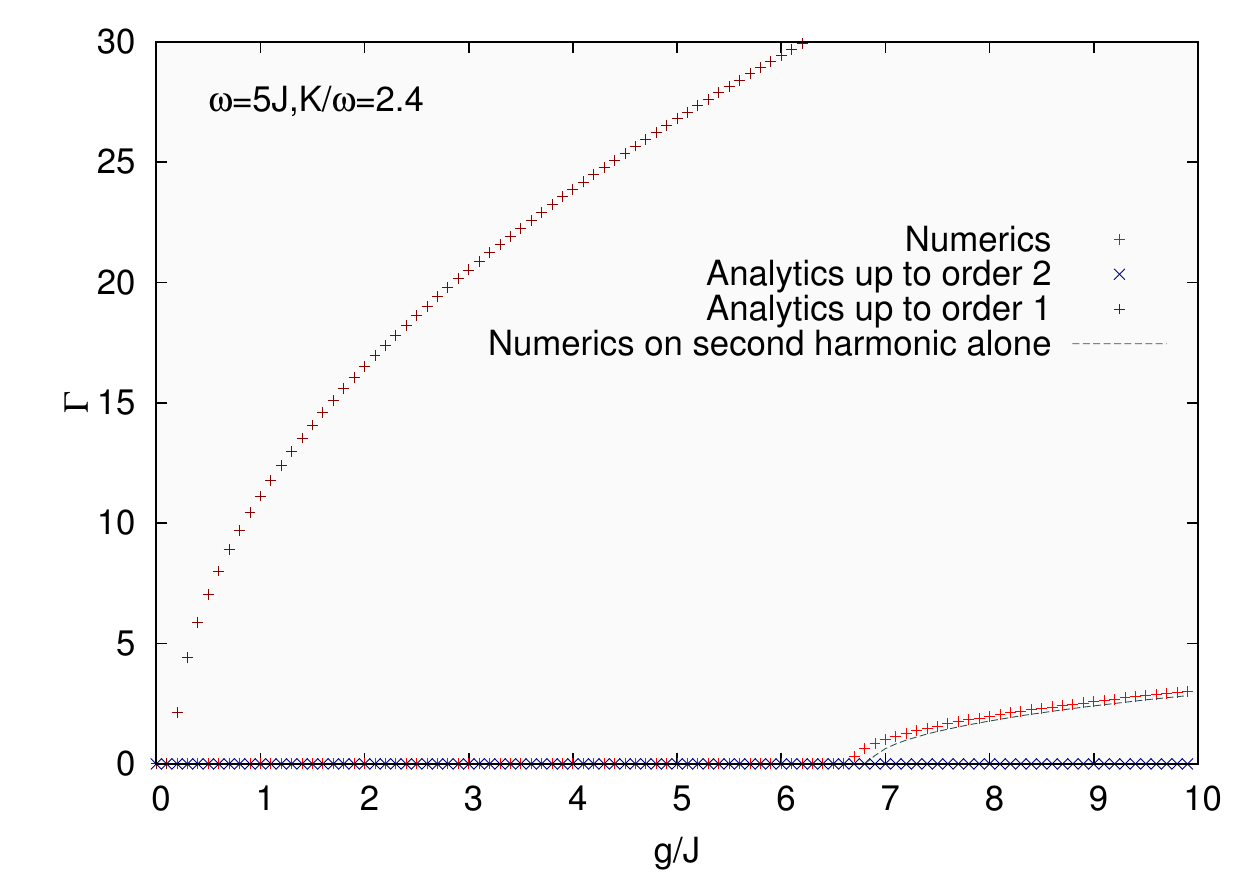}
			\end{center}
			\vspace{0.1cm}
			\caption{Comparison between the numerically and analytically computed instability rate $\Gamma$ in the case $K/\omega\!=\!2.4$ [zone C on the stability diagram of Fig.~\ref{fig:2Diags}, where $\mathcal{J}_0(K/\omega)\approx0$ and $\mathcal{J}_2(K/\omega)$ is large]. Since the effective Bogoliubov dispersion is flat, the analytical perturbative approach breaks down.
				\label{fig:2.4}
			}
		\end{figure}
		The analytical approach is based on a perturbative solution in the detuning and the amplitude of the perturbation, which is performed on an effective model restricted to the second harmonic [see Sect.~\ref{sec:anaEff}]. We observe that it breaks down in two main regions: \\ 
	\begin{enumerate}	
		\item The main discrepancy with the numerical results occurs around the first zero of $\mathcal{J}_0$ (the only one where $\mathcal{J}_2$ is not small as well, zones C on Fig.~\ref{fig:2Diags}), the analytics fails to capture the numerically observed instability, and we also notice a completely different behavior between the first order and second order analytics [see Fig.~\ref{fig:2.4}]. This is coming from the fact that in this case the effective dispersion $\Eav$ is flat, and the perturbative approach breaks down. More precisely, the first order analytical solution gives a resonance domain centered on $\omega=\Eav(q)\approx 0$ but of infinite width ($\propto \Eav(q)^{-1}$), displaying thus a strong instability. The second order correction displaces the center of the resonance domain as $\propto \Eav(q)^{-3}$, sweeping away all instability in the considered parameter range. More generally, the perturbative approach does not converge and cannot capture the observed behavior (which seems though to be due to the second harmonic alone).\\ 
		
		\item Slight discrepancies also occur ``between" the lobes of the stability diagram, as the numerics and the analytics disagree on the way those "black zones" close (or not) at large $g$. Firstly, near the zeros of $\mathcal{J}_0$ (i.e. at the left border of each lobe, zones D on Fig.~\ref{fig:2Diags}), the numerics displays a transition to instability which is not captured by the analytics (similarly to what was observed in case C). However, this failure of the analytics is less marked than in case C (in the sense that it appears for much larger $g$) for two main reasons: on the one hand, the fact that $\mathcal{J}_2(K/\omega)$ is also small can partly kill the divergences due to the vanishing of $\mathcal{J}_0(K/\omega)$; on the other hand, the stable regime extends here to much larger $g$ coherently with the fact that the second harmonic has a very small amplitude (indeed, as visible on Fig.~\ref{fig:defavo}(a), the numerical instability seems rather due to higher harmonics). More precisely, at points where $\mathcal{J}_2(K/\omega)=0$, the analytics, which is essentially built on a second-harmonic-based model, is always stable and cannot reproduce the numerical instabilities due to higher harmonics (which nevertheless show up only at large $g$). Near such points (zones E), the numerical instability seems due to a joined effect of the second and next harmonics, and is therefore not accurately captured by the analytical approach [see Fig.~\ref{fig:defavo}(b)].
		
		\begin{figure}[h!]
			
			\begin{minipage}{8cm}
				
				\includegraphics[width=8cm]{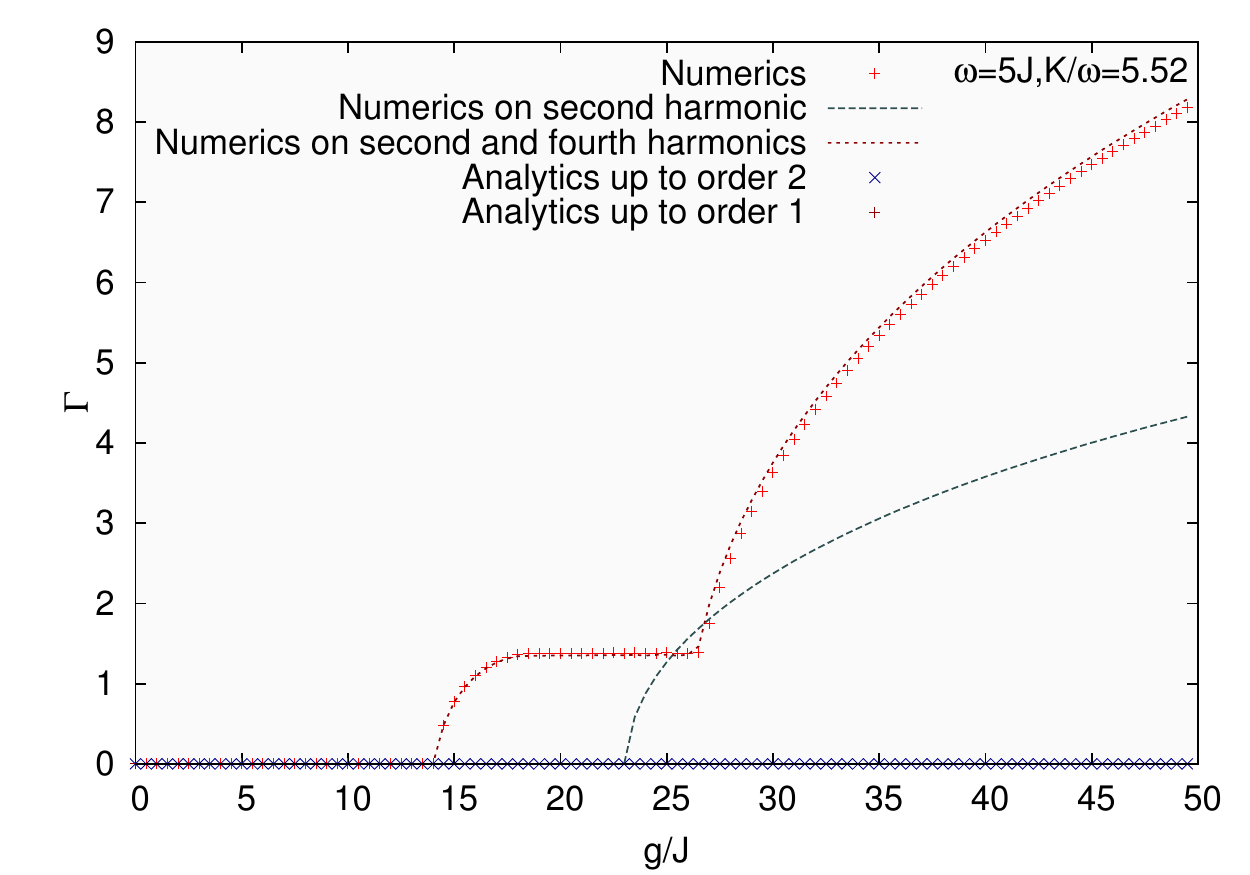}
				\centerline{(a)}
			\end{minipage}
			\begin{minipage}{8cm}
				
				\includegraphics[width=8cm]{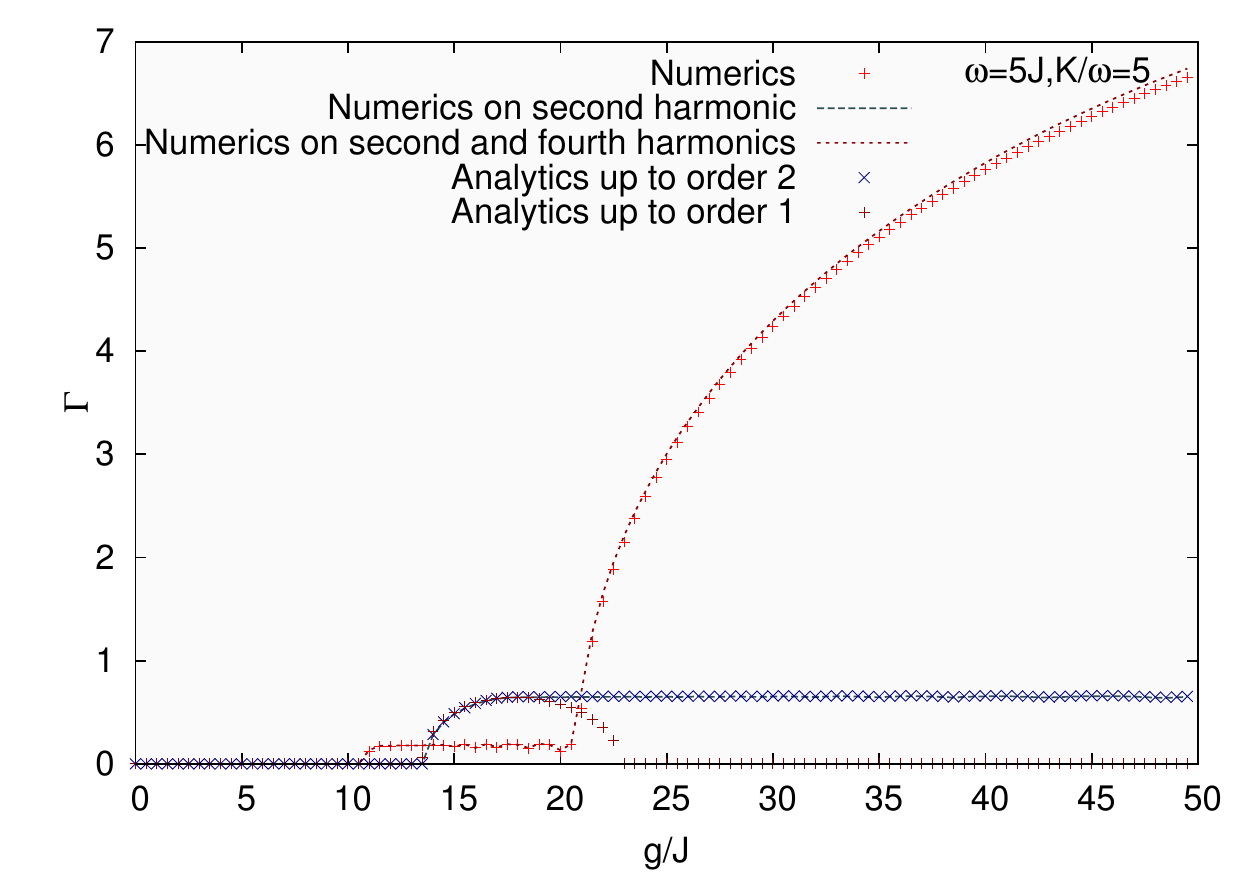}
				\centerline{(b)}
			\end{minipage}
			\vspace{0.2cm}
			\caption{Comparison between the numerically and analytically computed instability rate $\Gamma$ in the cases $K/\omega\!=\!5.52$ [(left), zone D of the stability diagram of Fig.~\ref{fig:2Diags}, where $\mathcal{J}_0(K/\omega)\approx0$] and $K/\omega\!=\!5$ [(right), zone E of the stability diagram of Fig.~\ref{fig:2Diags}, where $\mathcal{J}_2(K/\omega)\approx0$]. The disagreement in those cases can be due to a breakdown of the perturbative approach [when $\mathcal{J}_0(K/\omega)$ vanishes, (left)] and/or a vanishing of the second harmonic [when $\mathcal{J}_2(K/\omega)$ vanishes, (right)]. }
			\label{fig:defavo}
		\end{figure}
		
		\end{enumerate}
		


\end{document}